\documentclass[twocolumn,showpacs,preprintnumbers,amsmath,amssymb]{revtex4}

\usepackage{graphicx}
\usepackage{dcolumn}
\usepackage{bm}

\usepackage{amssymb}
\usepackage{amsmath}

\begin{document}


\title{Effective dissipative cosmology with $\Lambda$ from the first law of thermodynamics}

\author{Nobuyoshi {\sc Komatsu}}  \altaffiliation{E-mail: komatsu@se.kanazawa-u.ac.jp} 
\affiliation{Department of Mechanical Systems Engineering, Kanazawa University, Kakuma-machi, Kanazawa, Ishikawa 920-1192, Japan}

\date{\today}

\begin{abstract}
We phenomenologically derive a cosmological model that includes both a cosmological constant term $\Lambda/3$ and a dissipative driving term $\beta (2 H^{2} + \dot{H})$ by applying both the first law of thermodynamics and an effective entropy (that is proportional to the Bekenstein--Hawking entropy) to matter creation cosmology.
Here $H$, $\dot{H}$, and $\beta$ are the Hubble parameter, the time derivative of $H$, and a non-negative dimensionless coefficient used for the effective entropy, respectively.
The dissipative term is proportional to the Ricci scalar curvature, suggesting that the dynamic creation pressure has the same dependence.
We examine the model's background evolution in the late universe and its horizon thermodynamics.
The present model supports a transition from a decelerating universe to an accelerating universe when $\beta <0.5$. 
The second law of thermodynamics is always satisfied on the horizon, and maximization of entropy is satisfied in the final stage.
In addition, we study first-order density perturbations related to structure formation, by applying a neo-Newtonian approach to the present model.
We then examine constraints on the present model using three types of observational data and the transitional and thermodynamic constraints and find that a weakly dissipative universe with $\Lambda$ is likely favored and consistent with our Universe.
We also discuss irreversible entropy due to adiabatic particle creation, assuming a holographic-like matter creation cosmology.

\end{abstract}


\maketitle

\section{Introduction} 
\label{Introduction}

Lambda cold dark matter ($\Lambda$CDM) models, which assume a cosmological constant $\Lambda$ and dark energy, provide an elegant framework for describing an accelerated expansion of the late Universe \cite{PERL1998_Riess1998,Planck2018,Hubble2017,Riess20162019,Pantheon2018-2022,DES-SN5YR,DESI2024,H(z)57data,57dataOdin}.
However, $\Lambda$CDM models exhibit several theoretical problems \cite{Weinberg1989}.
To resolve those problems in standard cosmology, astrophysicists have proposed various cosmological models \cite{Bamba1,Frusciante1}, e.g., time-varying $\Lambda (t)$ cosmology \cite{FreeseOverduin,Sola_2009,Sola_2013,Valent2015,Sola2019,Sola_2020,Odintsov2025EPJC}, 
bulk viscous cosmology \cite{Weinberg0,Barrow1986,Lima1988,Brevik2002,Nojiri2005-Odintsov,Avelino2-Brevik,Floerchinger,Mathew2018,Yang2022,Paul2025}, 
creation of CDM (CCDM) models \cite{Prigogine_1988-1989,Lima1992-1992,Lima1996-2012,Lima2014b,Freaza2002,Others2001-2015,Sola2020,Singha2020,Pan2016,Jesus2017, Barrow2019Jesus2020,Lima2024,Lima_Newtonian_1997,Lima2011,Ramos_2014,Reis_2003},
and Ricci dark energy models \cite{Gao2009,Zhang2009,Cai2009,Richarte2011,Feng2009,Singh2018,Kumar2021,Wu2025}.
In addition, thermodynamic scenarios based on the holographic principle \cite{Hooft-Bousso} have been examined from various viewpoints, such as entropic cosmology \cite{EassonCai,Basilakos1,Basilakos2014,Koma45,Koma6,Koma7,Koma8,Koma9,Mohammadi2025}, holographic cosmology \cite{Padma2010,Padma2012AB,Cai2012,Mathew2022,Chen2022,Chen2024,Koma10,Koma11,Koma12,Koma18}, 
the first law of thermodynamics \cite{Cai2005,Dynamical-T-2007,Dynamical-T-20092014,Sheykhi1,Sheykhi2Karami,Santos2022,Sheykhia2018,ApparentHorizon2022,Cai2007,Cai2007B,Cai2008,Sanchez2023,Nojiri2024,Odintsov2023ab,Odintsov2024,Odintsov2025DU,Mohammadi2023,Koma21,Koma22}, 
the second law of thermodynamics, and maximization of entropy. \cite{Egan1,Bamba2018Pavon2019,Saridakis2021,Ashraf2025,Koma14,Koma15,Koma16,Koma17,Koma19,Koma20,Neto2022,Gohar2024}.
A notable example is the derivation of cosmological equations from a thermodynamic relation based on the first law of thermodynamics \cite{Cai2007,Cai2007B,Cai2008,Sanchez2023,Nojiri2024,Odintsov2023ab,Odintsov2024,Odintsov2025DU,Mohammadi2023,Koma21,Koma22}.

Formulations of cosmological models can be categorized into several types \cite{Koma14,Koma15,Koma16}.
From a dissipative viewpoint, the formulation of cosmological equations can be categorized into two types in a Friedmann--Lema\^{i}tre--Robertson--Walker (FLRW) universe \cite{Koma15}.
The first type is $\Lambda(t)$, which is similar to $\Lambda (t)$CDM models \cite{FreeseOverduin,Sola_2009,Sola_2013,Valent2015,Sola2019,Sola_2020,Odintsov2025EPJC}.
In $\Lambda(t)$ models, an extra driving term $f_{\Lambda}(t)$ is added to both the Friedmann equation and the Friedmann--Lema\^{i}tre acceleration equation \cite{Koma15}.
In this case, the right-hand side of the continuity equation is generally non-zero due to the time derivative of $f_{\Lambda}(t)$, unlike the standard continuity equation used for $\Lambda$CDM models.
The universe for a pure $\Lambda(t)$ model is non-dissipative because dissipative terms are not considered.
In the $\Lambda(t)$ model, a specific from of $f_{\Lambda}(t)$, which is composed of constant and small time-varying terms, is likely favored, see, e.g., Refs.\ \cite{Valent2015,Sola2019} and references therein.

The second type is BV, which is similar to both bulk viscous cosmology and matter creation cosmology \cite{Weinberg0,Barrow1986,Lima1988,Brevik2002,Nojiri2005-Odintsov,Avelino2-Brevik,Floerchinger,Mathew2018,Yang2022,Paul2025,Prigogine_1988-1989,Lima1992-1992,Lima1996-2012,Lima2014b,Freaza2002,Others2001-2015,Sola2020,Singha2020,Pan2016,Jesus2017,Barrow2019Jesus2020,Lima2024,Lima_Newtonian_1997,Lima2011,Ramos_2014}.
In bulk-viscous-cosmology-like (BV) models, the acceleration equation includes an extra driving term $h_{\textrm{B}}(t)$ related to dissipation, whereas the Friedmann equation does not \cite{Koma15}.
In this model, the right-hand side of the continuity equation is non-zero due to the dissipative driving term.
In fact, the background evolution of the universe in the $\Lambda(t)$ and BV models becomes the same when an equivalent driving term is assumed \cite{Koma15}.
However, the universe for the BV model is dissipative and, therefore, irreversible entropy is produced, unlike for a pure $\Lambda(t)$ model without $h_{\textrm{B}}(t)$.
In addition, density perturbations related to structure formation should be different in dissipative and non-dissipative universes, even if the background evolution of the two universes is the same \cite{Koma16}.
For example, a negative sound speed \cite{Lima2011} and the existence of clustered matter \cite{Ramos_2014} are necessary to properly describe the density perturbations in matter creation cosmology such as BV models without $f_{\Lambda}(t)$ \cite{Koma8,Koma16}.

Of course, we can consider BV models with $f_{\Lambda}(t)$, equivalent to $\Lambda(t)$ models with $h_{\textrm{B}}(t)$, which are similar to viscous dark energy models \cite{Feng2009,Singh2018,Kumar2021,Brevik2015,Nojiri-Odintsov2020,Silva2021,Silva2025}.
These dissipative models are expected to bridge the gap between observations and standard cosmology.
A previous work indicates that a constant $f_{\Lambda}(t)$ term plays important roles in the dissipative model \cite{Koma8}.
We have therefore studied BV models with constant $f_{\Lambda}(t)$
and found that the BV model with constant $f_{\Lambda}(t)$ can be derived from a thermodynamic relation based on the first law of thermodynamics using a general continuity equation, instead of the standard continuity equation.
The form of $h_{\textrm{B}}(t)$ derived is similar to the Ricci scalar curvature and is determined by the thermodynamic relation between quantities on a cosmological horizon and in the bulk, as will be examined in this paper.
This type of dissipative universe has not yet been examined from those viewpoints.
The present study should contribute to a better understanding of the dissipative universe, such as matter creation cosmology and bulk viscous cosmology.
(Note that a new mechanism for the origin of bulk viscosity has been recently discussed in Ref.\ \cite{Paul2025}.)

In the present study, we phenomenologically derive a modified thermodynamic relation using the first law of thermodynamics and a general formulation for cosmological equations.
Based on the thermodynamic relation, we formulate a BV model that includes both a constant term $f_{\Lambda}(t)$ and a dissipative term $h_{\textrm{B}}(t)$, and then formulate the present model by applying an effective entropy.
The background evolution of the late universe, first-order density perturbations, horizon thermodynamics, and constraints on the present model are examined to clarify the fundamental properties of the model.
(Inflation of the early universe is not discussed here.)

The remainder of the article is organized as follows.
In Sec.\ \ref{Matter creation and cosmological equations}, adiabatic particle creation and a general formulation for cosmological equations are reviewed.
In Sec.\ \ref{First law of thermodynamics and the present model}, a dissipative cosmological model is derived from the first law of thermodynamics.
For this, in Sec.\ \ref{Entropy and temperature}, entropy and temperature on the cosmological horizon are introduced.
In Sec.\ \ref{The first law}, a modified thermodynamic relation is derived using the first law and the general formulation for cosmological equations.
In Sec.\ \ref{BV models with constant fL}, a BV model with constant $f_{\Lambda}(t)$ is derived from the thermodynamic relation.
In addition, the present model is formulated by applying an effective entropy that is proportional to the Bekenstein--Hawking entropy.
%
In Sec.\ \ref{Background evolution of the universe}, the background evolution of the universe in the present model is examined.
In Sec.\ \ref{Entropy Sm}, irreversible entropy is discussed, assuming holographic-like matter creation cosmology.
In Sec.\ \ref{Entropy SBH on the Hubble horizon}, horizon thermodynamics is examined.
%
In Sec.\ \ref{First-order density perturbations in the present model}, first-order density perturbations are examined by applying a neo-Newtonian approach to the present model.
In Sec.\ \ref{Observational, transitional, and thermodynamic constraints}, constraints on the present model are discussed.
Finally, in Sec.\ \ref{Conclusions}, the conclusions of the study are presented.

\section{Matter creation and cosmological equations}
\label{Matter creation and cosmological equations}

A homogeneous, isotropic, and spatially flat universe, namely a flat FLRW universe, is considered in this paper.
Therefore, the apparent horizon of the universe is equivalent to the Hubble horizon.
An expanding universe is assumed from observations \cite{Hubble2017}.

We expect that matter (particle) creation cosmology should be related to cosmological equations based on the first law of thermodynamics.
Therefore, fundamental equations for adiabatic particle creation are reviewed in Sec.\ \ref{Adiabatic particle creation}.
A general formulation for cosmological equations is introduced in Sec.\ \ref{Cosmological equations}.
In addition, the relationship between the formulation and the fundamental equation for adiabatic particle creation is discussed.

Note that the first law of thermodynamics is not considered in this section.
Cosmological equations based on the first law are examined in Sec.\ \ref{First law of thermodynamics and the present model}.

\subsection{Adiabatic particle creation}
\label{Adiabatic particle creation}

Prigogine \textit{et al.} have proposed nonequilibrium thermodynamics of open systems and have examined the thermodynamics of cosmological matter (particle) creation \cite{Prigogine_1988-1989}. 
Based on this concept, matter creation cosmology such as CCDM models has been examined 
\cite{Lima1992-1992,Lima1996-2012,Freaza2002,Others2001-2015,Lima2014b,Sola2020,Singha2020,Pan2016,Jesus2017,Barrow2019Jesus2020,Lima2024,Lima_Newtonian_1997,Lima2011,Ramos_2014}.
In this section, fundamental equations for adiabatic particle creation are reviewed, in accordance with Ref.\ \cite{Lima2014b}.

We consider nonequilibrium thermodynamic states of cosmological fluids in a flat FLRW background \cite{Koma8,Koma15}, assuming adiabatic particle creation \cite{Lima1992-1992,Lima1996-2012,Lima2014b}.
The balance equations for the number of particles, entropy, and energy can be written as \cite{Lima2014b}
\begin{equation}
       \dot{n} + 3 H n   = n \Gamma   ,    
\label{eq:NonEquil_1}
\end{equation}
\begin{equation}
       \dot{s} + 3 H s   = s \Gamma   ,    
\label{eq:NonEquil_2}
\end{equation}
\begin{equation}
       \dot{\varepsilon} + 3 H ( \varepsilon + p + p_{c} )   = 0   ,
\label{eq:NonEquil_3}
\end{equation}
and the Hubble parameter $H$ is defined by
\begin{equation}
   H \equiv  \frac{ da/dt } {a(t)}  = \frac{ \dot{a}(t) } {a(t)}  ,
\label{eq:Hubble}
\end{equation}
where $a(t)$ is the scale factor at time $t$.
Also, $n$, $s$, $\varepsilon$, and $p$ are the particle number density, entropy density, energy density, and pressure, respectively. 
$\Gamma$ and $p_{c}$ are the particle production rate and the dynamic creation pressure, respectively.
The entropy considered in this section is irreversible entropy due to adiabatic particle creation.
(Horizon entropy is discussed in Sec.\ \ref{First law of thermodynamics and the present model}.)

The total number $N$ of particles and the entropy $S$ in the comoving volume is given by \cite{Lima2014b}
\begin{equation}
        N = n a^3   \quad \textrm{and} \quad  S = s a^3   .
\label{eq:NS-a3}
\end{equation}
Using $N = n a^3$, Eq.\ (\ref{eq:NonEquil_1}) can be written as 
\begin{equation}
\frac{\dot{N}}{N}  = \Gamma   .
\label{eq:N-Gamma}
\end{equation}
We assume that the entropy per particle $\sigma \equiv S/N$ is constant \cite{Lima1992-1992,Lima1996-2012,Lima2014b,Sola2020}:
\begin{equation}
    \sigma \equiv \frac{S}{N} = \textrm{cst.} \quad \textrm{or equivalently} \quad   \dot{\sigma} =0. 
\label{eq:sigma_cst}
\end{equation}
From Eqs.\ (\ref{eq:NS-a3}), (\ref{eq:N-Gamma}), and (\ref{eq:sigma_cst}), Eq.\ (\ref{eq:NonEquil_2}) can be written as 
\begin{equation}
\frac{\dot{S}}{S}  = \frac{\dot{N}}{N} + \frac{\dot{\sigma}}{\sigma}  = \Gamma+ \frac{\dot{\sigma}}{\sigma} =\Gamma   , 
\label{eq:S-Gamma}
\end{equation}
where $N \neq 0$ and $S \neq 0$ are assumed \cite{Sola2020}.
Note that the constant $\sigma$ has been used in Eq.\ (\ref{eq:NonEquil_2}).

We now discuss the balance equation for the energy density, given by Eq.\ (\ref{eq:NonEquil_3}).
For this, the local Gibbs relation is considered to be valid in the nonequilibrium thermodynamic states discussed here \cite{Lima2014b}.
The local Gibbs relation can be written as
\begin{equation}
         n k_{B} T d \left ( \frac{s}{n}    \right ) \equiv  n k_{B} T d\sigma  = d \varepsilon - \frac{\varepsilon + p}{n} dn      ,
\label{eq:localGibbs}
\end{equation}
where $k_{B}$ and $T$ are the Boltzmann constant and the temperature, respectively.
Substituting $\dot{\sigma} =d\sigma/dt =0$ into Eq.\ (\ref{eq:localGibbs}) and applying the resultant equation and Eq.\ (\ref{eq:NonEquil_1}) to Eq.\ (\ref{eq:NonEquil_3}) yields the dynamic creation pressure \cite{Koma8}:
\begin{equation}
        p_{c}  = - (\varepsilon + p) \frac{\Gamma}{3 H}  .    
\label{eq:pc}
\end{equation}
From this relation, Eq.\ (\ref{eq:NonEquil_3}) is written as
\begin{equation}
       \dot{\varepsilon} + 3 H ( \varepsilon + p  )   = (\varepsilon + p) \Gamma   .
\label{eq:NonEquil_3_sigma-cst}
\end{equation}
Substituting $p=0$ into the above equation and applying the mass density $\rho =\varepsilon /c^{2}$ yields
\begin{equation}
       \dot{\rho} + 3 H  \rho   = \rho  \Gamma   ,
\label{eq:NonEquil_3_sigma-cst_p=0_mass}
\end{equation}
where  $p=0$ is used for a matter-dominated universe and $c$ is the speed of light.
When BV models are considered, Eq.\ (\ref{eq:NonEquil_3_sigma-cst_p=0_mass}) should correspond to the continuity equation for cosmological equations.
To discuss this, a general formulation for cosmological equations is reviewed in the next subsection.

\subsection{General formulation for cosmological equations} 
\label{Cosmological equations}

We introduce a general formulation for cosmological equations used for various cosmological models, in accordance with Refs.\ \cite{Koma6,Koma9,Koma14,Koma15,Koma16,Koma21,Koma22}.
The formulation is independent of the first law of thermodynamics.

The general Friedmann, acceleration, and continuity equations are written as 
\begin{equation}
 H^2      =  \frac{ 8\pi G }{ 3 } \rho    + f_{\Lambda}(t)            ,                                                 
\label{eq:General_FRW01} 
\end{equation} 
\begin{align}
  \frac{ \ddot{a}}{ a}             &= -  \frac{ 4\pi G }{ 3 }  \left ( \rho +  \frac{3 p}{c^2} \right )                   +   f_{\Lambda}(t)    +  h_{\textrm{B}}(t)  , 
\label{eq:General_FRW02}
\end{align}
\begin{equation}
       \dot{\rho} + 3  H \left ( \rho +  \frac{p}{c^2} \right )       =    -  \frac{3}{8 \pi G }   \dot{f}_{\Lambda}(t)      +    \frac{3 }{4 \pi G}     H h_{\textrm{B}}(t)              , 
\label{eq:drho_General}
\end{equation}
and, in addition, subtracting Eq.\ (\ref{eq:General_FRW01}) from Eq.\ (\ref{eq:General_FRW02}) yields 
\begin{equation}
    \dot{H} = - 4\pi G  \left ( \rho +  \frac{p}{c^2} \right )  + h_{\textrm{B}}(t)   .
\label{eq:dotH}
\end{equation}
Here $G$ is the gravitational constant.
Two extra driving terms, $f_{\Lambda}(t)$ and $h_{\textrm{B}}(t)$, are phenomenologically assumed \cite{Koma14}.
Specifically, $f_{\Lambda}(t)$ is used for a $\Lambda (t)$ model, similar to $\Lambda(t)$CDM models, whereas $h_{\textrm{B}}(t)$ is used for a BV model, similar to bulk viscous models and CCDM models \cite{Koma14,Koma16}.
The general continuity equation given by Eq. (\ref{eq:drho_General}) can be derived from Eqs.\ (\ref{eq:General_FRW01}) and (\ref{eq:General_FRW02}), because only two of these three equations are independent \cite{Ryden1}.

Equation\ (\ref{eq:drho_General}) indicates that the right-hand side of the general continuity equation is non-zero, unlike for the standard continuity equation, which is given by $\dot{\rho} + 3  H [ \rho +  (p/c^2)] =0$.
A similar non-zero term appears in various cosmological models, such as energy exchange cosmology \cite{Barrow22,Dynamical20052013}, bulk viscous cosmology, and matter creation cosmology  
\cite{Weinberg0,Barrow1986,Lima1988,Brevik2002,Nojiri2005-Odintsov,Avelino2-Brevik,Floerchinger,Mathew2018,Yang2022,Paul2025,Prigogine_1988-1989,Lima1992-1992,Lima1996-2012,Lima2014b,Freaza2002,Others2001-2015,Sola2020,Singha2020,Pan2016,Jesus2017, Barrow2019Jesus2020,Lima2024,Lima_Newtonian_1997,Lima2011,Ramos_2014}.

When both $f_{\Lambda} (t) = \Lambda / 3$ and $ h_{\textrm{B}} (t) \neq 0$ are considered, the general continuity equation reduces to 
\begin{equation}
       \dot{\rho} + 3  H \left ( \rho +  \frac{p}{c^2} \right )       =     \frac{3 }{4 \pi G}     H h_{\textrm{B}}(t)              .
\label{eq:drho_BV_fL-cst}
\end{equation}
This is the continuity equation for BV models with constant $f_{\Lambda}(t)$.
From Eqs.\ (\ref{eq:NonEquil_3_sigma-cst_p=0_mass}) and (\ref{eq:drho_BV_fL-cst}), the particle production rate $\Gamma$ is written as \cite{Koma15}
\begin{equation}
       \Gamma  =    \frac{3 H }{4 \pi G}    \frac{ h_{\textrm{B}}(t) }{ \rho }                , 
\label{eq:Gamma_w=0_hb_0}
\end{equation}
where a matter-dominated universe ($p=0$) is considered.
Substituting Eq.\ (\ref{eq:Gamma_w=0_hb_0}), $p=0$, and $\varepsilon = \rho c^{2}$ into Eq.\ (\ref{eq:pc}) yields the dynamic creation pressure, given by 
\begin{equation}
        p_{c} =   - \frac{c^2 h_{\textrm{B}}(t) }{4 \pi G}    .
\label{eq:pc_w=0_hb}
\end{equation}
These two equations indicate that $\Gamma$ and $p_{c}$ depend on the dissipative driving term $h_{\textrm{B}}(t)$.
In general, we select forms of $\Gamma$, $p_{c}$, and $h_{\textrm{B}}(t)$, and then discuss a dissipative universe.
However, in the present study, $h_{\textrm{B}}(t)$ is derived from the first law of thermodynamics, without selecting these forms.
To derive $h_{\textrm{B}}(t)$, we review the first law of thermodynamics in the next section.

Before proceeding further, a cosmological equation is derived.
Coupling Eq.\ (\ref{eq:General_FRW01}) with Eq.\ (\ref{eq:General_FRW02}) yields the cosmological equation \cite{Koma14,Koma15,Koma16}:
\begin{equation}
    \dot{H} = - \frac{3}{2} (1+w)  H^{2}  +  \frac{3}{2}   (1+w)  f_{\Lambda}(t)     + h_{\textrm{B}}(t)   ,
\label{eq:Back2w}
\end{equation}
where $w$ represents the equation for the state parameter for a generic component of matter, given as $w = p/(\rho  c^2)$.
For a matter-dominated universe and a radiation-dominated universe, the values of $w$ are $0$ and $1/3$, respectively.
We consider a matter-dominated universe, i.e., $w =0$, although $w$ is retained for generality.

\section{Dissipative cosmology with $\Lambda$ from the first law of thermodynamics}
\label{First law of thermodynamics and the present model}

In this section, cosmological equations are phenomenologically derived from the first law of thermodynamics.
For this, in Sec.\ \ref{Entropy and temperature}, entropy and temperature on the Hubble horizon are introduced.
In Sec.\ \ref{The first law}, a modified thermodynamic relation is derived using the first law of thermodynamics and the general formulation for cosmological equations.
In Sec.\ \ref{BV models with constant fL}, a BV model with constant $f_{\Lambda}(t)$ is derived from the thermodynamic relation.
In addition, the present model is formulated by applying an effective entropy that is proportional to the Bekenstein--Hawking entropy.
(Note that adiabatic particle creation is not used for the derivation.)

\subsection{Entropy and temperature on the horizon} 
\label{Entropy and temperature}

In thermodynamic scenarios, a cosmological horizon is assumed to have an associated entropy and an approximate temperature \cite{EassonCai}.
The entropy $S_{H}$ and the temperature $T_{H}$ on the horizon are introduced in accordance with previous works \cite{Koma17,Koma18,Koma19,Koma20,Koma21,Koma22}.
In general, the cosmological horizon is examined by replacing the event horizon of a black hole by the cosmological horizon \cite{Koma17,Koma18,Koma19,Koma20,Koma21,Koma22}. 
This replacement method has been widely accepted \cite{Padma2010,Padma2012AB,Cai2012,Cai2007,Cai2007B,Cai2008,Sanchez2023,Nojiri2024,Odintsov2023ab,Odintsov2024} and we use it here.

First, we review a form of the Bekenstein--Hawking entropy as an associated entropy on the Hubble horizon.
Based on the form, the Bekenstein--Hawking entropy $S_{\rm{BH}}$ is written as \cite{Hawking1,Bekenstein1}  
\begin{equation}
S_{\rm{BH}}  = \frac{ k_{B} c^3 }{  \hbar G }  \frac{A_{H}}{4}   ,
\label{eq:SBH}
\end{equation}
where $\hbar$ is the reduced Planck constant.
The reduced Planck constant is defined by $\hbar \equiv h/(2 \pi)$, where $h$ is the Planck constant \cite{Koma11,Koma12}.
$A_{H}$ is the surface area of the sphere with a Hubble horizon (radius) $r_{H}$ given by
\begin{equation}
     r_{H} = \frac{c}{H}   .
\label{eq:rH}
\end{equation}
Substituting $A_{H}=4 \pi r_{H}^2 $ into Eq.\ (\ref{eq:SBH}) and applying Eq.\ (\ref{eq:rH}) yields
\begin{equation}
S_{\rm{BH}}  = \frac{ k_{B} c^3 }{  \hbar G }   \frac{A_{H}}{4}       
                  =  \left ( \frac{ \pi k_{B} c^5 }{ \hbar G } \right )  \frac{1}{H^2}  
                  =    \frac{K}{H^2}    , 
\label{eq:SBH2}      
\end{equation}
where $K$ is a positive constant given by
\begin{equation}
  K =  \frac{  \pi  k_{B}  c^5 }{ \hbar G } . 
\label{eq:K-def}
\end{equation}
Differentiating Eq.\ (\ref{eq:SBH2}) with regard to $t$ yields \cite{Koma11,Koma12}
\begin{equation}
\dot{S}_{\rm{BH}}  
                          = \frac{d}{dt}   \left ( \frac{K}{H^{2}} \right )  =  \frac{-2K \dot{H} }{H^{3}}                  .
\label{eq:dSBH}      
\end{equation}
Cosmological observations indicate $H >0$ and $\dot{H} < 0$ (see, e.g., Ref.\ \cite{Hubble2017}) and, therefore, $\dot{S}_{\rm{BH}}$ should be positive in our Universe \cite{Koma11,Koma12}.

Various forms of horizon entropy \cite{Das2008,Radicella2010,Tsallis2012,Czinner1Czinner2,Barrow2020,Nojiri2022} have been proposed and these entropies can be interpreted as an extended version of $S_{\rm{BH}}$.
For generality, we consider an arbitrary form of entropy $S_{H}$ on the Hubble horizon.
The horizon entropy $S_{H}$ can be written as \cite{Koma22}
\begin{align}
S_{H} = S_{\rm{BH}} +  S_{\Delta}   ,
\label{eq:SH_SBH_dS}      
\end{align}
where $S_{\Delta}$ represents a deviation from $S_{\rm{BH}}$.
Using this equation, $(\partial {S}_{H}/\partial S_{\rm{BH}})$ can be written as \cite{Koma22}
\begin{align}
\left (  \frac{\partial {S}_{H}}{\partial S_{\rm{BH}}} \right ) = 1 + \left ( \frac{\partial S_{\Delta} }{\partial S_{\rm{BH}}} \right )  ,
\label{eq:dSHdSBH_1-dS}      
\end{align}
where $(\partial {S}_{H}/\partial S_{\rm{BH}}) = \dot{S}_{H} / \dot{S}_{\rm{BH}}$. Also, $\dot{S}_{H}$ is given by
\begin{align}
              \dot{S}_{H}     &= \dot{S}_{\rm{BH}}     \left [ 1 + \left (   \frac{\partial S_{\Delta} }{\partial S_{\rm{BH}}} \right ) \right ]  .
\label{eq:dSH_dSBH_DeltaS}      
\end{align}
In this study, we use a symbol with brackets, namely $(\partial S_{H}/\partial S_{\rm{BH}})$, according to Ref.\ \cite{Odintsov2024}.
When $S_{H} = S_{\rm{BH}}$, $(\partial S_{\Delta} / \partial S_{\rm{BH}})$ reduces to $0$.
We can calculate $(\partial S_{\Delta} / \partial S_{\rm{BH}})$ from various entropies on the horizon.
For example, when an effective entropy is given by $S_{H} =S_{\rm{BH}}(1-\beta)$, $(\partial S_{\Delta} / \partial S_{\rm{BH}})$ reduces to $-\beta$, where $\beta$ is a dimensionless constant.
The effective entropy and the constant $(\partial S_{\Delta} / \partial S_{\rm{BH}})$ are discussed in Sec.\ \ref{BV models with constant fL}.

Next, we introduce an approximate temperature $T_{H}$ on the Hubble horizon \cite{Koma21,Koma22}.
For this, we review the dynamical Kodama--Hayward temperature \cite{Dynamical-T-1998,Dynamical-T-2008}, which is interpreted as an extended version of the Gibbons--Hawking temperature given by \cite{GibbonsHawking1977}
\begin{equation}
T_{\rm{GH}}  = \frac{ \hbar H}{   2 \pi  k_{B}  }   .
\label{eq:T_GH}
\end{equation}
The Kodama--Hayward temperature $T_{\rm{KH}}$ for a flat FLRW universe can be written as 
\begin{equation}
 T_{\rm{KH}} = \frac{ \hbar H}{   2 \pi  k_{B}  }  \left ( 1 + \frac{ \dot{H} }{ 2 H^{2} }\right )  .
\label{eq:T_KH}
\end{equation}
Here $H>0$ and $2 H^{2} +\dot{H} \ge 0$ are assumed for a non-negative temperature in an expanding universe \cite{Koma19,Koma20,Koma21,Koma22}.
In the present paper, $T_{\rm{KH}}$ is used for the temperature $T_{H}$ on the horizon \cite{Cai2007,Cai2007B,Cai2008,Sanchez2023,Nojiri2024,Odintsov2023ab,Odintsov2024}.

Note that, in a flat FLRW universe, the Ricci scalar curvature is proportional to $2 H^{2} + \dot{H}$ and should be related to $T_{\rm{KH}}$.
The Ricci scalar curvature is also used for Ricci dark energy models \cite{Gao2009,Zhang2009,Cai2009,Richarte2011,Feng2009,Singh2018,Kumar2021,Wu2025}.  

In addition, we calculate $T_{\rm{GH}}  \dot{S}_{\rm{BH}}$ and $T_{\rm{KH}}  \dot{S}_{\rm{BH}}$ to examine a modified thermodynamic relation.
Substituting Eqs.\ (\ref{eq:T_GH}) and (\ref{eq:dSBH}) into $T_{\rm{GH}}  \dot{S}_{\rm{BH}}$ yields \cite{Koma21}
\begin{align}
T_{\rm{GH}}  \dot{S}_{\rm{BH}}  
                                             &=  \frac{ \hbar H}{   2 \pi  k_{B}  }   \left ( \frac{-2 \left ( \frac{  \pi  k_{B}  c^5 }{ \hbar G } \right ) \dot{H} }{H^{3}}   \right )        
                                               =  \left ( \frac{c^5 }{G} \right ) \left ( \frac{ - \dot{H} }{H^{2}} \right )     ,
\label{eq:TGH-dSBH}      
\end{align}
using $ K = \pi k_{B} c^5 /(\hbar G)$ given by Eq.\ (\ref{eq:K-def}).
From Eqs.\ (\ref{eq:T_GH})--(\ref{eq:TGH-dSBH}), $T_{\rm{KH}}  \dot{S}_{\rm{BH}}$ is given by 
\begin{align}
T_{\rm{KH}}  \dot{S}_{\rm{BH}}  
                                            &=  \left ( \frac{c^5 }{G} \right ) \left ( \frac{ - \dot{H} }{H^{2}} \right )  \left ( 1 + \frac{ \dot{H} }{ 2 H^{2} } \right )     .
\label{eq:TKH-dSBH}       
\end{align}
The obtained $T_{\rm{GH}}  \dot{S}_{\rm{BH}}$ and $T_{\rm{KH}}  \dot{S}_{\rm{BH}}$ are used later.

\subsection{A modified thermodynamic relation} 
\label{The first law}

We phenomenologically derive a modified thermodynamic relation using the first law of thermodynamics and the general formulation for cosmological equations.
The first law of thermodynamics, reviewed in Refs.\ \cite{Cai2007,Cai2007B,Cai2008,Sanchez2023,Nojiri2024,Odintsov2023ab,Odintsov2024,Koma21,Koma22},
can be written as
\begin{align}
-dE_{\rm{bulk}}    + W dV  &  = T_{H} dS_{H}  ,
\label{eq:1stLaw}      
\end{align}
where $E_{\rm{bulk}}$ is the total internal energy of the matter fields inside the horizon, given by
\begin{align}
E_{\rm{bulk}}  &= \rho c^{2} V .
\label{Ebulk}
\end{align}
$W$ represents the density of work done by the matter fields \cite{Nojiri2024} and is written as
\begin{align}
W  &= \frac{\rho c^{2} - p }{2}   ,
\label{eq:W1}      
\end{align}
and $V$ is the Hubble volume, written as
\begin{equation}
V = \frac{4 \pi}{3} r_{H}^{3} =  \frac{4 \pi}{3} \left ( \frac{c}{H} \right )^{3}   ,
\label{eq:V}
\end{equation}
using $r_{H} = c/H$ given by Eq.\ (\ref{eq:rH}).
In addition, Eq.\ (\ref{eq:1stLaw}) can be written as 
\begin{align}
-\dot{E}_{\rm{bulk}}    + W \dot{V}  &=  T_{H} \dot{S}_{H}  =  T_{\rm{KH}} \dot{S}_{H}   ,
\label{eq:1stLaw_dt1}      
\end{align}
using $T_{H} = T_{\rm{KH}}$.
The right-hand side of Eq.\ (\ref{eq:1stLaw_dt1}) corresponds to thermodynamic quantities on the horizon, whereas the left-hand side corresponds to those in the bulk.

We first calculate the left-hand side of Eq.\ (\ref{eq:1stLaw_dt1}).
Substituting Eqs.\ (\ref{Ebulk}) and  (\ref{eq:W1}) into $-\dot{E}_{\rm{bulk}}    + W \dot{V}$ yields \cite{Nojiri2024}
\begin{align}
-\dot{E}_{\rm{bulk}}    + W \dot{V}  &= - \frac{d (\rho c^{2} V)}{dt}                    + \left ( \frac{\rho c^{2} - p }{2} \right )  \dot{V}    \notag \\
                                                   &= -\dot{\rho} c^{2} V  - \left ( \frac{\rho c^{2} + p }{2} \right )  \dot{V}    .
\label{eq:Left1stLaw}      
\end{align}
In addition, substituting Eqs.\ (\ref{eq:drho_General}) and (\ref{eq:dotH}) into Eq.\ (\ref{eq:Left1stLaw}), applying $V= (4 \pi/3)(c/H)^{3}$ given by Eq.\ (\ref{eq:V}) and $\dot{V}= -4 \pi c^{3} H^{-4} \dot{H}$, and performing several operations yields \cite{Koma21}
\begin{align}
      &    -\dot{E}_{\rm{bulk}}    + W \dot{V}   =  -\dot{\rho} c^{2} V  - \left ( \frac{\rho c^{2} + p }{2} \right )  \dot{V}                                                                                                              \notag \\   
                                                         &=   \left ( \frac{c^5 }{G} \right ) \left ( \frac{ - \dot{H} }{H^{2}} \right )  \left ( 1 + \frac{ \dot{H} }{ 2 H^{2} } \right )                                                   \notag \\      
                                                         &   \quad    + \frac{1}{2} \left ( \frac{c^5 }{G} \right ) \left ( \frac{ \dot{f}_{\Lambda}(t)  }{H^{3}}  + \frac{ \dot{H} h_{\textrm{B}}(t) }{ H^{4} } \right )     \notag \\                   
                                                         &=   T_{\rm{KH}}  \dot{S}_{\rm{BH}}                 
                                                             + T_{\rm{GH}}  \dot{S}_{\rm{BH}}     \left ( - \frac{ \dot{f}_{\Lambda}(t)  }{2 H \dot{H}}  - \frac{ h_{\textrm{B}}(t) }{ 2 H^{2} } \right )     ,   
\label{eq:Left1stLaw_2}      
\end{align}
where $T_{\rm{GH}}  \dot{S}_{\rm{BH}}$ given by Eq.\ (\ref{eq:TGH-dSBH}) and $T_{\rm{KH}}  \dot{S}_{\rm{BH}}$ given by Eq.\ (\ref{eq:TKH-dSBH}) are also used.
Equation\ (\ref{eq:Left1stLaw_2}) corresponds to the left-hand side of Eq.\ (\ref{eq:1stLaw_dt1}).
Therefore, from Eq.\ (\ref{eq:Left1stLaw_2}), the first law based on Eq.\ (\ref{eq:1stLaw_dt1}) can be written as \cite{Koma21}
\begin{align}
     -\dot{E}_{\rm{bulk}}    + W \dot{V}                
                                                        &=   T_{\rm{KH}}  \dot{S}_{\rm{BH}}                 
                                                             + T_{\rm{GH}}  \dot{S}_{\rm{BH}}     \left ( - \frac{ \dot{f}_{\Lambda}(t)  }{2 H \dot{H}}  - \frac{ h_{\textrm{B}}(t) }{ 2 H^{2} } \right )               \notag \\
                                                       &= T_{\rm{KH}}     \dot{S}_{H}                                                                                                                                                  ,
\label{eq:1stLaw_general}      
\end{align}
where $S_{H}$ is a type of the effective entropy examined in, e.g., Ref.\ \cite{Cai2007B}.
Equation\ (\ref{eq:1stLaw_general}) is a modified thermodynamic relation between quantities on the horizon and in the bulk.
Substituting Eq.\ (\ref{eq:dSH_dSBH_DeltaS}) into Eq.\ (\ref{eq:1stLaw_general}) yields 
\begin{align}
     -\dot{E}_{\rm{bulk}}    + W \dot{V}                
                                                        &=   T_{\rm{KH}}  \dot{S}_{\rm{BH}}                 
                                                             + T_{\rm{GH}}  \dot{S}_{\rm{BH}}     \left ( - \frac{ \dot{f}_{\Lambda}(t)  }{2 H \dot{H}}  - \frac{ h_{\textrm{B}}(t) }{ 2 H^{2} } \right )               \notag \\
                                                        &= T_{\rm{KH}}  \dot{S}_{\rm{BH}}     \left [ 1 + \left (   \frac{\partial S_{\Delta} }{\partial S_{\rm{BH}}} \right ) \right ]  ,
\label{eq:1stLaw_general_DeltaS}      
\end{align}
where $ S_{\Delta}= S_{H} - S_{\rm{BH}}$ is given by Eq.\ (\ref{eq:SH_SBH_dS}).
The additional $\dot{f}_{\Lambda} (t)$ and $h_{\textrm{B}}(t)$ terms in Eqs.\ (\ref{eq:1stLaw_general}) and (\ref{eq:1stLaw_general_DeltaS}) are based on two driving terms included in the general cosmological equations.
From Eq.\ (\ref{eq:1stLaw_general_DeltaS}), we can discuss the relationship among the two terms and $( \partial S_{\Delta} / \partial S_{\rm{BH}})$.
Using Eq.\ (\ref{eq:1stLaw_general_DeltaS}), the relation is written as
\begin{align}
                            T_{\rm{GH}}  \dot{S}_{\rm{BH}}   \left ( - \frac{ \dot{f}_{\Lambda}(t)  }{2 H \dot{H}}  - \frac{ h_{\textrm{B}}(t) }{ 2 H^{2} } \right )   &= T_{\rm{KH}}  \dot{S}_{\rm{BH}}   \left (   \frac{\partial S_{\Delta} }{\partial S_{\rm{BH}}} \right )  , 
\label{eq:1stLaw_general_fL_hB_DeltaS_0}      
\end{align}
or equivalently, 
\begin{align}
                     - \frac{ \dot{f}_{\Lambda}(t)  }{2 H \dot{H}}  - \frac{ h_{\textrm{B}}(t) }{ 2 H^{2} }  &= \frac{ T_{\rm{KH}}}{T_{\rm{GH}} }   \left (   \frac{\partial S_{\Delta} }{\partial S_{\rm{BH}}} \right )   \notag \\
                                                                                                                                           &= \left (1+  \frac{ \dot{H} }{ 2 H^{2} } \right )  \left (   \frac{\partial S_{\Delta} }{\partial S_{\rm{BH}}} \right )   , 
\label{eq:1stLaw_general_fL_hB_DeltaS}      
\end{align}
using $T_{\rm{GH}}$ given by Eq.\ (\ref{eq:T_GH}) and $T_{\rm{KH}}$ given by Eq.\ (\ref{eq:T_KH}).

As examined in Refs.\ \cite{Cai2007,Cai2007B,Cai2008,Sanchez2023,Nojiri2024,Odintsov2023ab,Odintsov2024,Koma21,Koma22}, the first law of thermodynamics can phenomenologically lead to cosmological equations with a constant term $ f_{\Lambda}(t)=\Lambda/3$.
This constant term does not affect the continuity equation given by Eq.\ (\ref{eq:drho_General}), because $\dot{f}_{\Lambda}(t)=0$.
Hereafter, we consider a BV model with constant $f_{\Lambda}(t)$, to examine a dissipative universe.
In the next subsection, we derive a dissipative driving term $h_{\textrm{B}}(t)$ for this model, which is a kind of dissipative $\Lambda$CDM model, namely dissipative cosmology with $\Lambda$.

It should be noted that when $h_{\textrm{B}}(t)     =   \dot{f}_{\Lambda}(t) /(2H)$, Eq.\ (\ref{eq:drho_General}) reduces to the standard continuity equation $\dot{\rho} + 3  H [ \rho +  (p/c^2)] =0$.
This case has been examined in the previous works and the result is summarized in, e.g., Ref.\ \cite{Koma22}.

Note that we call Eq.\ (\ref{eq:1stLaw_general}) a `modified thermodynamic relation', because Eq.\ (\ref{eq:1stLaw_general}) has not yet been established \cite{Koma21}.
The modified thermodynamic relation is phenomenologically derived from the first law of thermodynamics.
The two extra driving terms included in the general cosmological equations, namely $f_{\Lambda} (t)$ and $h_{\textrm{B}}(t)$, are phenomenologically applied to the first law.
In addition, $f_{\Lambda} (t)$ and $h_{\textrm{B}}(t)$ are assumed to be related to reversible exchange of energy and irreversible dissipative processes, respectively.
Nonequilibrium entropies are not considered here.
Consequently, the complexities of the nonequilibrium thermodynamics should be phenomenologically described by these two terms.
In this sense, Eq.\ (\ref{eq:1stLaw_general}) may not truly explain the fundamental physical principles of energy exchange.
In the present paper, accepting these uncertainties, we assume the modified thermodynamic relation and examine a BV model with constant $f_{\Lambda}(t)$.

\subsection{The present model} 
\label{BV models with constant fL}

We consider a BV model with constant $f_{\Lambda}(t)$.
The constant $f_{\Lambda}(t)$ is assumed to be given by
\begin{align}
                        f_{\Lambda}(t)  &=  \frac{\Lambda}{3}  .
\label{eq:fL(t)_1stLaw_zero}      
\end{align}
When $f_{\Lambda}(t)$ is constant, Eq.\ (\ref{eq:1stLaw_general_fL_hB_DeltaS}) is written as
\begin{align}
                      - \frac{ h_{\textrm{B}}(t) }{ 2 H^{2} }  &= \left (1+  \frac{ \dot{H} }{ 2 H^{2} } \right )  \left (   \frac{\partial S_{\Delta} }{\partial S_{\rm{BH}}} \right )   , 
\label{eq:1stLaw_fL=cst_hB_DeltaS}      
\end{align}
and the dissipative driving term $h_{\textrm{B}}(t)$ is given by
\begin{align}
                      h_{\textrm{B}}(t)   &=  - (2 H^{2} + \dot{H})  \left ( \frac{\partial S_{\Delta} }{\partial S_{\rm{BH}}} \right )   ,
\label{eq:hB_1stLaw_fL=cst}      
\end{align}
where $ S_{\Delta}= S_{H} - S_{\rm{BH}}$ from Eq.\ (\ref{eq:SH_SBH_dS}).
As discussed in the previous section, $2 H^{2} + \dot{H} \ge 0$ is assumed for a non-negative temperature in an expanding universe.
Therefore, $(2 H^{2} + \dot{H})$ is non-negative and should be related to the Kodama--Hayward temperature $T_{\rm{KH}}$.

The dissipative driving term $h_{\textrm{B}}(t)$ given by Eq.\ (\ref{eq:hB_1stLaw_fL=cst}) includes $(\partial S_{\Delta} / \partial S_{\rm{BH}})$.
As described in Sec.\ \ref{Entropy and temperature}, an effective entropy proportional to the Bekenstein--Hawking entropy, $S_{H} =S_{\rm{BH}}(1-\beta)$, leads to a constant $(\partial S_{\Delta} / \partial S_{\rm{BH}})$, namely $(\partial S_{\Delta} / \partial S_{\rm{BH}}) = -\beta$.
Here $\beta$ is a dimensionless non-negative constant. 
When $\beta=0$ is considered, $h_{\textrm{B}}(t)$ is reduced to $0$ from Eq.\ (\ref{eq:hB_1stLaw_fL=cst}).
This result implies that $\beta$ used for the effective entropy is related to dissipation.
To discuss the origin of $\beta$, we propose two entropies proportional to the Bekenstein--Hawking entropy.

The first is a scaled Bekenstein--Hawking entropy using an effective horizon radius $r_{e} = r_{H} \sqrt{1-\beta}$.
Based on the holographic principle \cite{Hooft-Bousso}, the horizon has an entropy due to the information holographically stored there and, therefore, $S_{\rm{BH}}$ proportional to $r_{H}^{2}$ can be calculated from the number of degrees of freedom on the horizon.
Similarly, $S_{H}$ proportional to $r_{e}^{2}$ should be calculated by replacing $r_{H}$ by $r_{e}$.
Consequently, we can obtain $S_{H} =S_{\rm{BH}}(1-\beta)$.
The effective radius may be related to a scale, e.g., a cutoff radius discussed in Ricci dark energy models \cite{Cai2009} and ultraviolet/infrared cutoff radii used for an entanglement entropy \cite{Das_2008_review}.
In particular, this type of scale effect due to the effective radius is likely similar to a scale effect based on the kinetic theory discussed in Ref.\ \cite{Paul2025}, where a mechanism for the origin of bulk viscosity is proposed using the principles of the kinetic theory of gas in the comoving expansion of the universe. 
Therefore, the origin of $\beta$ may be interpreted as a scale effect based on the kinetic theory.

The second is an extended power-law-corrected entropy.
In fact, $S_{H} =S_{\rm{BH}}(1-\beta)$ should be obtained from a power-law-corrected entropy \cite{Das2008} by assuming two independent parameters $\alpha$ and $\beta$ and setting $\alpha=2$, as examined in Appendix\ \ref{The present model and a power-law-corrected entropy}.
The original form of the power-law-corrected entropy is based on the entanglement of quantum fields between the inside and outside of the horizon and is computed by tracing over its degrees of freedom inside its sphere \cite{Das2008}.
In this sense, the origin of $\beta$ may be related to the entanglement of quantum fields.
The microphysical justification of the extended form has not yet been established.
However, this type of effective entropy based on the entanglement of quantum fields is considered to be a viable scenario.

In this way, the constant $(\partial S_{\Delta} / \partial S_{\rm{BH}})$ can be obtained from these two entropies that satisfy $S_{H} =S_{\rm{BH}}(1-\beta)$.
Again, no microphysical justification yet exists for this construction, but the use of the two entropies in this way is reasonable phenomenological approach.
Hereafter, the two entropies are used for the effective entropy.
The same form is applied to a BV model with constant $f_{\Lambda}(t)$.
We note that the origin of $\beta$ for the two entropies may be interpreted as (1) a scale effect based on the kinetic theory and (2) an entanglement of quantum fields.

In the present paper, based on the above considerations, we assume $S_{H} =S_{\rm{BH}}(1-\beta)$, where $\beta$ is a dimensionless non-negative constant.
Consequently, $(\partial S_{\Delta} / \partial S_{\rm{BH}})$ is constant:
\begin{align}
                    \left ( \frac{\partial S_{\Delta} }{\partial S_{\rm{BH}}}  \right )     &=   - \beta   .
\label{eq:dSpl_dSBH_cst_0}      
\end{align}
Substituting Eq.\ (\ref{eq:dSpl_dSBH_cst_0}) into Eq.\ (\ref{eq:hB_1stLaw_fL=cst}) yields
\begin{align}
                      h_{\textrm{B}}(t)   &=  \beta (2 H^{2} + \dot{H})  .
\label{eq:hB_beta_fL=cst}      
\end{align}
This is the dissipative driving term $h_{\textrm{B}}(t)$ for the present model, where $h_{\textrm{B}}(t)$ should be non-negative.
The range of $\beta$ is given by $0 \le \beta < 3(1+w)/4$, as discussed in the next section.
A matter-dominated universe, i.e., $w =0$, is considered although $w$ is retained for generality.

The Friedmann, acceleration, and continuity equations for the present model are written as 
\begin{equation}
 H^2      =  \frac{ 8\pi G }{ 3 } \rho    + \frac{\Lambda}{3}           ,                                                 
\label{eq:hB_FRW01} 
\end{equation} 
\begin{align}
  \frac{ \ddot{a}}{ a}    &= -  \frac{ 4\pi G }{ 3 }  \left ( \rho +  \frac{3 p}{c^2} \right )                   +   \frac{\Lambda}{3}   +  h_{\textrm{B}}(t)  , 
\label{eq:hB_FRW02}
\end{align}
\begin{equation}
       \dot{\rho} + 3  H \left ( \rho +  \frac{p}{c^2} \right )       =       \frac{3 }{4 \pi G}     H h_{\textrm{B}}(t)              , 
\label{eq:drho_hB}
\end{equation}
where $h_{\textrm{B}}(t) = \beta (2 H^{2} + \dot{H}) $ is given by Eq.\ (\ref{eq:hB_beta_fL=cst}).
We expect that small $h_{\textrm{B}}(t)$, corresponding to small $\beta$, should be favored, as for $\Lambda(t)$ models examined in, e.g., Refs.\ \cite{Valent2015,Sola2019}.
Small $\beta$ is discussed later.

The following are assumed for the present model:
Firstly, we assume a modified thermodynamic relation phenomenologically derived from the first law of thermodynamics.
Secondly, a BV model with constant $f_{\Lambda}(t)$ is assumed, that is, a kind of dissipative $\Lambda$CDM model is assumed.
Thirdly, $S_{H} =S_{\rm{BH}}(1-\beta)$ is assumed for constant $(\partial S_{\Delta} / \partial S_{\rm{BH}})$, by applying an effective entropy such as a scaled Bekenstein--Hawking entropy and an extended power-law-corrected entropy.
These assumptions have not been established but are considered to be viable and are used for the present model.
In next sections, we examine fundamental properties of the present model.

It should be noted that $(2 H^{2} + \dot{H})$ included in $h_{\textrm{B}}(t)$ is equivalent to the form of the Ricci scalar curvature \cite{Gao2009,Zhang2009,Cai2009}, which is also used for Ricci dark energy models.
The Ricci dark energy model generally assumes the standard continuity equation and a dark energy density proportional to the Ricci scalar curvature \cite{Gao2009,Zhang2009,Cai2009,Richarte2011,Feng2009,Singh2018,Kumar2021,Wu2025}.
As examined in the previous section, the dynamic creation pressure $p_{c}$ given by Eq.\ (\ref{eq:pc_w=0_hb}) is proportional to $h_{\textrm{B}}(t)$.
In this sense, the present model implies that $p_{c}$ is proportional and related to the Ricci scalar curvature.

\section{Background evolution of the universe in the present model} 
\label{Background evolution of the universe}

This section examines the background evolution of the universe in the present model.

Coupling Eq.\ (\ref{eq:hB_FRW01}) with Eq.\ (\ref{eq:hB_FRW02}) using $w = p/(\rho  c^2)$, and substituting Eq.\ (\ref{eq:hB_beta_fL=cst}) into the resultant equation yields 
\begin{align}
    \dot{H} &= - \frac{3}{2} (1+w)  H^{2}  +  \frac{3}{2}   (1+w)  f_{\Lambda}(t)     + h_{\textrm{B}}(t)   \notag \\ 
               &= - \frac{3}{2} (1+w)  H^{2}  +  \frac{3}{2}   (1+w)  \frac{\Lambda}{3}     + \beta (2 H^{2} + \dot{H})             .
\label{eq:Back_Present}
\end{align}
This cosmological equation corresponds to Eq.\ (\ref{eq:Back2w}) and
can be written as
\begin{align}
(1-\beta)  \dot{H}  &= - \frac{3(1+w)}{2} H^{2}  \left ( 1 -  \frac{\Lambda}{3 H^{2}}  \right )  +  2 \beta H^{2}       \notag \\ 
                           &= - \frac{3(1+w)}{2} H^{2}  \left ( 1 -  \frac{\Lambda}{3 H^{2}}  -  \frac{2 \beta }{\frac{3(1+w)}{2}} \right )     \notag \\ 
                           &= - \frac{3(1+w)}{2} H^{2}  \left ( 1 -  \gamma - \frac{\Lambda}{3 H^{2}} \right )     ,  
\label{eq:Back_Present_2}
\end{align}
where $\gamma$ is given by
\begin{equation}  
                                 \gamma  = \frac{2 \beta }{\frac{3(1+w)}{2}}  = \frac{4 \beta}{3(1+w)}  .
\label{eq:gamma}
\end{equation}
In this study, $\beta$ and $\gamma$ are used, for simplicity.
Using Eq.\ (\ref{eq:gamma}), we can convert $\beta$ into $\gamma$ and vice versa.

The general solution for the present model can be derived by applying a method examined in Refs.\ \cite{Koma45,Koma6,Koma10,Koma14}. 
The derivation of the solution is summarized in Appendix\ \ref{Solution1}.
For simplicity, the normalized Hubble parameter $\tilde{H}$ and the normalized scale factor $\tilde{a}$ are defined by
\begin{equation}
 \tilde{H} \equiv \frac{H}{H_{0}}  ,
\label{def_H_H0}
\end{equation}
\begin{equation}
   \tilde{a} \equiv \frac{a} { a_{0}}      ,
\label{def_a_a0}
\end{equation}
where $H_{0}$ and $a_{0}$ represent the Hubble parameter and the scale factor at the present time, respectively.
From Eq.\ (\ref{eq:Sol_HH0_ap}), the solution for the present model is written as
\begin{align}
       \tilde{H}^{2}      &=  \left (  1-  \frac{\Omega_{\Lambda}}{1-\gamma} \right )  \tilde{a}^{ - 3 (1+w) \frac{1-\gamma}{1-\beta}  }  + \frac{\Omega_{\Lambda}}{1-\gamma}   \notag \\
                              &=          (  1-  \Omega_{\Lambda,\gamma}  )  \tilde{a}^{ - D}  + \Omega_{\Lambda,\gamma}  .
\label{eq:Sol_HH0}
\end{align}
Here $\Omega_{\Lambda}$, $\Omega_{\Lambda,\gamma}$, and $D$ are the density parameter for $\Lambda$, an effective density parameter for both $\Lambda$ and $\gamma$, and a coefficient, respectively, given by
\begin{equation}
   \Omega_{\Lambda} = \frac{\Lambda}{3 H_{0}^{2}},    \quad   \Omega_{\Lambda,\gamma}  = \frac{\Omega_{\Lambda}}{1-\gamma}   ,
\label{eq:Omega_Lgamma}
\end{equation}
\begin{equation}
   D = 3 (1+w) \frac{1-\gamma}{1-\beta}  .
\label{eq:D}
\end{equation}
The ranges of these parameters are given by 
\begin{equation}
  0 \le \Omega_{\Lambda} \leq 1, \quad   0 \leq \Omega_{\Lambda,\gamma}  \leq 1        ,        \quad   0 < D \leq 3(1+w)    ,
\label{eq:Omega_Lgamma-D_range}
\end{equation}
\begin{equation}
  0 \le \gamma < 1, \quad   0 \leq \beta < \frac{3 (1+w) }{4}   ,     
\label{eq:gamma-beta_range}
\end{equation}
for $1+w>0$.
When $w=0$, the ranges of $D$ and $\beta$ reduce to $0 < D \leq 3$ and $0 \leq \beta < 3/4$, respectively.
The background evolution of the universe in the present model is calculated from Eq.\ (\ref{eq:Sol_HH0})
and the solution depends on $\Omega_{\Lambda,\gamma}$ and $D$.
Here $D$ is determined using Eq.\ (\ref{eq:D}), after $\beta$ and $w$ are set and $\gamma$ is calculated from Eq.\ (\ref{eq:gamma}).
In the present study, a matter-dominated universe ($w=0$) is considered and, therefore, $\beta$ and $\Omega_{\Lambda,\gamma}$ are free parameters.

We note that $\Omega_{\Lambda}$ is calculated from $\Omega_{\Lambda} = (1-\gamma) \Omega_{\Lambda,\gamma} $ and, therefore, $\Omega_{\Lambda}$ is not a free parameter in this paper.
Also, when $\beta =0$ and $w=0$ are considered, the general solution reduces to $\tilde{H}^{2} =  ( 1-  \Omega_{\Lambda} )  \tilde{a}^{ - 3}  + \Omega_{\Lambda} $ for $\Lambda$CDM models, because $\gamma =0$, $\Omega_{\Lambda,\gamma} =\Omega_{\Lambda}$, and $D=3$.
Here the density parameter for matter is given by $1- \Omega_{\Lambda}$, neglecting the influence of radiation, in a late flat FLRW universe.

The temporal deceleration parameter $q$ is useful for examining the background evolution of the universe. 
The deceleration parameter is defined by  
\begin{equation}
q \equiv  - \left ( \frac{\ddot{a} } {a H^{2}} \right )  , 
\label{eq:q_def}
\end{equation}
where a positive and negative $q$ represent deceleration and acceleration, respectively \cite{Koma14}. 
Substituting $\ddot{a}/a = \dot{H} + H^{2}$ into Eq.\ (\ref{eq:q_def}), applying Eq.\ (\ref{eq:Back_Present_2}), $\tilde{H} =H/H_{0}$, $\Omega_{\Lambda} = \Lambda /(3 H_{0}^{2})$, and Eq.\ (\ref{eq:Sol_HH0}) to the resultant equation, and performing several operations yields
\begin{align}
q  &=   -  \frac{ \dot{H} } {H^{2}}   -1     = \frac{3(1+w)}{2(1-\beta)} \left ( 1 -  \gamma - \frac{\Lambda}{3 H^{2}} \right )             -1     \notag \\
    &= \frac{3(1+w)}{2(1-\beta)} \left ( 1 -  \gamma - \frac{\Lambda}{3 H_{0}^{2}} \frac{1}{\tilde{H}^{2}} \right )             -1     \notag \\
    &= \frac{3(1+w)}{2(1-\beta)} \left ( 1 -  \gamma - \frac{ \Omega_{\Lambda}}{ (  1-  \Omega_{\Lambda,\gamma}  )  \tilde{a}^{ - D}  + \Omega_{\Lambda,\gamma}  }  \right )             -1     \notag \\
    &= \frac{3(1+w)}{2(1-\beta)}  (1 -  \gamma) \left ( 1  - \frac{ \frac{\Omega_{\Lambda}}{1 -  \gamma} }{ (  1-  \Omega_{\Lambda,\gamma}  )  \tilde{a}^{ - D}  + \Omega_{\Lambda,\gamma}  }  \right )             -1     \notag \\
    &= \frac{ \frac{1}{2} D (1-  \Omega_{\Lambda,\gamma}) \tilde{a}^{-D}    }{  (  1-  \Omega_{\Lambda,\gamma}  ) \tilde{a}^{-D}     + \Omega_{\Lambda,\gamma}}             -1     ,
\label{eq:q_present}
\end{align}
using $\Omega_{\Lambda} = (1-\gamma) \Omega_{\Lambda,\gamma} $ given by Eq.\ (\ref{eq:Omega_Lgamma}) and  $D = 3 (1+w) \frac{1-\gamma}{1-\beta}$ given by Eq.\ (\ref{eq:D}).
When both $\beta=0.5$ and $w=0$ are considered, $D=2$ is obtained from Eqs.\ (\ref{eq:D}) and (\ref{eq:gamma}) and, in this case, $q$ is always non-positive.
Consequently, an initially decelerating and then accelerating universe (hereafter `decelerating and accelerating universe') should be satisfied when both $\beta < 0.5$ and $w=0$.

To examine such a transition from a decelerating universe to an accelerating universe, we calculate the boundary for $q = 0$.
From Eq.\ (\ref{eq:q_present}), the boundary for $q = 0$ can be written as 
\begin{align}  
  \tilde{a} = \left [  \frac{ (\frac{1}{2} D-1)(  1-  \Omega_{\Lambda,\gamma}  )   }{   \Omega_{\Lambda,\gamma}  }   \right ]^{  \frac{1}{D}   }       ,
\label{eq:q=0_a_present}
\end{align}
or equivalently
\begin{align}  
\Omega_{\Lambda,\gamma} =\frac{D-2}{2  \tilde{a}^{D}    + D-2}   .
\label{eq:q=0_OmegaLG_present}
\end{align}
Using the boundary, we can discuss a decelerating and accelerating universe in the present model.
(A similar boundary for a different dissipative model was discussed in Ref.\ \cite{Koma15}.)

\begin{figure} [t] 
\begin{minipage}{0.495\textwidth}
\begin{center}
\scalebox{0.33}{\includegraphics{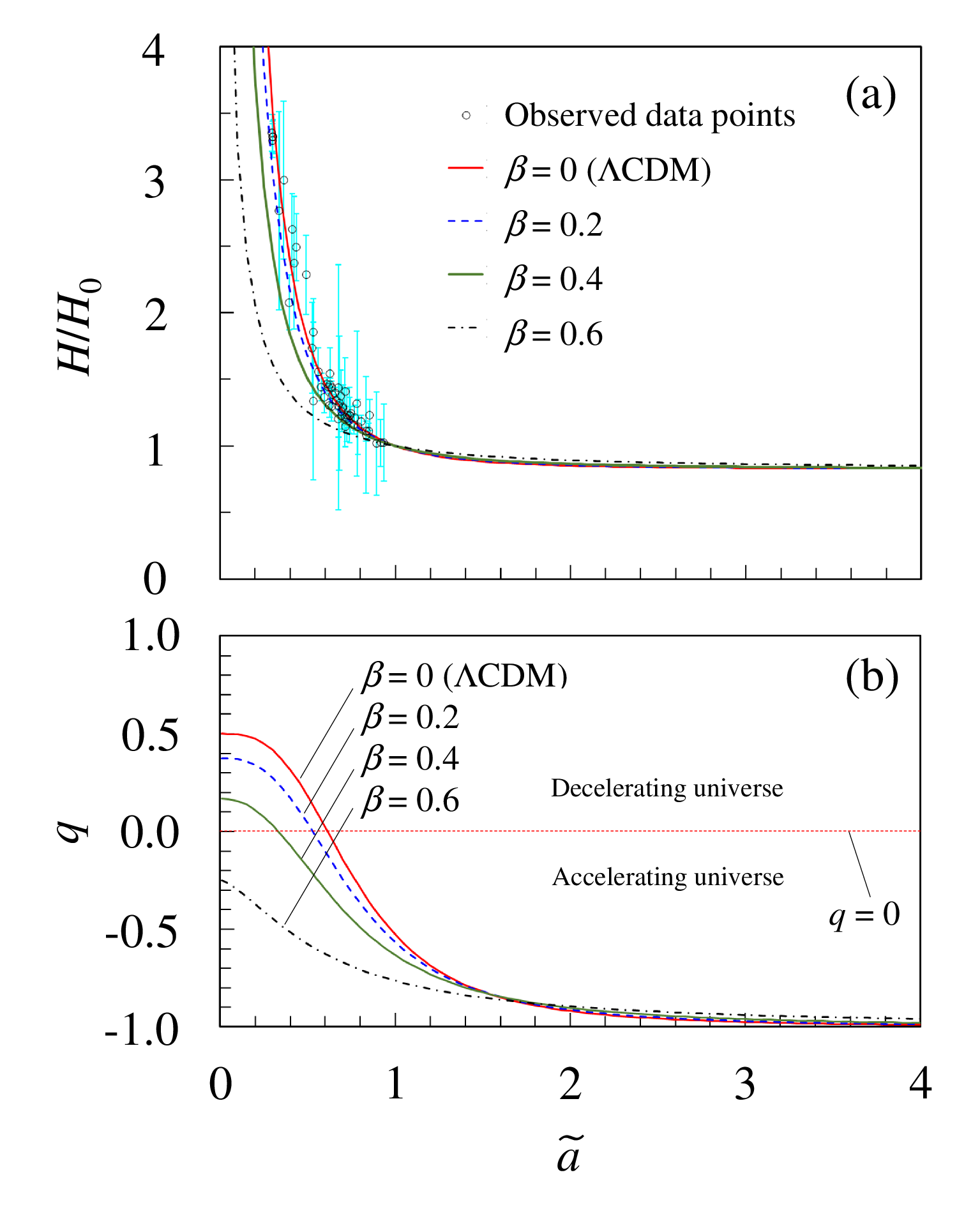}}
\end{center}
\end{minipage}
\caption{ (Color online). Evolution of the universe for the present model for $\Omega_{\Lambda,\gamma} =0.685$.
(a) Normalized Hubble parameter $H/H_{0}$.
(b) Deceleration parameter $q$.
In (a), the open circles with error bars are observed points ($57$ recently compiled data points) taken from Ref.\ \cite{57dataOdin}. 
To normalize the data points, $H_{0}$ is set to $67.4$ km/s/Mpc based on  Ref.\ \cite{Planck2018}. 
In (b), the horizontal break line represents $q=0$.
A positive and negative $q$ represent deceleration and acceleration, respectively
 }
\label{Fig-H-a}
\end{figure}

\begin{figure} [t] 
\begin{minipage}{0.495\textwidth}
\begin{center}
\scalebox{0.34}{\includegraphics{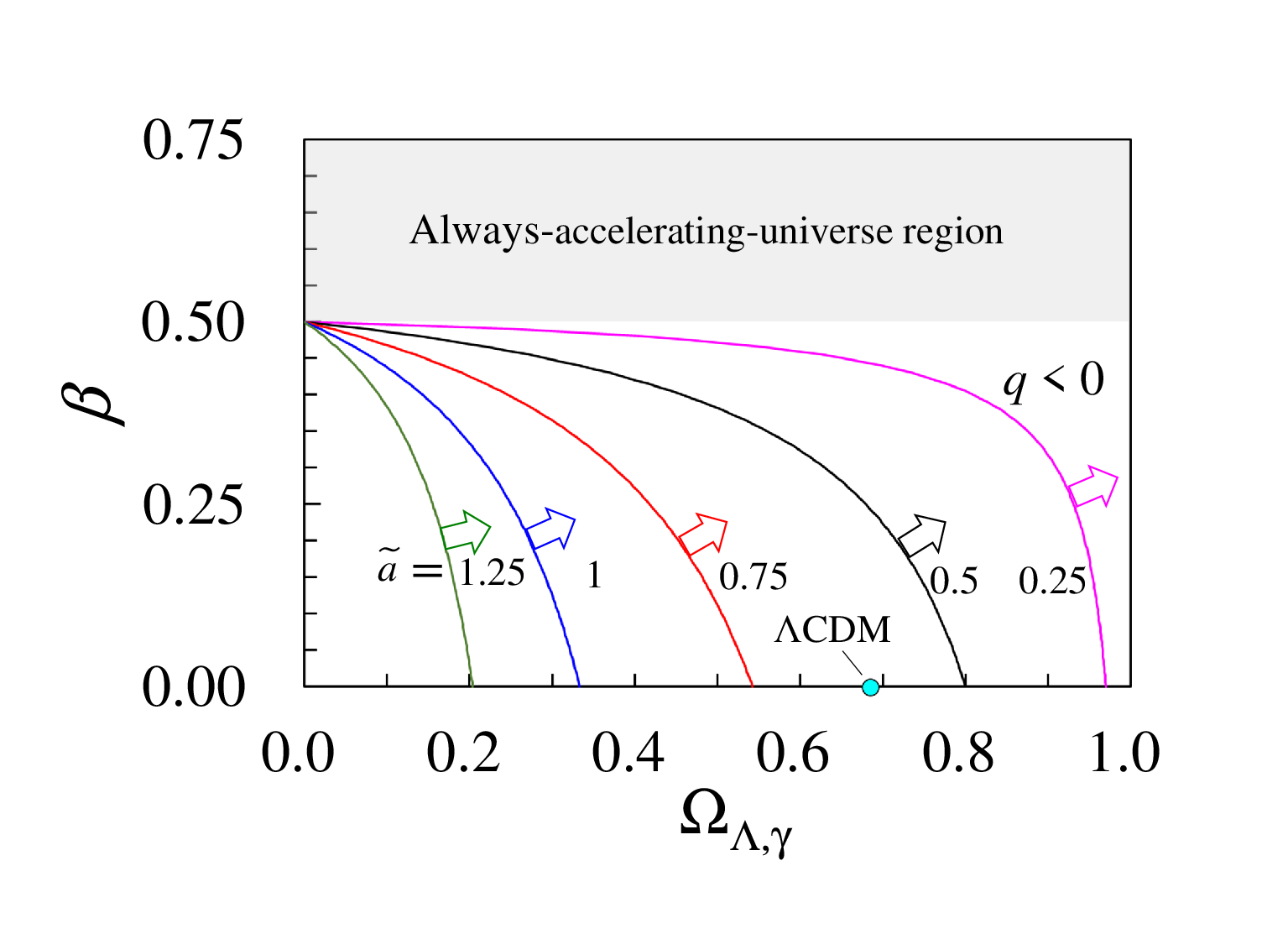}}
\end{center}
\end{minipage}
\caption{Boundary of $q= 0$ in the $(\Omega_{\Lambda,\gamma}, \beta)$ plane for various values of $\tilde{a}=a/a_{0}$.
The arrow attached to each boundary indicates an accelerating-universe region that satisfies $q< 0$.
Each numerical value (from $0.25$ to $1.25$) near each boundary represents the value of $\tilde{a}$.
The region in gray (namely $\beta >0.5$) represents an always-accelerating-universe region.
This region and $\beta=0.5$ do not satisfy a decelerating and accelerating universe.
The closed circle represents $(\Omega_{\Lambda,\gamma}, \beta) = (0.685, 0)$ for the $\Lambda$CDM model. 
We can convert $\beta$ into $\gamma = 4 \beta/3$, using Eq.\ (\ref{eq:gamma}) and $w=0$.
}
\label{Fig-q_plane}
\end{figure}

We now examine the background evolution of the universe in the present model.
For this, the evolution of the Hubble parameter and that of the deceleration parameter are shown in Fig.\ \ref{Fig-H-a}.
Here we set $w=0$ for a matter-dominated universe.
To examine typical results, $\beta$ is set to $0$, $0.2$, $0.4$, and $0.6$. 
From Eqs.\ (\ref{eq:gamma}) and (\ref{eq:D}), $\beta=0$, $0.2$, $0.4$, and $0.6$ lead to $D=3$, $2.75$, $7/3$, and $1.5$, respectively.
In addition, based on the Planck 2018 results \cite{Planck2018}, $\Omega_{\Lambda,\gamma}$ is set to $0.685$.
The results for $\beta =0$ correspond to those for the $\Lambda$CDM model.
In this study, the normalized scale factor $\tilde{a}$ increases with time because an expanding universe is considered.

As shown in Fig.\ \ref{Fig-H-a}(a), $H/H_{0}$ for all $\beta$ decreases with $\tilde{a}$ and gradually approaches a positive value, corresponding to $\sqrt{\Omega_{\Lambda,\gamma}}$.
Similarly, $q$ for all $\beta$ decreases with $\tilde{a}$ and gradually approaches $-1$ [Fig.\ \ref{Fig-H-a}(b)].
At the present time ($\tilde{a}=1$), $q$ for all $\beta$ is negative.
However, in the early stage ($\tilde{a} \ll 1$), $q$ is positive, except for $\beta=0.6$.
This result indicates that a decelerating and accelerating universe is satisfied, when $\beta=0$, $0.2$, and $0.4$.

To examine a decelerating and accelerating universe systematically, we plot the boundary of $q = 0$ in the $(\Omega_{\Lambda,\gamma}, \beta)$ plane.
In Fig.\ \ref{Fig-q_plane}, $\tilde{a}$ is set to $0.25$, $0.5$, $0.75$, $1$, and $1.25$, to examine typical boundaries.
The arrow attached to each boundary indicates an accelerating-universe region that satisfies $q< 0$.
The upper side of each boundary corresponds to this region.
The region displayed in gray (namely $\beta > 0.5$) represents an always-accelerating-universe region.
This region and $\beta=0.5$ (namely $\beta \geq 0.5$) do not satisfy a decelerating and accelerating universe.
In contrast, we can expect that $\beta < 0.5$ satisfies a decelerating and accelerating universe.
Therefore, we focus on $\beta <0.5$.
Note that we can convert $\beta$ into $\gamma = 4 \beta/3$, using Eq.\ (\ref{eq:gamma}) and $w=0$.

As shown in Fig.\ \ref{Fig-q_plane}, the accelerating-universe region varies with $\tilde{a}$ when $\beta < 0.5$. 
The boundary for $\tilde{a}=0.75$ indicates that a large-$\Omega_{\Lambda,\gamma}$ and large-$\beta$ region tends to be an accelerating universe.
A decelerating and accelerating universe is further expected with increasing $\tilde{a}$.
Of course, a non-dissipative universe for $(\Omega_{\Lambda,\gamma}, \beta) = (0.685, 0)$, corresponds to the $\Lambda$CDM model, satisfies a decelerating and accelerating universe at the present time.
In this sense, a weakly dissipative universe (namely small $\beta$) with $\Lambda$ is likely favored.

In this way, the transition from a decelerating universe to an accelerating universe can be systematically examined using the $(\Omega_{\Lambda,\gamma}, \beta)$ plane.
Observational constraints on the present model are discussed later.
In the next section, we examine irreversible entropy due to adiabatic particle creation, by assuming that the irreversible entropy is related to the present model.

\section{Irreversible entropy due to adiabatic particle creation} 	
\label{Entropy Sm}

The Bekenstein--Hawking entropy on the Hubble horizon is dozens of orders of magnitude larger than the other entropies related to matter, radiation, etc. \cite{Egan1}.
Accordingly, irreversible entropy due to adiabatic particle creation is expected to be far smaller than the horizon entropy.
However, the present model includes a dissipative driving term $h_{\textrm{B}}(t)$, which should be related to adiabatic particle creation.
Therefore, in this section, we examine the irreversible entropy, to discuss fundamental properties of the present model.
For this, we assume that the irreversible entropy is related to the dissipative term $h_{\textrm{B}}(t)$ in the present model.
That is, holographic-like matter creation cosmology is assumed.
Also, a matter-dominated universe, i.e., $w =p/(\rho c^2)=0$, is considered.
(Holographic-like matter creation cosmology should provide a new insight into the discussion of the irreversible entropy and the Bekenstein--Hawking entropy. 
We will discuss this in the next section.)

To examine the irreversible entropy, we use the particle production rate $\Gamma$ given by Eq.\ (\ref{eq:Gamma_w=0_hb_0}) and an entropy density relation given by Eq.\ (\ref{eq:NonEquil_2}).
From Eq.\ (\ref{eq:Gamma_w=0_hb_0}), $\Gamma$ is written as
\begin{equation}
       \Gamma  =    \frac{3 H }{4 \pi G}  \frac{ h_{\textrm{B}}(t) }{ \rho }  .
\label{eq:Gamma_w=0_hb}
\end{equation}
From Eq.\ (\ref{eq:NonEquil_2}), $\dot{s}/s$ is written as
\begin{equation}
\frac{\dot{s}}{s}  = \Gamma -3H  .
\label{eq:s-Gamma}
\end{equation}
Here $s$ is the entropy density due to adiabatic particle creation.
Integrating Eq.\ (\ref{eq:s-Gamma}) from the present time $t_{0}$ to an arbitrary time $t$ gives \cite{Koma15}
\begin{equation}
     \int_{s_{0}}^{s}  \frac{d s }{ s }  =  \int_{t_{0}}^{t}  ( \Gamma (t)  - 3H(t)   )  dt  ,
\label{eq:dots_s_int1}
\end{equation}
and solving this equation yields
\begin{equation}
     \frac{s}{s_{0}}  = \exp  \left [  \int_{t_{0}}^{t}   ( \Gamma (t)  - 3H(t)   )  dt \right ]  ,
\label{eq:dots_s_s-t}
\end{equation}
where $s_{0}$ is $s$ at the present time.
Applying $\dot{H} = dH/dt$ to the above equation yields 
\begin{align}
    \frac{s}{s_{0}}    &= \exp  \left [  \int_{H_{0}}^{{H}}  \frac{ \Gamma (H) -3 {H}  }  {  \dot{H}    }    dH \right ]    .
\label{eq:dots_s_s-H}
\end{align}
Equations\ (\ref{eq:dots_s_s-t}) and (\ref{eq:dots_s_s-H}) are the entropy density relation for adiabatic particle creation.
For simplicity, the same symbol is used for both the integrating variable and the variable for the integration interval.

We now calculate $\Gamma$ and $s/s_{0}$ for the present model.
Substituting $ h_{\textrm{B}}(t) $ given by Eq.\ (\ref{eq:hB_beta_fL=cst}) and $\rho$ given by Eq.\ (\ref{eq:hB_FRW01}) into Eq.\ (\ref{eq:Gamma_w=0_hb}) yields 
\begin{align}
       \Gamma                     =    \frac{3H}{4 \pi G}  \frac{ \beta (2 H^{2} + \dot{H}) }{ \frac{3 (H^{2} - \frac{\Lambda}{3})  }{ 8\pi G }  }    
                                      &=  \frac{ 2 \beta  H (2 H^{2} + \dot{H}) }{ H^{2} - \frac{\Lambda}{3}  }    .
\label{eq:Gamma-H_Present}
\end{align}
Using Eq.\ (\ref{eq:Back_Present_2}), $\dot{H}$ can be written as 
\begin{align}
\dot{H}  &= - \frac{3(1+w)}{2(1-\beta)} H^{2}  \left ( 1 -  \gamma - \frac{\Lambda}{3 H^{2}} \right )     \notag \\
            &= - \frac{3(1+w)(1 -  \gamma)}{2(1-\beta)} H^{2}  \left ( 1 - \frac{\Lambda}{3(1 -  \gamma) H^{2}} \right )  \notag \\
            &= - \frac{3(1+w)(1 -  \gamma)}{2(1-\beta)}  \left ( H^{2}  - \frac{\Lambda}{3 H_{0}^{2}} \frac{ H_{0}^{2} }{ (1 -  \gamma) } \right )  \notag \\
            &= - \frac{D}{2}  ( H^{2}  - \Omega_{\Lambda,\gamma} H_{0}^{2} )  ,
\label{eq:dotH_present}
\end{align}
using $\Omega_{\Lambda,\gamma} = \Lambda/(3H_{0}^{2})/(1-\gamma)$ given by Eq.\ (\ref{eq:Omega_Lgamma}) and  $D = 3 (1+w) \frac{1-\gamma}{1-\beta}$ given by Eq.\ (\ref{eq:D}).
Note that $w$ is retained although $w=0$ is considered.
Substituting $\dot{H}$ given by Eq.\ (\ref{eq:dotH_present}) into the numerator of Eq.\ (\ref{eq:Gamma-H_Present}) yields
\begin{align}
   2 \beta H (2 H^{2} + \dot{H}) &=     \beta H [4 H^{2}  - D ( H^{2}  - \Omega_{\Lambda,\gamma} H_{0}^{2} ) ]  \notag \\
                                           &=        \beta H [(4-D) H^{2}   +D \Omega_{\Lambda,\gamma} H_{0}^{2} ]  .
\label{eq:numerator_Gamma-H_Present}
\end{align}
Substituting Eq.\ (\ref{eq:numerator_Gamma-H_Present}) into Eq.\ (\ref{eq:Gamma-H_Present}) and applying $\tilde{H}=H/H_{0}$ and $\Omega_{\Lambda} = \Lambda/(3H_{0}^{2})$ yields
\begin{align}
     \Gamma          &=  \frac{  \beta H  [(4-D) H^{2}      +D \Omega_{\Lambda,\gamma} H_{0}^{2} ]  }{ H^{2}           - \frac{\Lambda}{3}    }    \notag \\
                           &=  \beta H  \left ( \frac{    (4-D) \tilde{H}^{2}   +D \Omega_{\Lambda,\gamma}        }{\tilde{H}^{2} - \Omega_{\Lambda}   } \right )   .
\label{eq:Gamma_Present}
\end{align}
The above form for the present model is slightly different from that for other dissipative models, such as $\Gamma \propto H$ and $\Gamma \propto H^2$, typically examined in CCDM models.

Substituting Eqs.\ (\ref{eq:dotH_present}) and (\ref{eq:Gamma_Present}) into Eq.\ (\ref{eq:dots_s_s-H}) and performing several calculations yields 
\begin{align}
    \frac{s}{s_{0}}   =  \frac{ \tilde{H}^{2} - \Omega_{\Lambda}  }{1- \Omega_{\Lambda} }  .     
\label{eq:dots_s_s-H_Present_w=0}
\end{align}
The derivation of $s/s_{0}$ is summarized in Appendix\ \ref{Derivation of entropy density}.
Substituting Eq.\ (\ref{eq:Sol_HH0}) into the above equation yields
\begin{align}
    \frac{s}{s_{0}}   &=  \frac{ \tilde{H}^{2} - \Omega_{\Lambda}  }{1- \Omega_{\Lambda} }  = \frac{ (  1-  \Omega_{\Lambda,\gamma}  )  \tilde{a}^{ - D}  + \Omega_{\Lambda,\gamma} - \Omega_{\Lambda}  }{1- \Omega_{\Lambda} }  .
\label{eq:dots_s_s-a_Present_w=0}
\end{align}

The irreversible entropy $S_{mc}$ in the comoving volume is calculated from Eq.\ (\ref{eq:dots_s_s-a_Present_w=0}). 
The normalized entropy $S_{mc}/S_{mc,0}$ in the comoving volume is written as
\begin{align}
        \frac{S_{mc}}{S_{mc,0}} &=  \frac{ s  a^3 }{ s_{0}  a_{0}^3}   =  \left (   \frac{s}{s_{0}} \right )    \tilde{a}^{3}                                                                                         \notag \\
                                     &=  \left (  \frac{ (  1-  \Omega_{\Lambda,\gamma}  )  \tilde{a}^{ - D}  + \Omega_{\Lambda,\gamma} - \Omega_{\Lambda}  }{1- \Omega_{\Lambda} }  \right )   \tilde{a}^{3}    ,
\label{eq:SmcSmc0_comoving}
\end{align}
where $S_{mc,0}$ is $S_{mc}$ at the present time.

In addition, we derive the irreversible entropy $S_{mH}$ in the Hubble volume from Eq.\ (\ref{eq:dots_s_s-H_Present_w=0}). 
Using both Eq.\ (\ref{eq:dots_s_s-H_Present_w=0}) and $r_{H} =c/H$ given by Eq.\ (\ref{eq:rH}), the normalized entropy $S_{mH}/S_{mH,0}$ in the Hubble volume is written as
\begin{align}
        \frac{S_{mH}}{S_{mH,0}} &= \frac{ s r_{H}^3 }{ s_{0} r_{H0}^3}  =  \frac{ s }{ s_{0} }  \frac{  {(c/H)}^3 }{ {(c/H_{0})}^3}     \notag \\
                                     &=  \frac{1}{\tilde{H}^{3}}   \left ( \frac{ \tilde{H}^{2} - \Omega_{\Lambda}  }{1- \Omega_{\Lambda} }  \right )     ,   
\label{eq:SmHSmH0}
\end{align}
where $S_{mH,0}$ is $S_{mH}$ at the present time.
This equation is discussed in the next section.
Substituting Eq.\ (\ref{eq:Sol_HH0}) into the above equation yields
\begin{align}
     \frac{S_{mH}}{S_{mH,0}} 
 &=  \frac{ (  1-  \Omega_{\Lambda,\gamma}  )  \tilde{a}^{ - D} + \Omega_{\Lambda,\gamma}   - \Omega_{\Lambda}   }{ (1-  \Omega_{\Lambda})  [ (  1-  \Omega_{\Lambda,\gamma}  )  \tilde{a}^{ - D}     + \Omega_{\Lambda,\gamma}  ]^{3/2} }     .
\label{eq:SmHSmH0_a}
\end{align}

\begin{figure} [t] 
\begin{minipage}{0.495\textwidth}
\begin{center}
\scalebox{0.33}{\includegraphics{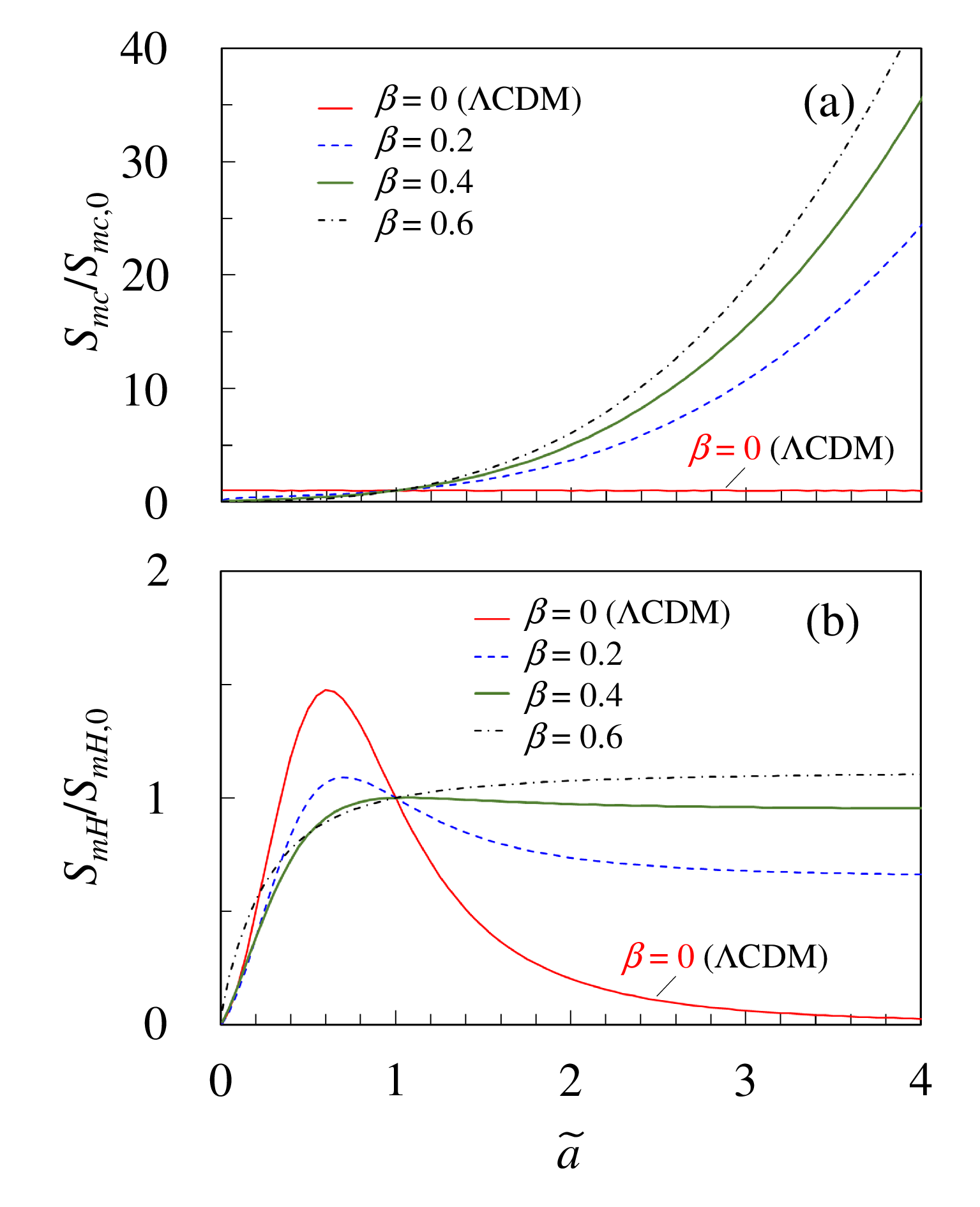}}
\end{center}
\end{minipage}
\caption{Evolution of the normalized $S_{mc}$ in the comoving volume and the normalized $S_{mH}$ in the Hubble volume for the present model for $\Omega_{\Lambda,\gamma} =0.685$. 
(a) $S_{mc} / S_{mc,0}$ in the comoving volume.
(b) $S_{mH} / S_{mH,0}$ in the Hubble volume.
The evolution for $\beta =0$ corresponds to that for the $\Lambda$CDM model in a non-dissipative universe.
}
\label{Fig-Smc-a_SmH-a}
\end{figure}

We now observe the evolution of $S_{mc}$ and $S_{mH}$ for the present model.
To examine typical results, $\beta$ is set to $0$, $0.2$, $0.4$, and $0.6$, and $\Omega_{\Lambda,\gamma}$ is set to $0.685$, as set in the previous section.
The results for $\beta =0$ correspond to those for the $\Lambda$CDM model in a non-dissipative universe.

We first observe the normalized $S_{mc}$ in the comoving volume.
As shown in Fig.\ \ref{Fig-Smc-a_SmH-a}(a), the normalized $S_{mc}$ for $\beta >0$ increases with $\tilde{a}$, whereas $S_{mc}$ for $\beta =0$ is constant.
However, as shown in Fig.\ \ref{Fig-Smc-a_SmH-a}(b), when $\beta=0$, the normalized $S_{mH}$ in the Hubble volume rapidly varies with $\tilde{a}$ in the early stage and then gradually approaches a constant value in the final stage. 
This result indicates that the normalized $S_{mH}$ in the Hubble volume varies with $\tilde{a}$, even if the normalized $S_{mc}$ in the comoving volume is constant.
The difference between the comoving volume and the Hubble volume causes this result.

Also, as shown in Fig.\ \ref{Fig-Smc-a_SmH-a}(b), the normalized $S_{mH}$ for each $\beta$ gradually approaches each constant value in the final stage.
For example, from Eq.\ (\ref{eq:SmHSmH0_a}), $S_{mH}$ for $\beta=0$ finally approaches zero, because $\Omega_{\Lambda,\gamma}   = \Omega_{\Lambda}$.
The approached value increases with $\beta$.
The normalized $S_{mH}$ in the Hubble volume is discussed again, in the next section.

In this section, we examined irreversible entropy due to adiabatic particle creation, assuming that the irreversible entropy is related to the present model, that is, holographic-like matter creation cosmology is assumed. 
The irreversible entropy should be dozens of orders of magnitude smaller than the Bekenstein--Hawking entropy $S_{\rm{BH}}$ on the Hubble horizon.
In fact, $S_{\rm{BH}}$ should be approximately equivalent to the total entropy of the universe \cite{Egan1}.
In the next section, we focus on $S_{\rm{BH}}$ and study horizon thermodynamics.

\section{Entropy on the Hubble horizon} 
\label{Entropy SBH on the Hubble horizon}

In this section, we examine horizon thermodynamics for the present model.
For this, we focus on the Bekenstein--Hawking entropy $S_{\rm{BH}}$ and examine evolution of $S_{\rm{BH}}$, $\dot{S}_{\rm{BH}}$, and $\ddot{S}_{\rm{BH}}$.
From Eq.\ (\ref{eq:SBH2}), the normalized $S_{\rm{BH}}$ is written as
\begin{equation}
\frac{S_{\rm{BH}}}{ S_{\rm{BH},0} } =  \left ( \frac{ H }{ H_{0} } \right )^{-2}  = \tilde{H}^{-2} ,
\label{eq:SBHSBH0_0a}      
\end{equation}
where $S_{\rm{BH},0}$ is $S_{\rm{BH}}$ at the present time and is given by $K/H_{0}^{2}$.
Substituting Eq.\ (\ref{eq:Sol_HH0}) into Eq.\ (\ref{eq:SBHSBH0_0a}) yields  
\begin{align}  
\frac{S_{\rm{BH}}}{ S_{\rm{BH},0} } &=    \left [    (  1-  \Omega_{\Lambda,\gamma}  )  \tilde{a}^{ - D}  + \Omega_{\Lambda,\gamma}  \right ]^{-1}      .
\label{eq:SBH_present}      
\end{align}  

We calculate $\dot{S}_{\rm{BH}}$ for the present model.
Arranging $\dot{S}_{\rm{BH}}$ given by Eq.\ (\ref{eq:dSBH}), substituting $- \dot{H}/H^{2}$ given by Eq.\ (\ref{eq:q_present}) into the resultant equation, and applying Eq.\ (\ref{eq:Sol_HH0}) yields 
\begin{align}  
\dot{S}_{\rm{BH}}   
&=  \frac{-2K \dot{H} }{H^{3}}  =  \frac{2K}{H_{0}}  \left ( \frac{- \dot{H} }{H^{2}} \right )    \frac{H_{0}}{H}                                    \notag \\
&= \frac{2K}{H_{0}}        \frac{ \frac{1}{2} D (1-  \Omega_{\Lambda,\gamma})  \tilde{a}^{ - D}      }{  (  1-  \Omega_{\Lambda,\gamma}  )  \tilde{a}^{ - D}     + \Omega_{\Lambda,\gamma}  }   \frac{H_{0}}{H}        \notag\\
&=    \frac{K}{H_{0}}       \frac{ D (1-  \Omega_{\Lambda,\gamma})  \tilde{a}^{ - D}  }{ \left [ (  1-  \Omega_{\Lambda,\gamma}  )  \tilde{a}^{ - D}     + \Omega_{\Lambda,\gamma}  \right ]^{3/2} }      .
\label{eq:dSBH_present_0}      
\end{align}   
Using Eq.\ (\ref{eq:dSBH_present_0}) and $S_{\rm{BH},0}= K/H_{0}^{2}$, the normalized $\dot{S}_{\rm{BH}}$ is written as 
\begin{align}  
\frac{\dot{S}_{\rm{BH}}}{ S_{\rm{BH},0} H_{0}}  &=  \frac{ D (1-  \Omega_{\Lambda,\gamma})  \tilde{a}^{ - D}  }{ \left [ (  1-  \Omega_{\Lambda,\gamma}  )  \tilde{a}^{ - D}     + \Omega_{\Lambda,\gamma}  \right ]^{3/2} }      ,
\label{eq:dSBH_present_aa0}      
\end{align}  
where $D = 3 (1+w) \frac{1-\gamma}{1-\beta}$ is given by Eq.\ (\ref{eq:D}). 
Equation\ (\ref{eq:dSBH_present_aa0}) indicates that $\dot{S}_{\rm{BH}} \geq 0$ is satisfied when $w > -1$ and $0 \le  \Omega_{\Lambda,\gamma}  \le 1$ are considered.
The generalized second law of thermodynamics should be also satisfied because $S_{\rm{BH}}$ is approximately equivalent to the total entropy of the universe.

Using Eq.\ (\ref{eq:Sol_HH0}), Eq.\ (\ref{eq:dSBH_present_aa0}) can be written as
\begin{align}  
\frac{\dot{S}_{\rm{BH}}}{ S_{\rm{BH},0} H_{0}}  &=  \frac{D}{\tilde{H}^{3} }  (\tilde{H}^{2} -\Omega_{\Lambda,\gamma}  )      .
\label{eq:dSBH_present_HH0}      
\end{align}  
The form of the normalized $\dot{S}_{\rm{BH}}$ given by Eq.\ (\ref{eq:dSBH_present_HH0}) is similar to that of the normalized $S_{mH}$ given by Eq.\ (\ref{eq:SmHSmH0}).
For example, when a pure dissipative universe without $\Lambda$, namely $\Omega_{\Lambda,\gamma}=\Omega_{\Lambda}=0$, is considered, Eqs.\ (\ref{eq:SmHSmH0}) and (\ref{eq:dSBH_present_HH0}) reduce to $\tilde{H}^{-1}$ and $D \tilde{H}^{-1}$, respectively.
This similarity may imply that $S_{mH}$ in the Hubble volume is related not to $S_{\rm{BH}}$ but to $\dot{S}_{\rm{BH}}$, through the thermodynamic relation between quantities on the horizon and in the bulk.
Of course, $S_{mH}$ itself should be dozens of orders of magnitude smaller than $S_{\rm{BH}}$.

It should be noted that holographic-like matter creation cosmology is based on the modified thermodynamic relation between quantities on the horizon and in the bulk, in contrast to other dissipative cosmologies.
Therefore, the similarity mentioned above is a new physical insight because it can serve as a bridge connecting the irreversible entropy and the Bekenstein--Hawking entropy through the modified thermodynamic relation.

Next, we calculate the normalized $\ddot{S}_{\rm{BH}}$ for the present model.
Differentiating Eq.\ (\ref{eq:dSBH_present_aa0}) with respect to $t$ and performing several calculations yields 
\begin{align} 
   \frac{ \ddot{S}_{\rm{BH}}  }{S_{\rm{BH},0} H_{0}^{2} } &=  \frac{ \frac{D^2}{2} (1-  \Omega_{\Lambda,\gamma})  \tilde{a}^{ - D}    [ (  1-  \Omega_{\Lambda,\gamma}  )  \tilde{a}^{ - D}  -2 \Omega_{\Lambda,\gamma}  ]  }{ [ (  1-  \Omega_{\Lambda,\gamma}  )  \tilde{a}^{ - D}     + \Omega_{\Lambda,\gamma}  ]^{2} }     .
\label{eq:d2SBH2SBH0_present}
\end{align}
Equation\ (\ref{eq:d2SBH2SBH0_present}) indicates that $ \ddot{S}_{\rm{BH}} < 0$ should be satisfied in the final stage.
That is, the universe for the present model should approach a kind of equilibrium state in the final stage.
To examine this relaxation, the boundary required for $\ddot{S}_{\rm{BH}} = 0$ is calculated.
From Eq.\ (\ref{eq:d2SBH2SBH0_present}), the boundary of $\ddot{S}_{\rm{BH}} = 0$ is given by
\begin{align}  
  \tilde{a} = \left (  \frac{  1-  \Omega_{\Lambda,\gamma}     }{  2 \Omega_{\Lambda,\gamma}  }   \right )^{  \frac{1}{D}   }       ,
\label{eq:d2Sdt2=0_a}
\end{align}
or equivalently
\begin{align}  
   \Omega_{\Lambda,\gamma}   =  \frac{ 1   }{  2 \tilde{a}^{D}  + 1    }  .
\label{eq:d2Sdt2=0}
\end{align}
Similar boundaries for different models were discussed in Refs.\ \cite{Koma14,Koma16}.

\begin{figure} [t] 
\begin{minipage}{0.495\textwidth}
\begin{center}
\scalebox{0.33}{\includegraphics{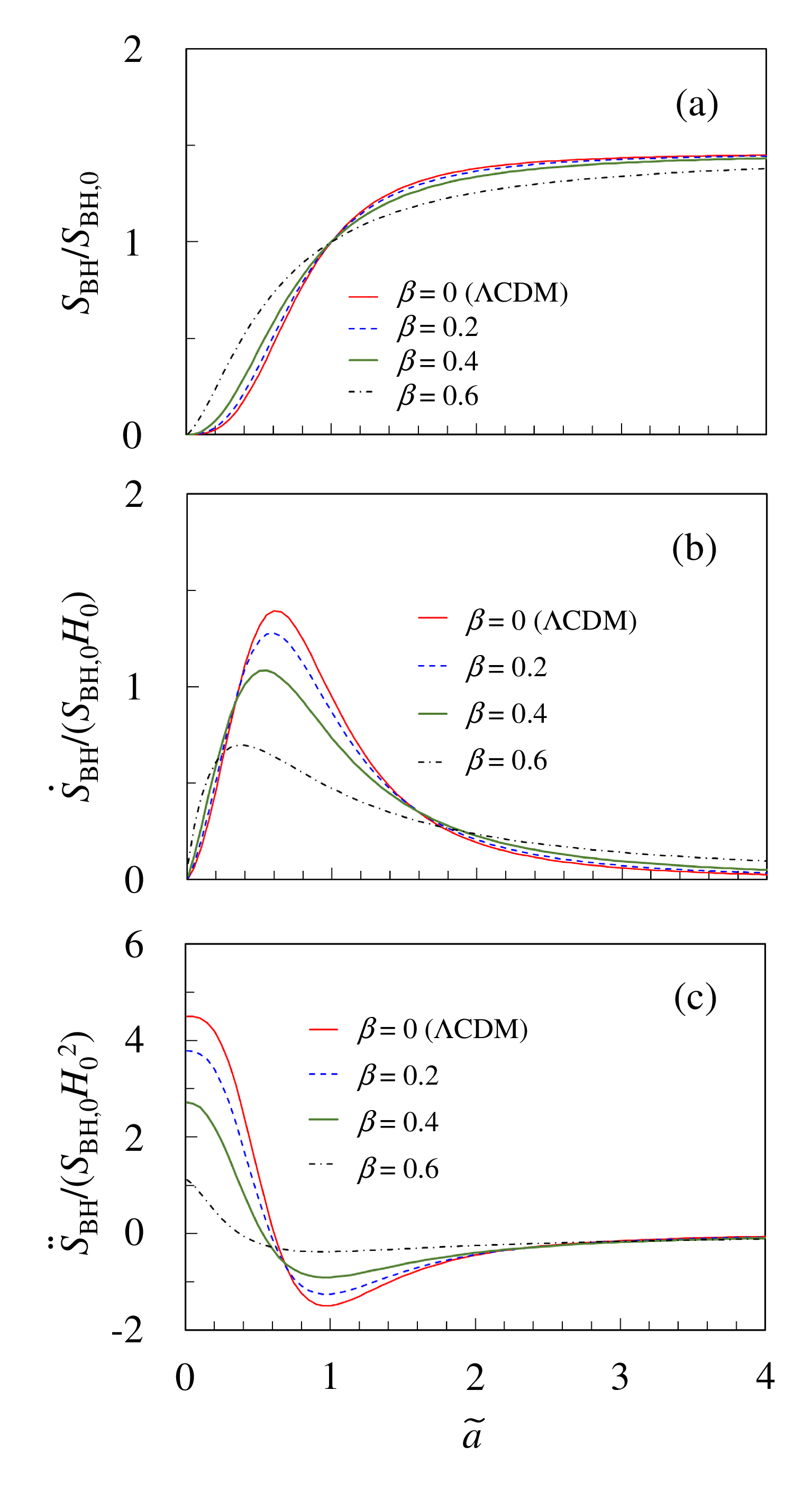}}
\end{center}
\end{minipage}
\caption{Evolution of the normalized $S_{\rm{BH}}$,  $\dot{S}_{\rm{BH}}$, and $\ddot{S}_{\rm{BH}}$ for the present model for $\Omega_{\Lambda,\gamma} =0.685$. 
(a) $S_{\rm{BH}} / S_{\rm{BH},0}$.
(b) $\dot{S}_{\rm{BH}}/ (S_{\rm{BH},0} H_{0})$.
(c) $\ddot{S}_{\rm{BH}}/ (S_{\rm{BH},0} H_{0}^{2})$.
}
\label{Fig-SBH_dSBH_d2SBH-a}
\end{figure}

\begin{figure} [t] 
\begin{minipage}{0.495\textwidth}
\begin{center}
\scalebox{0.30}{\includegraphics{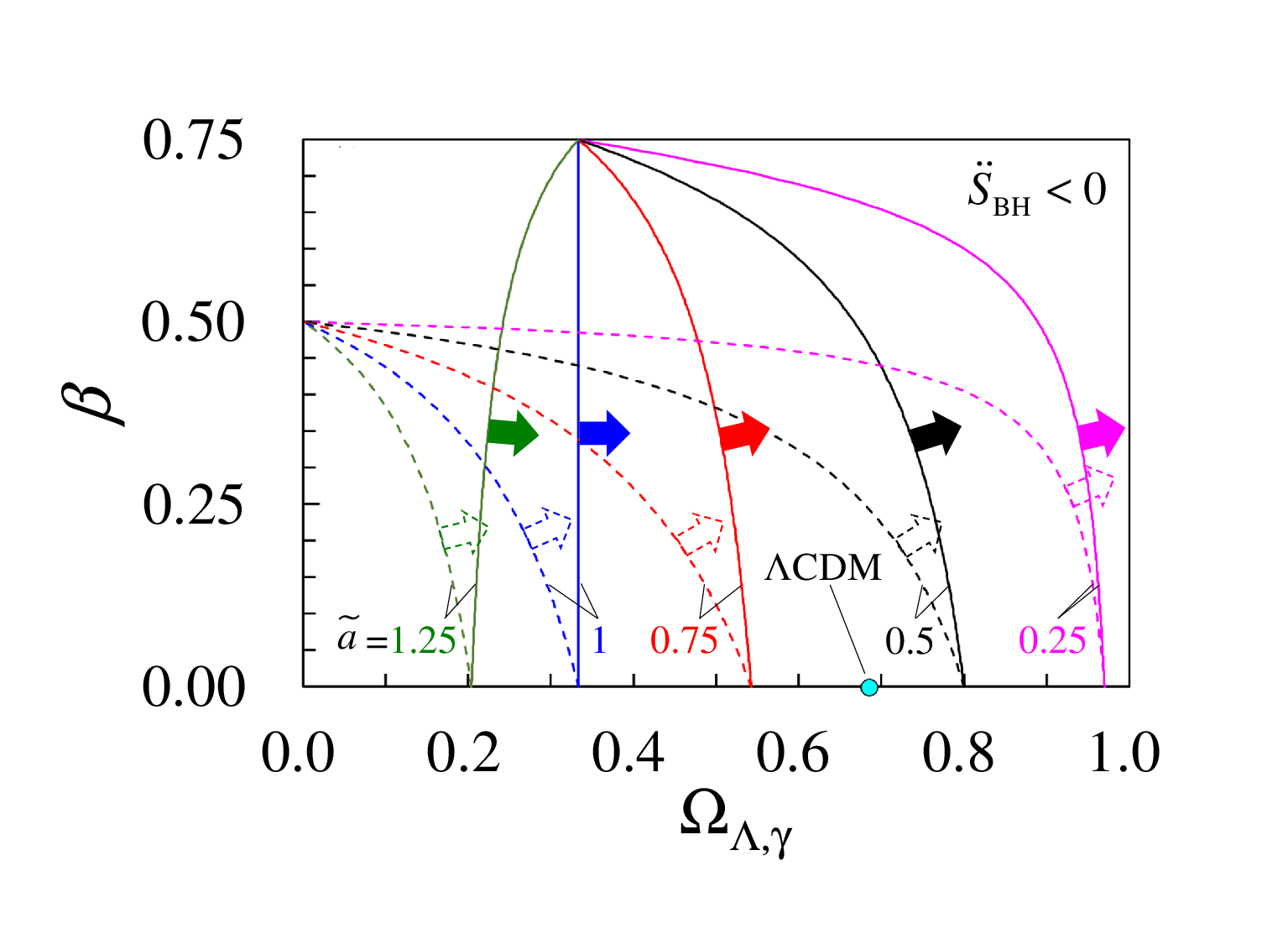}}
\end{center}
\end{minipage}
\caption{Boundary of $\ddot{S}_{\rm{BH}} = 0$ in the $(\Omega_{\Lambda,\gamma}, \beta)$ plane for various values of $\tilde{a}=a/a_{0}$. 
The solid lines represent the boundary of $\ddot{S}_{\rm{BH}} = 0$. 
The arrow attached to each solid-line boundary indicates a region that satisfies $\ddot{S}_{\rm{BH}} < 0$.
The dashed lines represent the boundary of $q = 0$, plotted from Fig.\ \ref{Fig-q_plane}.
The arrow attached to each dashed-line boundary indicates an accelerating-universe region that satisfies $q< 0$.
The region for $\beta < 0.5$ satisfies a decelerating and accelerating universe.
The closed circle represents $(\Omega_{\Lambda,\gamma}, \beta) = (0.685, 0)$ for the $\Lambda$CDM model.
See also the caption of Fig.\ \ref{Fig-q_plane}.
 }
\label{Fig-d2SBHdt2_plane}
\end{figure}

We now investigate the evolution of $S_{\rm{BH}}$, $\dot{S}_{\rm{BH}}$, and $\ddot{S}_{\rm{BH}}$ for the present model.
To examine typical results, $\beta$ is set to $0$, $0.2$, $0.4$, and $0.6$, and $\Omega_{\Lambda,\gamma}$ is set to $0.685$, as examined in Figs.\ \ref{Fig-H-a} and \ref{Fig-Smc-a_SmH-a}.
As shown in Fig.\ \ref{Fig-SBH_dSBH_d2SBH-a}(a), $S_{\rm{BH}}$ increases with $\tilde{a}$.
Therefore, the second law of thermodynamics, $\dot{S}_{\rm{BH}} \geq 0$, is satisfied [Fig.\ \ref{Fig-SBH_dSBH_d2SBH-a}(b)].
We can find that the evolution of $\dot{S}_{\rm{BH}}$ for $\beta=0$ is similar to that of $S_{mH}$ shown in Fig.\ \ref{Fig-Smc-a_SmH-a}(b), because of the similarity between Eqs.\ (\ref{eq:dSBH_present_HH0}) and (\ref{eq:SmHSmH0}).
The similarity is satisfied in both a non-dissipative universe, i.e., $\beta=0$, and a pure dissipative universe without $\Lambda$, i.e., $\Omega_{\Lambda,\gamma}= \Omega_{\Lambda}=0$.

In addition, $S_{\rm{BH}}$ rapidly increases in the early stage and gradually approaches a positive value in the final stage, as shown in Fig.\ \ref{Fig-SBH_dSBH_d2SBH-a}(a).
[The value is $S_{\rm{BH}} / S_{\rm{BH},0} = \Omega_{\Lambda,\gamma}^{-1}$, which is obtained by applying $ \tilde{a} \rightarrow \infty$ to Eq.\ (\ref{eq:SBH_present}).]
Consequently, $\ddot{S}_{\rm{BH}}$ is positive in the early stage and negative in the final stage, as shown in Fig.\ \ref{Fig-SBH_dSBH_d2SBH-a}(c).
That is, maximization of entropy, $\ddot{S}_{\rm{BH}} < 0$, should be satisfied in the final stage.

As discussed above, the universe examined approaches a kind of equilibrium state in the final stage.
To study this relaxation-like-process, the boundary required for $\ddot{S}_{\rm{BH}} = 0$ is examined.
From Eq.\ (\ref{eq:d2Sdt2=0}), the boundary of $\ddot{S}_{\rm{BH}} = 0$ for various values of $\tilde{a}=a/a_{0}$ can be plotted on the $(\Omega_{\Lambda,\gamma}, \beta)$ plane.
In Fig.\ \ref{Fig-d2SBHdt2_plane}, $\tilde{a}$ is set to be from $0.25$ to $1.25$ for typical boundaries.
The solid lines represent the boundary of $\ddot{S}_{\rm{BH}} = 0$. 
The arrow attached to each solid-line boundary indicates a relaxation-like-process region that satisfies $\ddot{S}_{\rm{BH}} < 0$.
The dashed lines represent the boundary of $q= 0$, plotted from Fig.\ \ref{Fig-q_plane}.
The normalized scale factor $\tilde{a}$ increases with time.

We first examine the boundary of $\ddot{S}_{\rm{BH}} = 0$.
The region that satisfies $\ddot{S}_{\rm{BH}} < 0$ varies with $\tilde{a}$, as shown in Fig.\ \ref{Fig-d2SBHdt2_plane}.
For each boundary, the right-hand side, namely a large-$\Omega_{\Lambda,\gamma}$ region, satisfies $\ddot{S}_{\rm{BH}} < 0$ and is a relaxation-like-process region.
This region tends to approach thermodynamic equilibrium-like states quickly.
We can confirm that a non-dissipative universe for $(\Omega_{\Lambda,\gamma}, \beta) = (0.685, 0)$, corresponds to the $\Lambda$CDM model, satisfies $\ddot{S}_{\rm{BH}} < 0$ at the present time ($\tilde{a}=1$).
The relaxation-like-process region extends leftward with increasing $\tilde{a}$.
Note that all the boundaries for $\ddot{S}_{\rm{BH}} = 0$ intersect at the point $(\Omega_{\Lambda,\gamma}, \beta) = (1/3, 0.75)$, where $\Omega_{\Lambda,\gamma}=1/3$ is obtained from Eq.\ (\ref{eq:d2Sdt2=0}).

Next, we compare the boundaries of $\ddot{S}_{\rm{BH}} = 0$ and $q=0$,
focusing on the region for $\beta <0.5$, because this region satisfies a decelerating and accelerating universe.
As shown in Fig.\ \ref{Fig-d2SBHdt2_plane}, the boundaries of $\ddot{S}_{\rm{BH}} = 0$ and $q=0$ approach each other as $\beta$ decreases.
However, in general, an accelerating-universe region for $q<0$ extends earlier than a relaxation-like-process region for $\ddot{S}_{\rm{BH}} < 0$.
That is, when $\beta <0.5$, the universe in the present model is an initially decelerating and then accelerating universe and thereafter approaches a kind of equilibrium state.
This result implies that thermodynamic constraints on $\ddot{S}_{\rm{BH}} < 0$ are consistent with constraints on a transition from a decelerating universe to an accelerating universe.
In Sec.\ \ref{Observational, transitional, and thermodynamic constraints}, we examine observational constraints on the present model and discuss the result with the transitional and thermodynamic constraints.

\section{First-order density perturbations in the present model}
\label{First-order density perturbations in the present model}

In this section, we examine density perturbations in the present model, corresponding to a BV model with $\Lambda$.
Density perturbations in a pure BV model have in fact been studied in previous works \cite{Koma6,Koma9,Koma14,Koma15,Koma16}, based on those of Jesus \textit{et al.} \cite{Lima2011}, using the neo-Newtonian approach proposed by Lima \textit{et al.} \cite{Lima_Newtonian_1997}.
Formulations for the pure BV model have been extended to those for the BV model with $\Lambda$ \cite{Koma7}.
Therefore, in Sec.\ \ref{Formulations for the BV model with L}, the extended formulation is reviewed.
In Sec.\ \ref{Formulations for the present model}, the formulation is applied to the present model.
Evolutions of density perturbations in the present model are discussed in the next section, using the combined best-fit based on a joint chi-squared analysis.

Note that we examine first-order density perturbations in the linear approximation by assuming a matter-dominated universe, i.e., $w=p/(\rho c^{2})=0$ \cite{Koma7}.
We focus on density perturbations for matter and neglect the other perturbations.
A negative sound speed \cite{Lima2011} and the existence of clustered matter \cite{Ramos_2014} are not assumed.

\subsection{Formulations for the BV model with $\Lambda$}
\label{Formulations for the BV model with L}

We review density perturbations in BV models, in accordance with previous works \cite{Koma6,Koma7,Koma9,Koma14,Koma15,Koma16}.
The formulation for a pure BV model is essentially equivalent to that for both bulk viscous and CCDM models.
For example, Jesus \textit{et al.} examined density perturbations in the CCDM model, using a neo-Newtonian approach \cite{Lima2011}.
We apply the neo-Newtonian approach to a BV model with $\Lambda$, according to Ref.\ \cite{Koma7}.
Here ${f}_{\Lambda} (t) =\Lambda/3$ is considered.

Substituting $p=0$ into Eq.\ (\ref{eq:drho_BV_fL-cst}) yields 
\begin{equation}
      \dot{\rho} + 3  H   \rho   =   \frac{3}{4 \pi G}  H h_{\textrm{B}} (t)   ,
\label{eq:drho_General_BV}
\end{equation}
where $ h_{\textrm{B}} (t)$ is a general dissipative driving term.
For generality, a specific $h_{\textrm{B}}(t)$ given by $\beta (2 H^{2} + \dot{H})$ is not used in this subsection.
Equation\ (\ref{eq:drho_General_BV}) can be written as 
\begin{equation}
      \dot{\rho} + 3 H \rho   =   \Gamma \rho  ,
\label{eq:drho_General_BV_gamma}
\end{equation}
where the particle production rate $\Gamma$ is given by Eq.\ (\ref{eq:Gamma_w=0_hb_0}).

In fact, Eqs.\ (\ref{eq:drho_General_BV}) and (\ref{eq:drho_General_BV_gamma}) are equivalent to those for a pure BV model.
Consequently, adopting units where $c=1$,  the time evolution equation for the matter density contrast  $\delta \equiv \delta \rho_{m} /\rho_{m} $, namely the perturbation growth factor, is given by \cite{Lima2011} 
\begin{align}
\ddot{\delta}  & + \left [ 2 H + \Gamma + 3 c_{\rm{eff}}^{2} H  - \frac{ \Gamma \dot{H} - H \dot{\Gamma} }{ H (3H -\Gamma)}  \right ] \dot{\delta}        \notag \\
                     & +  \Bigg \{      3 (\dot{H} + 2 H^{2}) \left (   c_{\rm{eff}}^{2}  + \frac{ \Gamma }{ 3H } \right )     \notag \\
                     & + 3 H  \left [ \dot{c}_{\rm{eff}}^{2}   - ( 1 + c_{\rm{eff}}^{2}) \frac{ \Gamma \dot{H} - H \dot{\Gamma} }{ H (3H -\Gamma)}  \right ] \notag \\     
                     &  - 4 \pi G \rho \left ( 1 -  \frac{ \Gamma }{ 3H } \right ) ( 1 + 3 c_{\rm{eff}}^{2} )  + \frac{  k^{2} c_{\rm{eff}}^{2} }{ a^{2} }    \Bigg \}     \delta = 0              .
\label{eq:delta-t_BV}
\end{align}
Here $\rho$ is the mass density of matter. 
Specifically, $\rho_{m}$ is replaced by $\rho$ because a matter-dominated universe is considered \cite{Koma6,Koma16}.
In addition, $\rho$ in Eq.\ (\ref{eq:delta-t_BV}) represents $\bar{\rho}$, corresponding to a homogenous and isotropic solution for the unperturbed equations \cite{Koma6,Koma16}.
In this study, the effective sound speed, $c_{\rm{eff}}^{2} \equiv  \delta p_{c} /\delta \rho $, is set to 
\begin{equation}
 c_{\rm{eff}}^{2}  \equiv  \frac{\delta p_{c} }{\delta \rho } =0    ,
\label{eq:ceff2_BV_0}
\end{equation}
to ensure equivalence between the neo-Newtonian and general relativistic approaches \cite{Reis_2003}. 
The neo-Newtonian equation given by Eq.\ (\ref{eq:delta-t_BV}) is equivalent to the general relativistic equation for a single-fluid-dominated universe only when $c_{\rm{eff}}^{2} = 0$, as examined by Reis \cite{Reis_2003}. 
The equivalence is discussed by Ramos \textit{et al.} \cite{Ramos_2014}, as described in a previous work \cite{Koma16}.

For numerical purposes, we use an independent variable \cite{Lima2011} that is defined as 
\begin{equation}
\eta \equiv \ln [\tilde{a}(t)] ,
\label{eta_def}
\end{equation}
where $\tilde{a}(t)$ is the normalized scale factor $a/a_{0}$ given by Eq.\ (\ref{def_a_a0}).
From this definition, $\dot{\delta}$ and $\ddot{\delta}$ are written as \cite{Koma16}
\begin{equation}
     \dot{\delta} = H \frac{d \delta}{d \eta} = H \delta^{\prime}  \quad \textrm{and} \quad  \ddot{\delta} =  H^{2} \delta^{\prime \prime}  + H^{\prime} H \delta^{\prime}   ,
\label{delta12}
\end{equation}
where $^{\prime}$ represents the differential with respect to $\eta$, namely $d/d \eta$.
Equation\ (\ref{delta12}) can be written as $\dot{x}  = H x^{\prime}$ and  $ \ddot{x} =  H^{2} x^{\prime \prime}  + H^{\prime} H x^{\prime}$ by using an arbitrary variable $x$.
Also, from Eq.\ (\ref{eq:hB_FRW01}), the Friedmann equation for the BV model with $\Lambda$ is written as 
\begin{equation}
 4 \pi G \rho =   \frac{3}{2}  [H^{2} - {f}_{\Lambda} (t) ]  ,
\label{eq:rho_BV_Lambda}
\end{equation}
where ${f}_{\Lambda} (t)$ is a constant given by $\Lambda/3$.
Applying these equations and $c_{\rm{eff}}^{2} = 0$ to Eq.\ (\ref{eq:delta-t_BV}) and performing several operations yields \cite{Koma7} 
\begin{equation}
\delta^{\prime \prime}  + F_{\textrm{B}} (\eta) \delta^{\prime}  +  G_{\textrm{B}} (\eta) \delta =0, 
\label{eq:delta-eta_c=0_BV_1}
\end{equation}
where $F_{\textrm{B}} (\eta)$ and $G_{\textrm{B}} (\eta)$ are given by 
\begin{equation}
F_{\textrm{B}} (\eta) =  2  + \frac{ \Gamma + H^{\prime} }{ H }   -   \frac{  \Gamma H^{\prime} - H \Gamma^{\prime}  }{ H (3H -\Gamma) }   ,
\label{eq:FB(eta)_c=0_1}
\end{equation}
\begin{align}
G_{\textrm{B}} (\eta)  =  &   \left (  \frac{ H^{\prime} }{ H }  + 2   \right )     \frac{ \Gamma }{ H }       -     \frac{ 3( \Gamma H^{\prime} - H \Gamma^{\prime} ) }{ H (3H -\Gamma) }   \notag \\
                &    - \frac{ 4 \pi G \rho }{ H^{2} } \left ( 1 -  \frac{ \Gamma }{ 3H } \right )    .
\label{eq:G(eta)_c=0_PRD90}
\end{align}
Equations\ (\ref{eq:delta-eta_c=0_BV_1})--(\ref{eq:G(eta)_c=0_PRD90}) are used to examine density perturbations in a BV model with $\Lambda$ \cite{Koma7}.

$F_{\textrm{B}} (\eta)$ given by Eq.\ (\ref{eq:FB(eta)_c=0_1}) is equivalent to that for a pure BV model examined in Ref.\ \cite{Lima2011}.
However, $G_{\textrm{B}} (\eta)$ given by Eq.\ (\ref{eq:G(eta)_c=0_PRD90}) is slightly different from that for the pure BV model because the BV model with $\Lambda$ is considered.
In fact, from Eq.\ (\ref{eq:rho_BV_Lambda}), $4 \pi G \rho$ in Eq.\ (\ref{eq:G(eta)_c=0_PRD90}) includes ${f}_{\Lambda} (t)=\Lambda/3$.
When $\Lambda=0$ is considered, $G_{\textrm{B}} (\eta)$ reduces to that for the pure BV model \cite{Lima2011,Koma6,Koma16}:
\begin{equation}
G_{\textrm{B,pure},\Lambda=0} (\eta)  =  \left (  \frac{ \Gamma }{ H }  - 1   \right )   \left (  \frac{ \Gamma }{ 2H }  + \frac{3}{2}   \right )   -   \frac{ 3 ( \Gamma H^{\prime} - H \Gamma^{\prime} ) }{ H (3H -\Gamma) }   .
\label{eq:GB(eta)_c=0_1}
\end{equation}
Also, $F_{\textrm{B}} (\eta)$ given by Eq.\ (\ref{eq:FB(eta)_c=0_1}) and $G_{\textrm{B}} (\eta)$ given by Eq.\ (\ref{eq:G(eta)_c=0_PRD90}) reduce to those for $\Lambda$CDM models when $\Gamma=0$.

In this subsection, we reviewed density perturbations in a BV model with $\Lambda$, where a general dissipative driving term $ h_{\textrm{B}} (t)$ was considered.
In the next subsection, the formulation is applied to the present model.

\subsection{Formulations for the present model}
\label{Formulations for the present model}

We apply the formulation for a BV model with $\Lambda$ to the present model.
From Eq.\ (\ref{eq:hB_beta_fL=cst}), the dissipative driving term $h_{\textrm{B}}(t)$ for the present model is written as
\begin{align}
                      h_{\textrm{B}}(t)   &=  \beta (2 H^{2} + \dot{H})  .
\label{eq:hB_beta_fL=cst_2}      
\end{align}

We now calculate $F_{\textrm{B}} (\eta)$ and $G_{\textrm{B}} (\eta)$ for the present model, using Eqs.\ (\ref{eq:FB(eta)_c=0_1}) and (\ref{eq:G(eta)_c=0_PRD90}).
For this, we calculate four terms included in $F_{\textrm{B}} (\eta)$ and $G_{\textrm{B}} (\eta)$.
The four terms are written as $4 \pi G \rho / H^{2}$, $H^{\prime}/H$, $\Gamma / H$, and $\frac{  \Gamma H^{\prime} - H \Gamma^{\prime}  }{ H (3H -\Gamma) }$.
The former two terms include $H$, whereas the latter two terms include both $H$ and $\Gamma$.

To calculate the former two terms, we write the background evolution again.
From Eq.\ (\ref{eq:Sol_HH0}), the evolution of the Hubble parameter can be written as 
\begin{align}  
 \frac{H}{H_{0}}
                            &=  \left [   (  1-  \Omega_{\Lambda,\gamma}  )  \tilde{a}^{ - D}      + \Omega_{\Lambda,\gamma}  \right ]^{\frac{1}{2}}     \notag \\
                            &=  \left [   (1- \Omega_{\Lambda,\gamma}  )            e^{- D \eta}   + \Omega_{\Lambda,\gamma}   \right ]^{\frac{1}{2}}     ,  
\label{eq:Sol_HH0_3}
\end{align}
where $\tilde{a}^{- D}$ is replaced by $e^{- D \eta}$ using $\eta \equiv \ln [\tilde{a}(t)]$ given by Eq.\ (\ref{eta_def}).
Differentiating Eq.\ (\ref{eq:Sol_HH0_3}) with respect to $\eta$ yields
\begin{align}  
\frac{ H^{\prime} }{H_{0}} &= \frac{d}{d \eta} \left ( \frac{H}{H_{0}} \right )  =  \frac{d}{d \eta} \left [  (  1-  \Omega_{\Lambda,\gamma}  )   e^{- D \eta}   + \Omega_{\Lambda,\gamma}  \right ]^ { \frac{1}{2} }               \notag \\
                                    &=  \frac{(- D)  (  1-  \Omega_{\Lambda,\gamma}  )   e^{- D \eta}   }{2}    \left [  (  1-  \Omega_{\Lambda,\gamma}  )   e^{- D \eta}   + \Omega_{\Lambda,\gamma}  \right ]^ {- \frac{1}{2}}         . 
\label{eq:Hp-H0_present}
\end{align}
Dividing Eq.\ (\ref{eq:Hp-H0_present}) by Eq.\ (\ref{eq:Sol_HH0_3}) yields
\begin{align}  
\frac{H^{\prime} }{H}          &= \frac{ \frac{(- D)  (  1-  \Omega_{\Lambda,\gamma}  )   e^{- D \eta}   }{2}    \left [  (  1-  \Omega_{\Lambda,\gamma}  )   e^{- D \eta}   + \Omega_{\Lambda,\gamma}  \right ]^ {- \frac{1}{2}}         }
                                                     {  \left [   (1- \Omega_{\Lambda,\gamma}  )            e^{- D \eta}   + \Omega_{\Lambda,\gamma}   \right ]^{\frac{1}{2}}     }  \notag \\ 
                                         &= \frac{ \frac{- D}{2}  (  1-  \Omega_{\Lambda,\gamma}  )   e^{- D \eta}       }{    (1- \Omega_{\Lambda,\gamma}  )            e^{- D \eta}   + \Omega_{\Lambda,\gamma}   }                                           
                                         = \frac{ \frac{- D}{2}  (  1-  \Omega_{\Lambda,\gamma}  )         }{    (1- \Omega_{\Lambda,\gamma}  )              + \Omega_{\Lambda,\gamma} e^{D \eta}   }          .
\label{eq:Hp-H_present_a_eta_1}
\end{align}
Using Eq.\ (\ref{eq:rho_BV_Lambda}) and ${f}_{\Lambda} (t) =\Lambda/3$ and applying Eq.\ (\ref{eq:Sol_HH0_3}) yields $4 \pi G \rho / H^{2}$, given by
\begin{align}  
\frac{4 \pi G \rho}{H^{2}} &= \frac{3}{2} - \frac{3}{2} \frac{\frac{\Lambda}{3}}{H^2} = \frac{3}{2} - \frac{3}{2} \frac{  \frac{\Lambda}{3H_{0}^{2}} }{\frac{H^2}{H_{0}^{2}} }  \notag \\
                                      &=  \frac{3}{2} - \frac{3}{2} \frac{\Omega_{\Lambda}  }{ (1- \Omega_{\Lambda,\gamma}  )            e^{- D \eta}   + \Omega_{\Lambda,\gamma}   }  \notag \\
                                      &=  \frac{3}{2} - \frac{3}{2} \frac{\Omega_{\Lambda} e^{D \eta}  }{ (1- \Omega_{\Lambda,\gamma}  )        + \Omega_{\Lambda,\gamma} e^{D \eta}    }  ,
\label{eq:}
\end{align}
where $\Omega_{\Lambda}= \frac{\Lambda}{3 H_{0}^{2}}$ given by Eq.\ (\ref{eq:Omega_Lgamma}) is also used.

Next, we calculate the latter two terms, namely $\Gamma / H$ and $\frac{  \Gamma H^{\prime} - H \Gamma^{\prime}  }{ H (3H -\Gamma) }$.
From Eq.\ (\ref{eq:Gamma_Present}), $\Gamma / H$ is given by
\begin{align}
   \frac{\Gamma}{H}    &=  \beta   \left ( \frac{    (4-D) \tilde{H}^{2}   +D \Omega_{\Lambda,\gamma}        }{\tilde{H}^{2} - \Omega_{\Lambda}   } \right )   .
\label{eq:GammabyH_Present}
\end{align}
Substituting Eq.\ (\ref{eq:Sol_HH0_3}) into Eq.\ (\ref{eq:GammabyH_Present}) yields 
\begin{align}
     \frac{  \Gamma }{ H }  &= \beta   \left ( \frac{    (4-D) [ (1- \Omega_{\Lambda,\gamma}  )            e^{- D \eta}   + \Omega_{\Lambda,\gamma}   ]   +D \Omega_{\Lambda,\gamma}        }{ (1- \Omega_{\Lambda,\gamma}  )            e^{- D \eta}   + \Omega_{\Lambda,\gamma}    - \Omega_{\Lambda}   } \right )   \notag \\                          
                                      &=  \beta   \left ( \frac{ (4-D)  (1- \Omega_{\Lambda,\gamma}  )   + 4 \Omega_{\Lambda,\gamma} e^{D \eta}       }{ (1- \Omega_{\Lambda,\gamma}  )     +  (\Omega_{\Lambda,\gamma} - \Omega_{\Lambda}) e^{D \eta}    } \right )                                   .
\label{eq:Gamma-H_present_a_eta}
\end{align}
We calculate $\Gamma^{\prime} / H$ and $(\Gamma/H)^{\prime}$, which are used for determining $\frac{  \Gamma H^{\prime} - H \Gamma^{\prime}  }{ H (3H -\Gamma) }$.
Differentiating $\Gamma = (\Gamma/H) H$ with respect to $\eta$ and dividing the resultant equation by $H$ yields 
\begin{equation}
   \frac{ \Gamma^{\prime} }{H} =  \left ( \frac{\Gamma}{H} \right )^{\prime} + \frac{\Gamma}{H}  \frac{ H^{\prime} }{H}   .
\label{eq:Gammaprime-H_0}
\end{equation}
$\Gamma^{\prime} / H$ includes $(\Gamma/H)^{\prime}$.
Applying Eq.\ (\ref{eq:Gamma-H_present_a_eta}) to $(\Gamma/H)^{\prime}$ and performing several operations yields 
\begin{equation}
   \left (\frac{ \Gamma}{H}  \right )^{\prime}    = \frac{  \beta  D  e^{D \eta}  (1- \Omega_{\Lambda,\gamma}) [ 4 \Omega_{\Lambda} + D (\Omega_{\Lambda,\gamma} - \Omega_{\Lambda}) ]      }{ \left [ (1- \Omega_{\Lambda,\gamma}  )     +  (\Omega_{\Lambda,\gamma} - \Omega_{\Lambda}) e^{D \eta} \right]^{2}   }      .  
\label{eq:(Gamma-H)prime}
\end{equation}
We now calculate $\frac{  \Gamma H^{\prime} - H \Gamma^{\prime}  }{ H (3H -\Gamma) }$.
Arranging $\frac{  \Gamma H^{\prime} - H \Gamma^{\prime}  }{ H (3H -\Gamma) }$ and substituting Eq.\ (\ref{eq:Gammaprime-H_0}) into the resultant equation yields
\begin{align}
 \frac{  \Gamma H^{\prime} - H \Gamma^{\prime}  }{ H (3H -\Gamma) }  &=  \frac{  \frac{\Gamma}{H} \frac{ H^{\prime} }{H} -    \frac{ \Gamma^{\prime} }{H} }{ 3 -  \frac{\Gamma}{H} } 
                                                                                                          =  \frac{  \frac{\Gamma}{H} \frac{ H^{\prime} }{H} -    \left ( \frac{\Gamma}{H} \right )^{\prime} - \frac{\Gamma}{H}  \frac{ H^{\prime} }{H}   }{ 3 -  \frac{\Gamma}{H} }  \notag  \\
                                                                                                        &=  \frac{  - \left ( \frac{\Gamma}{H} \right )^{\prime} }{ 3 -  \frac{\Gamma}{H} }  .
\label{eq:Complex_present_0}
\end{align}
In addition, substituting $\Gamma/H$ given by Eq.\ (\ref{eq:Gamma-H_present_a_eta}) and $(\Gamma/H)^{\prime}$ given by Eq.\ (\ref{eq:(Gamma-H)prime}) into Eq.\ (\ref{eq:Complex_present_0}) yields
\begin{align}
 & \frac{     \Gamma H^{\prime} - H \Gamma^{\prime}  }{ H (3H -\Gamma) }  =   -   \frac{  \beta  D  e^{D \eta}  (1- \Omega_{\Lambda,\gamma}) [ 4 \Omega_{\Lambda} + D (\Omega_{\Lambda,\gamma} - \Omega_{\Lambda}) ]      }{   (1- \Omega_{\Lambda,\gamma}  )     +  (\Omega_{\Lambda,\gamma} - \Omega_{\Lambda}) e^{D \eta}    }    \notag  \\
                                                                                                              &\times   \frac{1}{ [ 3-\beta(4-D) ] (1- \Omega_{\Lambda,\gamma}  )    +  [ (3-4\beta)\Omega_{\Lambda,\gamma} - 3 \Omega_{\Lambda}] e^{D \eta}            }   .   
\label{eq:Complex_present_1}
\end{align}

In this way, we have obtained the four terms, $4 \pi G \rho / H^{2}$, $H^{\prime}/H$, $\Gamma / H$, and $\frac{  \Gamma H^{\prime} - H \Gamma^{\prime}  }{ H (3H -\Gamma) }$.
From these results, we calculate $F_{\textrm{B}} (\eta)$ given by Eq.\ (\ref{eq:FB(eta)_c=0_1}) and $G_{\textrm{B}} (\eta)$ given by Eq.\ (\ref{eq:G(eta)_c=0_PRD90}) and numerically solve the differential equation given by Eq.\ (\ref{eq:delta-eta_c=0_BV_1}).
To solve this differential equation, the initial conditions for the Einstein--de Sitter growing model are used \cite{Lima2011}. 
The initial conditions are given by \cite{Lima2011,Koma6,Koma7,Koma16}
\begin{align}
 \delta (\tilde{a}_{i}) = \tilde{a}_{i}  \quad \textrm{and} \quad  \delta^{\prime}  (\tilde{a}_{i}) = \tilde{a}_{i}, 
\label{eq:ICforSolve}
\end{align}
where $\tilde{a}_{i}$ is set to $10^{-3}$.

Evolutions of density perturbations in the present model are discussed in the next section.
For this, we introduce useful parameters to examine density perturbations.
From the obtained $\delta$, we calculate an indicator of clustering, namely the growth rate $f(z)$ \cite{Lima2011,Koma7,Koma16}.
Here $z$ is the redshift, given by $ z = \tilde{a}^{-1} -1$.
The growth rate $f(z)$ is given by \cite{Peebles_1993} 
\begin{equation}
 f(z) = \frac{d \ln \delta }{  d \ln a } = - (1 + z ) \frac{d \ln \delta }{  dz }     . 
\label{eq:f(z)}
\end{equation}
In addition, we calculate a combination value $f(z) \sigma_{8}(z)$ \cite{Koma16}.
Here $\sigma_{8} (z)$ is the redshift-dependent root-mean-square (rms) fluctuations of the linear density field within a sphere of radius $R = 8h^{-1}$ Mpc \cite{Nesseris2017} (where $h$ is the reduced Hubble constant defined by $h = H_{0} / 100$).
The redshift-dependent rms fluctuations $\sigma_{8}(z)$ can be written as \cite{Basilakos2014}
\begin{equation}
\sigma_{8}(z) =  \sigma_{8}  \left [ \frac{\delta (z)}{\delta (z=0)} \right ]      , 
\label{eq:sigma8(z)}
\end{equation}
where $\sigma_{8}$ is $\sigma_{8} (z)$ at redshift $z=0$.
We set $\sigma_{8} =0.811$ from the Planck 2018 results \cite{Planck2018}, as examined in Ref.\ \cite{Koma16}.
These parameters are used in the next section.

In this section, we examined first-order density perturbations by applying a neo-Newtonian approach to the present model.
Therefore, we can discuss observational constraints on the present model, using the background evolution of the universe and density perturbations.
In the next section, we perform a chi-squared analysis in the $(\Omega_{\Lambda,\gamma}, \beta)$ plane and systematically examine observational constraints on the present model.
In addition, we discuss the result with the transitional and thermodynamic constraints.
A likelihood analysis is also performed.

\section{Observational, transitional, and thermodynamic constraints}
\label{Observational, transitional, and thermodynamic constraints}

In this section, we provide an approximate estimate of the constraints on the present model.
For this, we consider observational, transitional, and thermodynamic constraints.

To examine the observational constraints, we perform a chi-squared analysis using a distance modulus $\mu(z)$, the Hubble parameter $H(z)$, and a combination value $f(z) \sigma_{8}(z)$.
The distance modulus $\mu(z)$ is used to examine the background evolution of the universe and is defined as 
\begin{equation}
 \mu (z) = 5 \log d_{L}(z) + 25   ,
\end{equation}
where the luminosity distance $d_{L}(z)$ is written as \cite{Sato1} 
\begin{equation}
  \left ( \frac{ H_{0} }{ c } \right )   d_{L}(z)      =   (1+z)  \int_{1}^{1+z}  \frac{dy} { F(y) }    .
\label{eq:dL}  
\end{equation}
The integrating variable $y$ and the function $F(y)$ are given by $y =  \tilde{a}^{-1}$ and $F(y)  = \tilde{H}$, respectively.

The observed distance modulus $\mu(z)$ is obtained from the Dark Energy Survey Supernova 5-year (DES-SN5YR) dataset \cite{DES-SN5YR}.
This dataset consists of $1635$ points ($z > 0.1$) and a small set of $194$ points ($z < 0.1$), a total of $1829$ points \cite{DES-SN5YR}.
For the Hubble parameter, we use observational $H(z)$ data with $57$ recently compiled data points, which are summarized in, e.g., Refs.\ \cite{H(z)57data,57dataOdin}.
The $57$ points consist of $31$ points from a differential age technique and $26$ points evaluated using baryon acoustic oscillations (BAO) and other methods \cite{H(z)57data}.
We also use observational $f(z) \sigma_{8}(z)$ data with $22$ points, which are summarized in, e.g., Ref.\ \cite{SPaulDas2025}.
The $22$ points are based on several different observational surveys \cite{SPaulDas2025}.

\begin{figure*}[htbp]
    \begin{tabular}{cc}
      \begin{minipage}[t]{0.45\hsize}
        \centering
        \includegraphics[width=\textwidth]{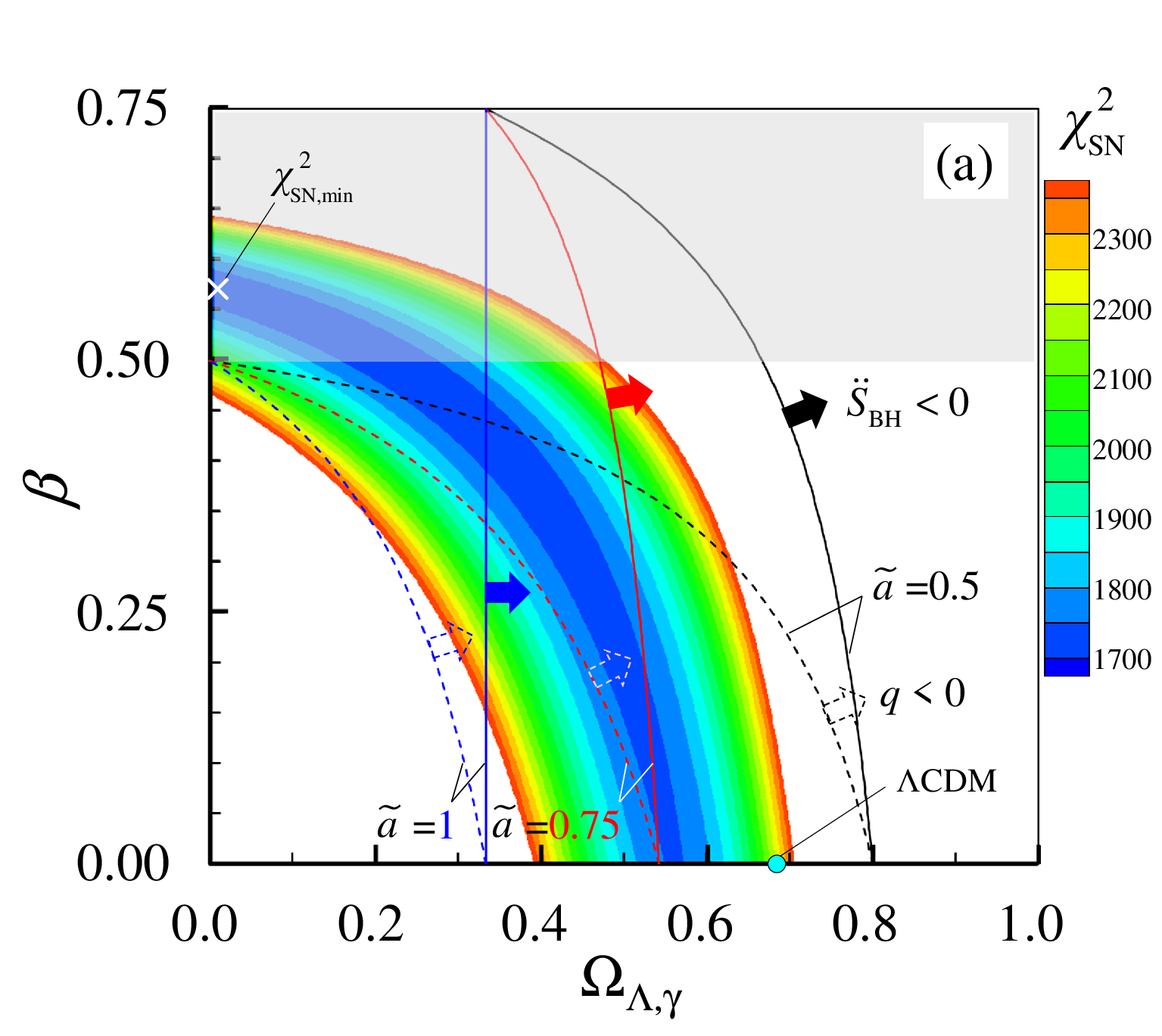}
(a) $\chi_{\textrm{SN}}^{2}$
      \end{minipage} &
      \begin{minipage}[t]{0.45\hsize}
        \centering
        \includegraphics[width=\textwidth]{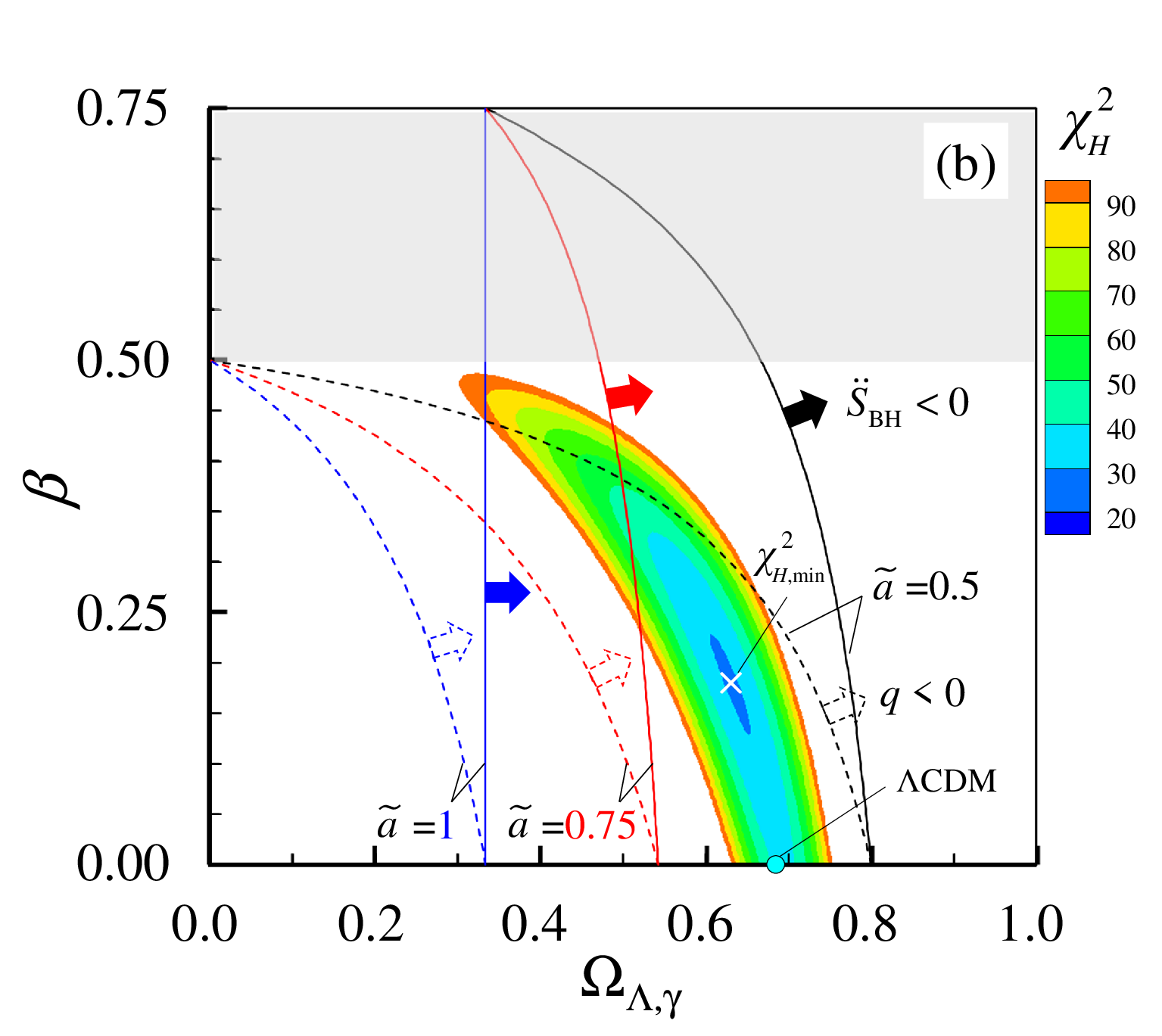}
(b) $\chi_{H}^{2}$
      \end{minipage} \\
   
      \begin{minipage}[t]{0.45\hsize}
        \centering
        \includegraphics[width=\textwidth]{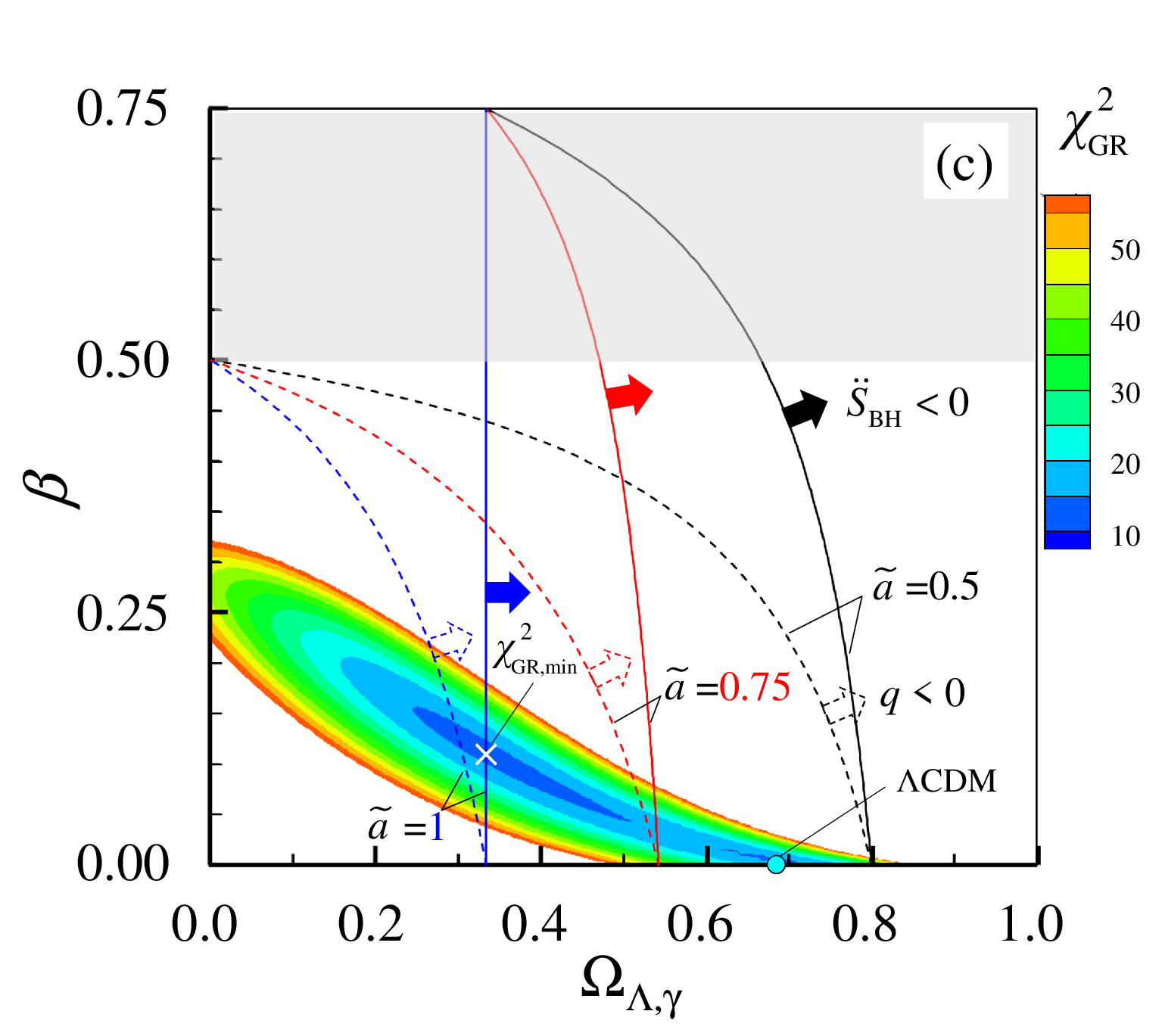}
(c) $\chi_{\textrm{GR}}^{2}$
      \end{minipage} &
      \begin{minipage}[t]{0.45\hsize}
        \centering
        \includegraphics[width=\textwidth]{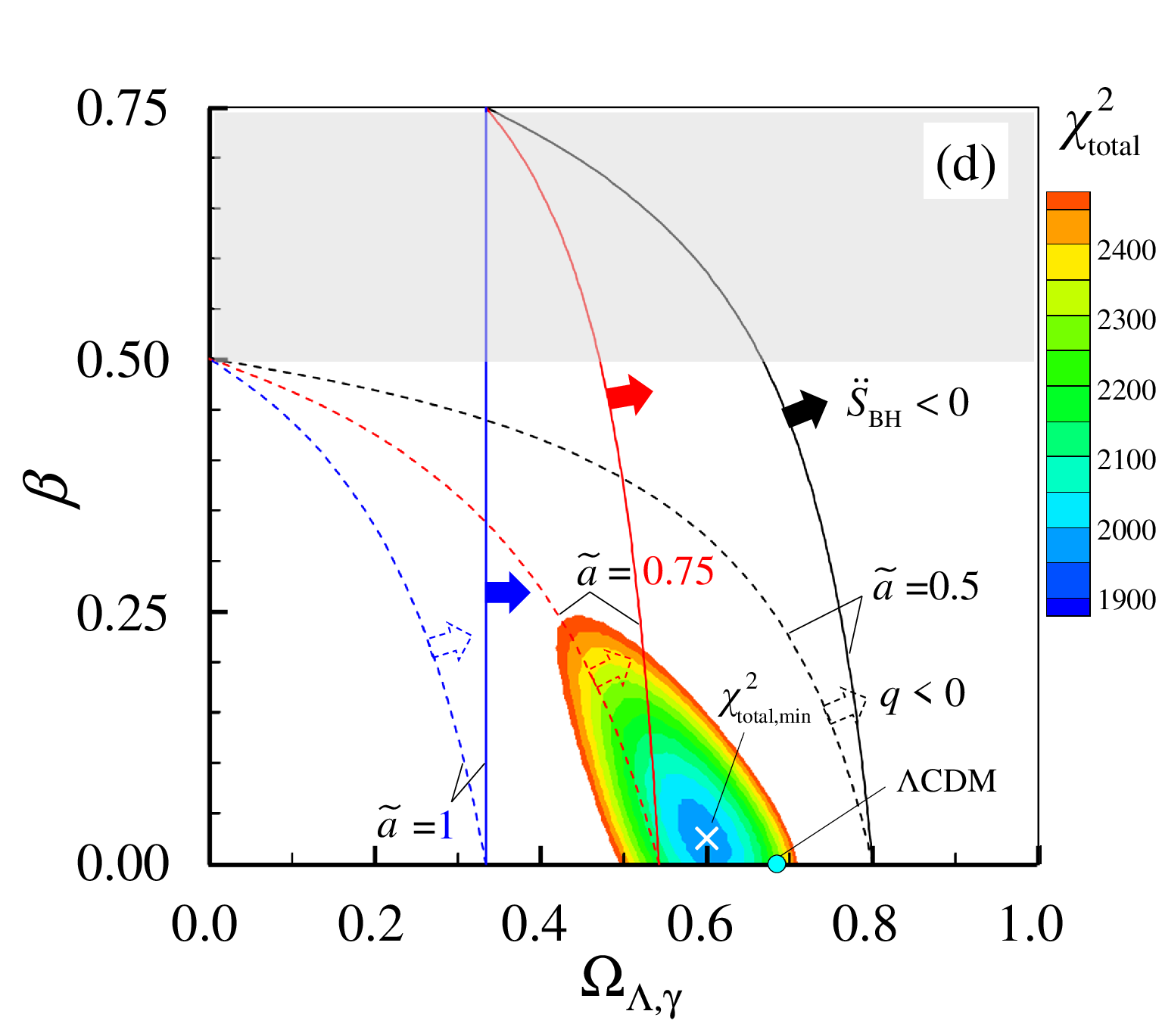}
(d) $\chi_{\textrm{total}}^{2}$ 
      \end{minipage} 
    \end{tabular}
\caption{
Contours of $\chi_{\textrm{SN}}^{2}$, $\chi_{H}^{2}$, $\chi_{\textrm{GR}}^{2}$, and $\chi_{\textrm{total}}^{2}$ in the $(\Omega_{\Lambda,\gamma}, \beta)$ plane.
Small-$\chi_{\textrm{SN}}^{2}$ $(\chi_{\textrm{SN}}^{2} < 2400)$, small-$\chi_{H}^{2}$ ($\chi_{H}^{2} < 100$), small-$\chi_{\textrm{GR}}^{2}$ $(\chi_{\textrm{GR}}^{2} < 60)$, and small-$\chi_{\textrm{total}}^{2}$ $(\chi_{\textrm{total}}^{2} < 2500)$ regions are displayed in (a), (b), (c), and (d), respectively.
The x mark indicates the location of the minimum value of each chi-squared function.
The closed circle on the $\Omega_{\Lambda,\gamma}$ axis indicates $(\Omega_{\Lambda,\gamma}, \beta) = (0.685, 0)$ for the $\Lambda$CDM model. 
The solid and dashed lines represent the boundaries of $\ddot{S}_{\rm{BH}} = 0$ and $q = 0$, respectively, from Fig.\ \ref{Fig-d2SBHdt2_plane}.
The arrow attached to each solid-line boundary indicates a region that satisfies $\ddot{S}_{\rm{BH}} < 0$. 
The arrow attached to each dashed-line boundary indicates a region that satisfies $q< 0$.
The region in gray ($\beta \geq 0.5$) does not satisfy a transition from a decelerating universe to an accelerating universe.
We can convert $\beta$ into $\gamma = 4 \beta/3$, using Eq.\ (\ref{eq:gamma}) and $w=0$.
The minimum value of each chi-squared function and its location $(\Omega_{\Lambda,\gamma}, \beta)$ are summarized in Table\ \ref{tab-chi2min}.  
}
\label{Fig-SN_H_GF_total_dS2BH=0_plane}
\end{figure*}

The chi-squared function $\chi_{\textrm{SN}}^{2}$ for the supernovae is given by
\begin{equation}
\chi_{\textrm{SN}}^{2} (\Omega_{\Lambda,\gamma}, \beta)
= \sum\limits_{i=1}^{1829} { \left[ \frac{   
                                                         \mu_{\textrm{obs}} (z_{i} ) - \mu_{\textrm{cal}} (z_{i}, \Omega_{\Lambda,\gamma}, \beta)  }{ \sigma_{\textrm{SN}}(z_{i}) }   
                                     \right]^{2}   }   ,
\label{eq:chi_SN}
\end{equation}
where $\mu_{\textrm{obs}} (z_{i} )$ and $ \mu_{\textrm{cal}} (z_{i}, \Omega_{\Lambda,\gamma}, \beta) $ are the observed and calculated distance moduli, respectively, and $\sigma_{\textrm{SN}}(z_{i})$ is the uncertainty in the observed distance modulus.
The observed data points (numbered $i=1$ to $1829$) are taken from Ref.\ \cite{DES-SN5YR}.

The chi-squared function $\chi_{H}^{2}$ for the Hubble parameter is given by
\begin{equation}
\chi_{H}^{2} (\Omega_{\Lambda,\gamma}, \beta)
= \sum\limits_{i=1}^{57} { \left[ \frac{   
                                                         H_{\textrm{obs}} (z_{i} ) - H_{\textrm{cal}} (z_{i}, \Omega_{\Lambda,\gamma}, \beta ) }{ \sigma_{H}(z_{i}) }   
                                     \right]^{2}   }   ,
\label{eq:chi_H}
\end{equation}
where $H_{\textrm{obs}} (z_{i} )$ and $H_{\textrm{cal}} (z_{i}, \Omega_{\Lambda,\gamma}, \beta ) $ are the observed and calculated Hubble parameters, respectively, and $\sigma_{H}(z_{i})$ is the uncertainty in the observed Hubble parameter.
The observed data points (numbered $i=1$ to $57$) are taken from the summary in Ref.\ \cite{57dataOdin} and shown in Fig.\ \ref{Fig-H-a}.
Each original data point and its reference are summarized in Ref.\ \cite{57dataOdin}.

\begin{table*}[tb]
\caption{Minimum value of each chi-squared function and its location $(\Omega_{\Lambda,\gamma}, \beta)$.
In this analysis, $\Omega_{\Lambda,\gamma}$ and $\beta$ are sampled in steps of $0.005$.
The location of $\chi_{\textrm{total,min}}^{2}$ corresponds to the combined best-fit.
The location of the minimum value for each chi-squared function is plotted in Fig.\ \ref{Fig-SN_H_GF_total_dS2BH=0_plane}.
For each location, $1\sigma$ ($2\sigma$) confidence levels are shown in increments of $0.005$.
Note that the contours of the $1 \sigma $, $2 \sigma$, and $3 \sigma$ confidence levels based on a likelihood analysis are plotted in Fig.\ \ref{Fig-L_plane}.
 }
\label{tab-chi2min}
\newcommand{\m}{\hphantom{$-$}}
\newcommand{\cc}[1]{\multicolumn{1}{c}{#1}}
\renewcommand{\tabcolsep}{1.2pc} 
\renewcommand{\arraystretch}{1.9} 
\begin{tabular}{lrc}
\hline
\hline
    Parameter                             &  Minimum value      &  $(\Omega_{\Lambda,\gamma}, \beta)$   \\
\hline
   $\chi_{\textrm{SN,min}}^{2}$       &  $1716.1$             &  $(0.010^{+0.155 (0.285)}_{-0.010 (0.010)},   0.570^{+0.010 (0.015)}_{-0.050 (0.115)})$      \\          
   $\chi_{H,\textrm{min}}^{2}$         &  $29.1$                &  $(0.630^{+0.040 (0.060)}_{-0.045 (0.075)},   0.180^{+0.075 (0.120)}_{-0.090 (0.160)})$      \\
   $\chi_{\textrm{GR,min}}^{2}$       &  $13.8$                &  $(0.335^{+0.420 (0.450)}_{-0.120 (0.180)},   0.110^{+0.065 (0.100)}_{-0.110 (0.110)})$      \\
   $\chi_{\textrm{total,min}}^{2}$    &  $1965.2$             &  $(0.600^{+0.010 (0.015)}_{-0.015 (0.020)},   0.025^{+0.015 (0.020)}_{-0.010 (0.015)})$      \\
\hline
\hline
\end{tabular}\\
\end{table*}
%

The chi-squared function $\chi_{\textrm{GR}}^{2}$ for the growth rate is given by
\begin{align}
& \chi_{\textrm{GR}}^{2}(\Omega_{\Lambda,\gamma}, \beta) = \notag \\
& \sum\limits_{i=1}^{22} { \left[ \frac{   
                                                         f_{\textrm{obs}}(z_{i} ) \sigma_{8}^{\textrm{obs}} (z_{i} ) - f_{\textrm{cal}} (z_{i}, \Omega_{\Lambda,\gamma}, \beta) \sigma_{8}^{\textrm{cal}} (z_{i}, \Omega_{\Lambda,\gamma}, \beta)  }{ \sigma_{\textrm{GR}}(z_{i}) }   
                                     \right]^{2}   }   .
\label{eq:chi_GR_fsigma8}
\end{align}
Here $f_{\textrm{obs}}(z_{i} ) \sigma_{8}^{\textrm{obs}} (z_{i} )$ and $  f_{\textrm{cal}} (z_{i}, \Omega_{\Lambda,\gamma}, \beta) \sigma_{8}^{\textrm{cal}} (z_{i}, \Omega_{\Lambda,\gamma}, \beta) $ are the observed and calculated values, respectively, and $\sigma_{\textrm{GR}}(z_{i})$ is the uncertainty in the observed value.
The observed data points (numbered $i=1$ to $22$) are taken from Ref.\ \cite{SPaulDas2025} and are shown in Fig.\ \ref{Fig-f(z)_f(z)sigma8-z_present}(b).
Each original data point and its reference are summarized in Ref.\ \cite{SPaulDas2025}.

In addition, a joint chi-squared analysis is performed using the three chi-squared functions.
For the joint chi-squared analysis, the combined chi-squared function $ \chi_{\textrm{total}}^{2}$ is defined by
\begin{equation}
 \chi_{\textrm{total}}^{2} =   \chi_{\textrm{SN}}^{2} +\chi_{H}^{2} + \chi_{\textrm{GR}}^{2}   .
\label{eq:chi_total}
\end{equation}
For these analyses, $H_{0}$ is set to $67.4$ km/s/Mpc from the Planck 2018 results \cite{Planck2018}, and both $\Omega_{\Lambda,\gamma}$ and $\beta$ are treated as free parameters.
$\Omega_{\Lambda,\gamma}$ is sampled in the range $0 \leq \Omega_{\Lambda,\gamma} < 1$ in steps of $0.005$ and $\beta$ is sampled in the range $0 \leq \beta < 0.75$ in steps of $0.005$.

We now examine the constraints on the present model.
Figure\ \ref{Fig-SN_H_GF_total_dS2BH=0_plane} shows the contours of $\chi_{\textrm{SN}}^{2}$, $\chi_{H}^{2}$, $\chi_{\textrm{GR}}^{2}$, and $\chi_{\textrm{total}}^{2}$ in the $(\Omega_{\Lambda,\gamma}, \beta)$ plane.
The x mark indicates the location of the minimum value of each chi-squared function.
The minimum value of each chi-squared function and its location $(\Omega_{\Lambda,\gamma}, \beta)$ are summarized in Table\ \ref{tab-chi2min}.  
Note that a likelihood analysis is performed later.

Firstly, we observe the contours of $\chi_{\textrm{SN}}^{2}$ for the supernovae.
Small-$\chi_{\textrm{SN}}^{2}$ regions (corresponding to $\chi_{\textrm{SN}}^{2} < 2400$) are displayed in Fig.\ \ref{Fig-SN_H_GF_total_dS2BH=0_plane}(a). 
The region surrounded by the contours indicates a favored region for $\chi_{\textrm{SN}}^{2}$.
The favored region includes a point $(\Omega_{\Lambda,\gamma}, \beta) = (0.685, 0)$, corresponding to the $\Lambda$CDM model.
In this analysis, the location for the minimum value of $\chi_{\textrm{SN}}^{2}$, namely $\chi_{\textrm{SN,min}}^{2}$, is $(\Omega_{\Lambda,\gamma}, \beta)=(0.010, 0.570)$.
The subscript `min' represents the minimum value.
The location for $\chi_{\textrm{SN,min}}^{2}$ deviates from the point $(\Omega_{\Lambda,\gamma}, \beta) = (0.685, 0)$.
In addition, the favored region for $\chi_{\textrm{SN}}^{2}$ partially overlaps with the region in gray ($\beta \geq 0.5$), which does not satisfy a transition from a decelerating universe to an accelerating universe.
However, when both $\Omega_{\Lambda,\gamma} \gtrapprox 0.2$ and $\beta < 0.5$ are considered, the favored region satisfies not only $q < 0$ and $\ddot{S}_{\rm{BH}} < 0$ but also the transitional and thermodynamic constraints at the present time ($\tilde{a}=1$).
Therefore, the favored region for $\chi_{\textrm{SN}}^{2}$ should be restricted to $\Omega_{\Lambda,\gamma} \approx 0.2$--$0.7$ and $\beta < 0.5$.
The restricted favored region includes a smaller-$\chi_{\textrm{SN}}^{2}$ region $(\chi_{\textrm{SN}}^{2} \leq 1750)$. 
We discuss this smaller-$\chi_{\textrm{SN}}^{2}$ region later.

Secondly, we observe the contours of $\chi_{H}^{2}$ for the Hubble parameter.
Small-$\chi_{H}^{2}$ regions (corresponding to $\chi_{H}^{2} < 100$) are displayed in Fig.\ \ref{Fig-SN_H_GF_total_dS2BH=0_plane}(b). 
In particular, $\Omega_{\Lambda,\gamma} \approx 0.6$--$0.7$ and $\beta \approx 0$--$0.3$ are likely favored.
The location for $\chi_{H,\textrm{min}}^{2}$ is $(\Omega_{\Lambda,\gamma}, \beta)=(0.630, 0.180)$ and slightly deviates from the point $(\Omega_{\Lambda,\gamma}, \beta) = (0.685, 0)$.
The favored region for $\chi_{H}^{2}$, $\Omega_{\Lambda,\gamma} \approx 0.6$--$0.7$, and $\beta \approx 0$--$0.3$ satisfies $q < 0$ and $\ddot{S}_{\rm{BH}} < 0$.
In fact, the favored region for $\chi_{H}^{2}$ satisfies the transitional and thermodynamic constraints not only at $\tilde{a}=1$ but also at $\tilde{a}=0.75$.

Thirdly, we observe the contours of $\chi_{\textrm{GR}}^{2}$ for the growth rate.
Small-$\chi_{\textrm{GR}}^{2}$ regions (corresponding to $\chi_{\textrm{GR}}^{2} < 60$) are displayed in Fig.\ \ref{Fig-SN_H_GF_total_dS2BH=0_plane}(c). 
In particular, $\Omega_{\Lambda,\gamma} \approx 0.1$--$0.8$ and $\beta \approx 0$--$0.2$ are likely favored.
The location for $\chi_{\textrm{GR,min}}^{2}$ is $(\Omega_{\Lambda,\gamma}, \beta)=(0.335, 0.110)$ and deviates from the point $(\Omega_{\Lambda,\gamma}, \beta) = (0.685, 0)$.
The favored region satisfies the transitional and thermodynamic constraints at $\tilde{a}=1$, except a small-$\Omega_{\Lambda,\gamma}$ region.
Consequently, the favored region for $\chi_{\textrm{GR}}^{2}$ should be restricted to $\Omega_{\Lambda,\gamma} \approx 0.3$--$0.8$ and $\beta \approx 0$--$0.2$.
Those results [namely Figs.\ \ref{Fig-SN_H_GF_total_dS2BH=0_plane}(a), (b), and (c)] imply that the present model for smaller $\beta$, close to the $\Lambda$CDM model, is consistent with the observations and satisfies the transitional and thermodynamic constraints.

Finally, to examine this expectation, we observe a combined chi-squared function $\chi_{\textrm{total}}^{2}$.
Small-$\chi_{\textrm{total}}^{2}$ regions (corresponding to $\chi_{\textrm{total}}^{2} < 2500$) are displayed in Fig.\ \ref{Fig-SN_H_GF_total_dS2BH=0_plane}(d).
In particular, $\Omega_{\Lambda,\gamma} \approx 0.5$--$0.7$ and $\beta \approx 0$--$0.1$ are likely favored.
The favored region satisfies the transitional and thermodynamic constraints at an earlier stage than $\tilde{a}=1$.
In addition, the location for $\chi_{\textrm{total,min}}^{2}$ is $(\Omega_{\Lambda,\gamma}, \beta)=(0.600, 0.025)$, corresponding to the combined best-fit, and slightly deviates from the point $(\Omega_{\Lambda,\gamma}, \beta)=(0.685, 0)$.
That is, a weakly dissipative universe (namely small $\beta$) with $\Lambda$ is consistent with the observations considered here and satisfies the transitional and thermodynamic constraints.

\begin{figure} [t] 
\begin{minipage}{0.495\textwidth}
\begin{center}
\scalebox{0.33}{\includegraphics{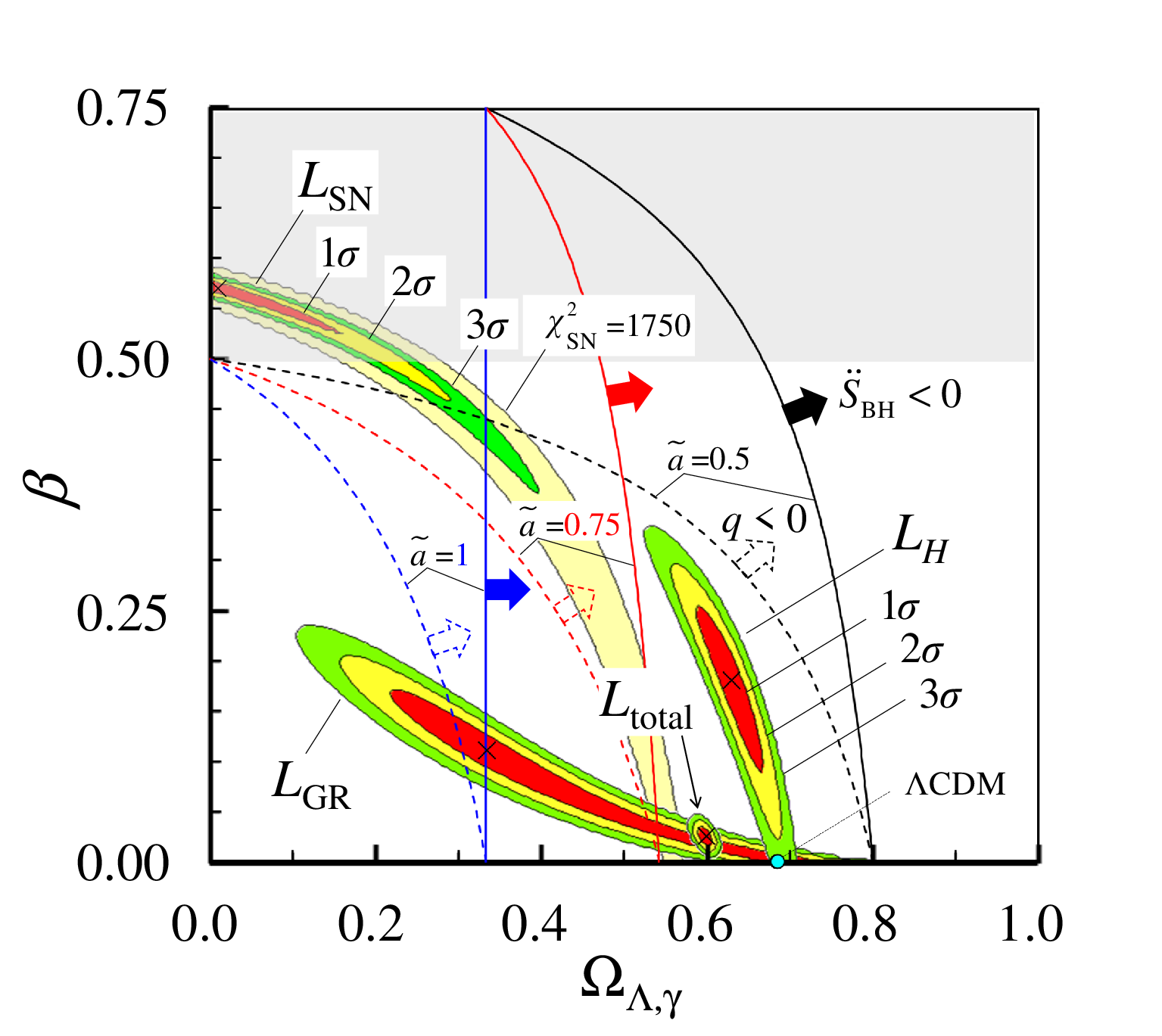}}
\end{center}
\end{minipage}
\caption{Contours of $L_{\textrm{SN}}$, $L_{H}$, $L_{\textrm{GR}}$, and $L_{\textrm{total}}$ in the $(\Omega_{\Lambda,\gamma}, \beta)$ plane.
The contours of the $1 \sigma$, $2 \sigma$, and $3 \sigma$ confidence levels are plotted.
The x mark indicates the location of the maximum value of each likelihood function. 
The location is equivalent to that for the minimum chi-squared function shown in Fig.\ \ref{Fig-SN_H_GF_total_dS2BH=0_plane} and Table \ref{tab-chi2min}.
The combined best-fit is $(\Omega_{\Lambda,\gamma}, \beta)=(0.600, 0.025)$.
The closed circle on the $\Omega_{\Lambda,\gamma}$ axis indicates $(\Omega_{\Lambda,\gamma}, \beta) = (0.685, 0)$ for the $\Lambda$CDM model. 
The lines outside the contours of $L_{\textrm{SN}}$ correspond to a smaller-$\chi_{\textrm{SN}}^{2}$ region $(\chi_{\textrm{SN}}^{2} \leq 1750)$ shown in Fig.\ \ref{Fig-SN_H_GF_total_dS2BH=0_plane}(a).
In the smaller-$\chi_{\textrm{SN}}^{2}$ region, the relative difference from $\chi_{\textrm{SN,min}}^{2}$ is approximately $2\%$ or less.
For the boundaries of $\ddot{S}_{\rm{BH}} = 0$ and $q = 0$, see the caption of Fig.\ \ref{Fig-SN_H_GF_total_dS2BH=0_plane}.
}
\label{Fig-L_plane}
\end{figure}

To study those results in detail, we perform a likelihood analysis.
Using a chi-squared function $\chi^{2}$, a likelihood function $L$ is given by
\begin{equation}
 L  \propto \exp ({- \chi^{2} / 2}) .
\label{eq:L-chi}
\end{equation}
In this study, $L$ is normalized by the maximum value of $L$ \cite{Koma8,Koma16}.
Based on both the above equation and its normalization, $L_{\textrm{SN}}$, $L_{H}$, and $L_{\textrm{GR}}$ are calculated from $\chi_{\textrm{SN}}^{2}$, $\chi_{H}^{2}$, and $\chi_{\textrm{GR}}^{2}$, respectively.
From these three normalized likelihood functions, a combined likelihood function $L_{\textrm{total}}$ is defined by
\begin{equation}
L_{\textrm{total}} = L_{\textrm{SN}}  \times  L_{H} \times   L_{\textrm{GR}}  .
\label{eq:L-joint}
\end{equation}
The obtained $L_{\textrm{total}}$ is normalized by the maximum value of $L_{\textrm{total}}$.
Using each normalized likelihood function, we plot the contours of the $1 \sigma $, $2 \sigma$, and $3 \sigma$ confidence levels, as shown in Fig.\ \ref{Fig-L_plane}.
Here $1 \sigma$, $2 \sigma$, and $3 \sigma$ correspond to the normalized $ L = 3.17 \times 10^{-1}$, $4.60 \times 10^{-2}$, and $2.73 \times 10^{-3}$, respectively \cite{Valent2015,Koma8,Koma16}.
The location of the maximum value of each normalized likelihood function is equivalent to that of the minimum value of each chi-squared function shown in Fig.\ \ref{Fig-SN_H_GF_total_dS2BH=0_plane} and Table \ref{tab-chi2min}.

We now examine the contours of the normalized likelihood functions in the $(\Omega_{\Lambda,\gamma}, \beta)$ plane.
We first observe the contours of the combined likelihood function $L_{\textrm{total}}$. 
As shown in Fig.\ \ref{Fig-L_plane}, the region surrounded by contours of $L_{\textrm{total}}$ satisfies the transitional and thermodynamic constraints at an earlier stage than $\tilde{a}=1$.
The region for $L_{\textrm{total}}$ slightly deviates from the point $(\Omega_{\Lambda,\gamma}, \beta)=(0.685, 0)$.
Based on this likelihood analysis, we can confirm that a weakly dissipative universe with $\Lambda$ is favored.
Note that the regions for $L_{\textrm{total}}$ and $L_{\textrm{GR}}$ partially overlap each other because, as examined later, $f(z) \sigma_{8}(z)$ for the combined best-fit agrees with the observed data points.

In addition, we observe the contours of $L_{\textrm{SN}}$, $L_{H}$, and $L_{\textrm{GR}}$.
As shown in Fig.\ \ref{Fig-L_plane}, the regions for $L_{H}$ and $L_{\textrm{GR}}$ partially overlap each other.
The overlapped region implies that a weakly dissipative universe with $\Lambda$, close to the $\Lambda$CDM model, is favored.
However, the region for $L_{\textrm{SN}}$ does not overlap with the regions for $L_{H}$ and $L_{\textrm{GR}}$.
These results may imply a tension among the constraints derived from the three types of observational data, especially for $L_{\textrm{SN}}$ for the supernovae.
In the present study, each likelihood function $L$ is calculated from each chi-squared function $\chi^{2}$, using the same equation given by Eq.\ (\ref{eq:L-chi}).
However, the number of data points used for $\chi_{\textrm{SN}}^{2}$ is greater than the number used for the other two and, therefore, $\chi_{\textrm{SN,min}}^{2} = 1716.1$ is larger than $\chi_{H,\textrm{min}}^{2}=29.1$ and $\chi_{\textrm{GR,min}}^{2}=13.8$.
The large value of $\chi_{\textrm{SN,min}}^{2}$ should be related to the tension even when each likelihood function $L$ is normalized by the maximum value of $L$, because Eq.\ (\ref{eq:L-chi}) is in the form of an exponential function.
To confirm this, a smaller-$\chi_{\textrm{SN}}^{2}$ region ($\chi_{\textrm{SN}}^{2} \leq 1750$), which is shown in Fig.\ \ref{Fig-SN_H_GF_total_dS2BH=0_plane}(a), is also plotted in Fig.\ \ref{Fig-L_plane}.
The outlines of the smaller-$\chi_{\textrm{SN}}^{2}$ region are indicated by $\chi_{\textrm{SN}}^{2} = 1750$ and are outside the contours of $L_{\textrm{SN}}$.
In the smaller-$\chi_{\textrm{SN}}^{2}$ region, the relative difference from $\chi_{\textrm{SN,min}}^{2}$ is approximately $2\%$ or less.
We note that when $L_{H}$ is considered, a corresponding region (in which the relative difference from $\chi_{H,\textrm{min}}^{2}$ is $2\%$ or less) is inside the contours of the $1 \sigma$ confidence level for $L_{H}$.

As shown in Fig.\ \ref{Fig-L_plane}, the smaller-$\chi_{\textrm{SN}}^{2}$ region is close to the region for $L_{H}$ and partially overlaps with the region for $L_{\textrm{GR}}$.
In this way, we can confirm that the tension considered here should be based on both Eq.\ (\ref{eq:L-chi}) and the large value of $\chi_{\textrm{SN,min}}^{2}$. 
In addition, the tension itself should be weak.
However, the location of $\chi_{\textrm{SN,min}}^{2}$, namely $(\Omega_{\Lambda,\gamma}, \beta)=(0.010, 0.570)$, deviates from that of the $\Lambda$CDM model, $(\Omega_{\Lambda,\gamma}, \beta) = (0.685, 0)$.
The deviation is examined in Appendix\ \ref{Evolution of the distance modulus}.
Consequently, it is found that, at high redshift ($z>0.1$), the distance modulus $\mu$ for [$\Omega_{\Lambda,\gamma}= 0.010,\beta =0.570$] is more consistent with the observed data points because the accelerated expansion of the universe is slower than that for [$\Omega_{\Lambda,\gamma}= 0.685,\beta =0$].
This difference of the accelerated expansion should lead to the deviation.
For details, see Appendix\ \ref{Evolution of the distance modulus}, in which the tension is also examined from a different viewpoint.

\begin{figure} [t] 
\begin{minipage}{0.495\textwidth}
\begin{center}
\scalebox{0.31}{\includegraphics{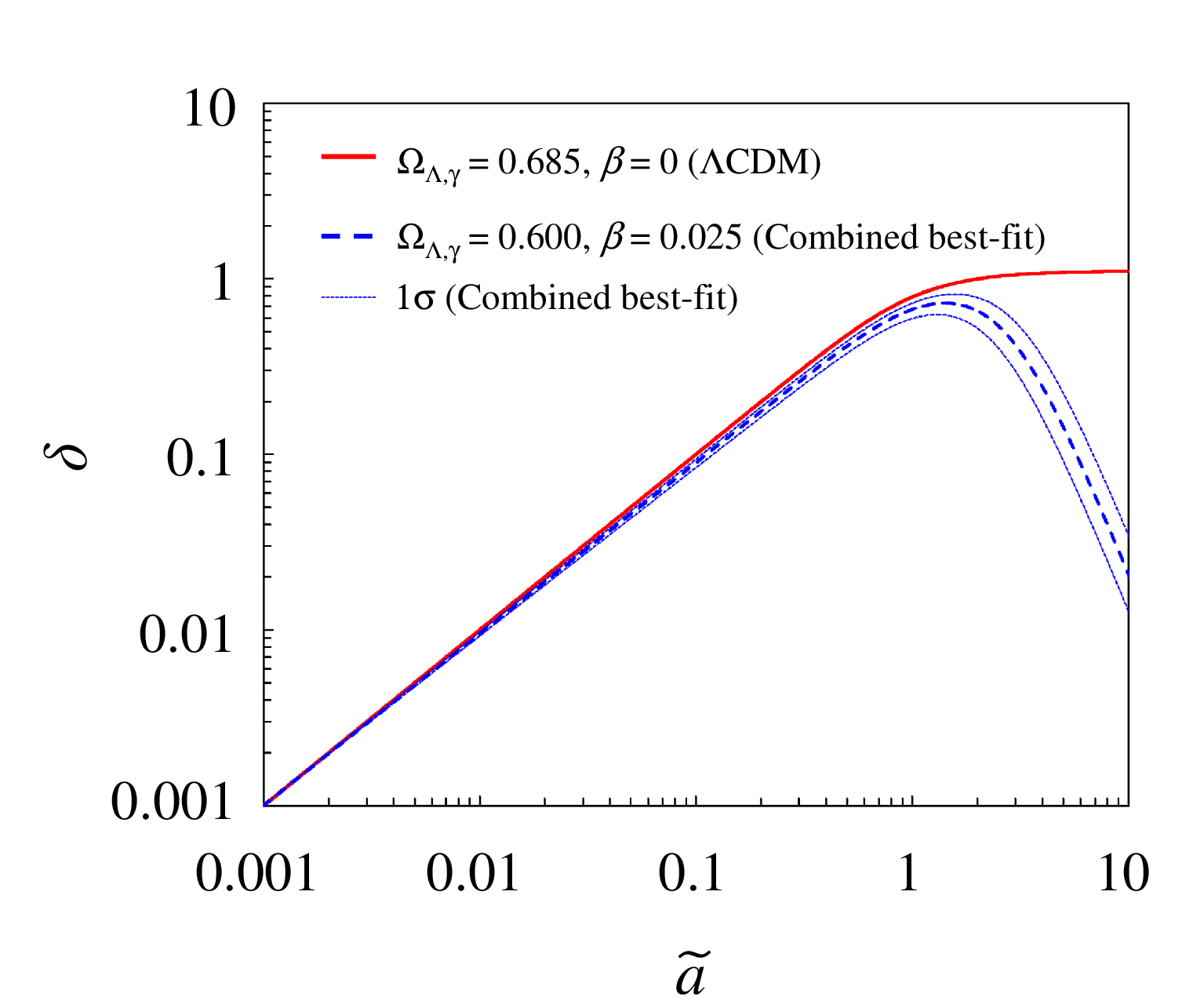}}
\end{center}
\end{minipage}
\caption{Evolution of the perturbation growth factor $\delta$ for the present model.
The bold solid-lines represents [$\Omega_{\Lambda,\gamma}= 0.685,\beta =0$], corresponding to the $\Lambda$CDM model.
The bold dashed-line represents [$\Omega_{\Lambda,\gamma}= 0.600,\beta =0.025$], corresponding to the combined best-fit in Table\ \ref{tab-chi2min}.
The two thin lines above and below the bold dashed-line correspond to the upper and lower limits of the $1 \sigma$ uncertainties for the combined best-fit, respectively.
The thin lines are plotted, based on the results calculated from several points around the contour of the $1 \sigma$ confidence level for $L_{\textrm{total}}$.
 }
\label{Fig-delta-a_present}
\end{figure}

\begin{figure} [t] 
\centering
\begin{minipage}{0.495\textwidth}
    \centering
\scalebox{0.33}{\includegraphics{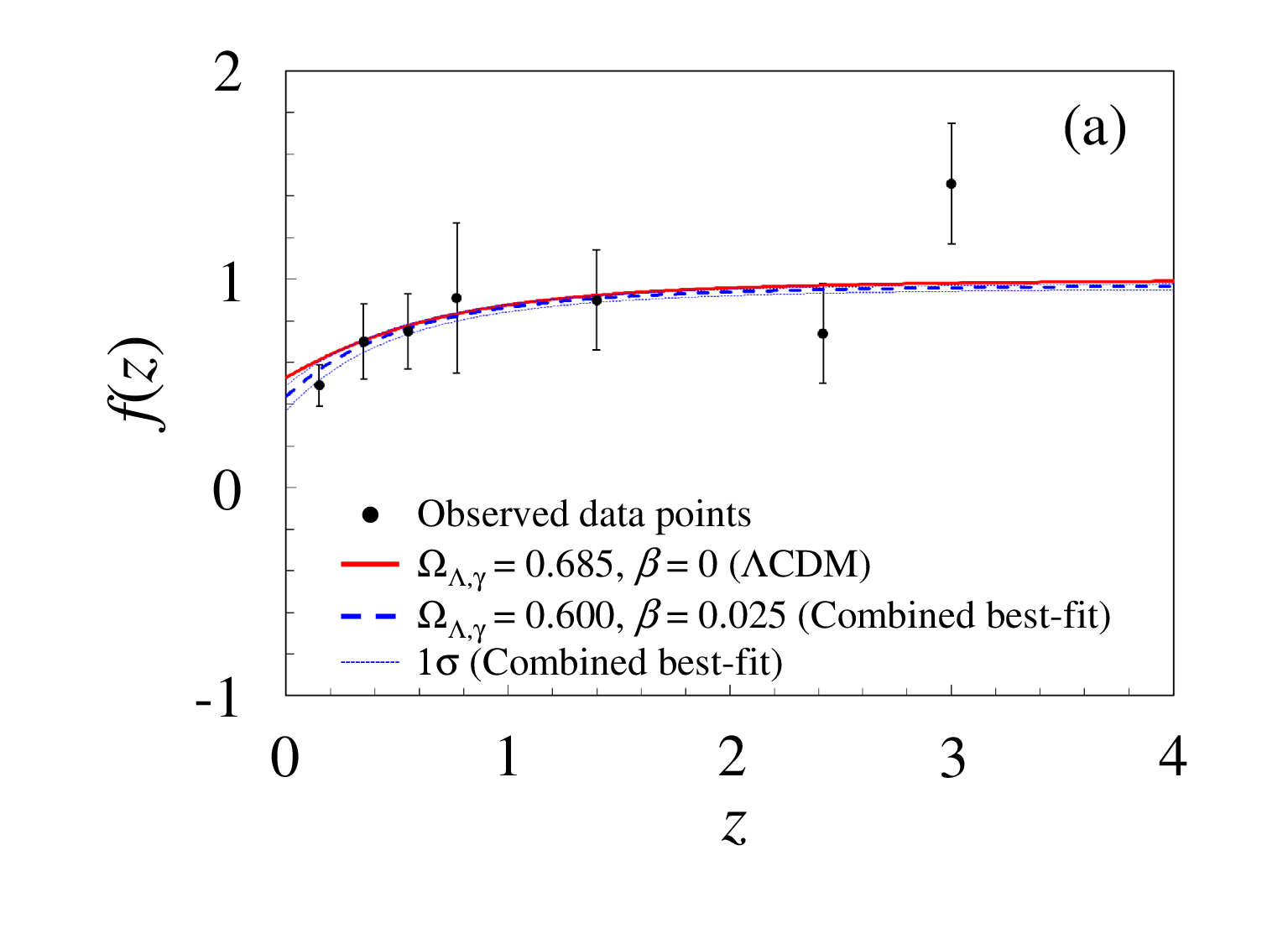}}
\end{minipage}
\begin{minipage}{0.495\textwidth}
    \centering
\scalebox{0.33}{\includegraphics{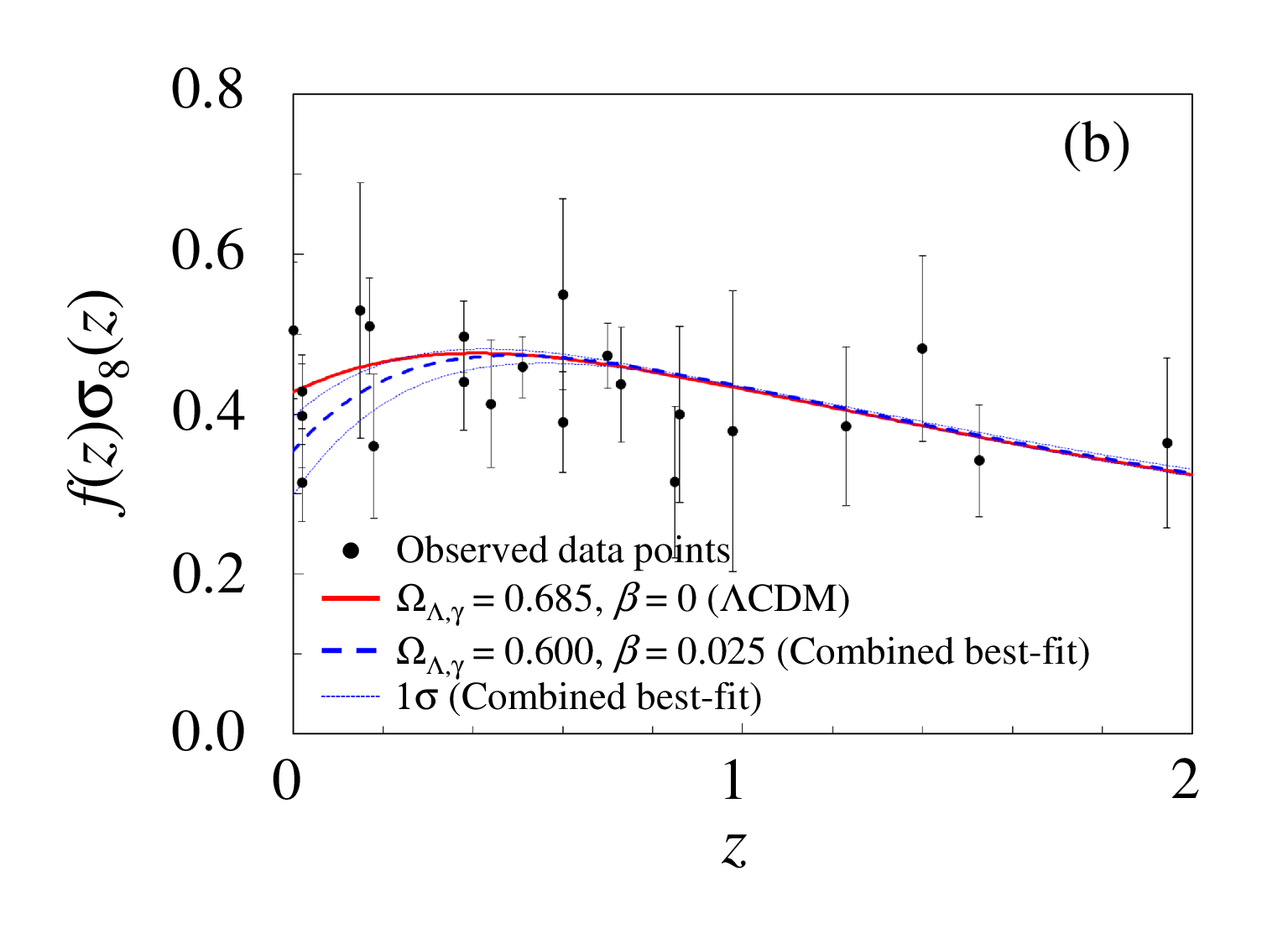}}
\end{minipage}
\caption{
Evolution of $f(z)$ and $f(z) \sigma_{8}(z)$ for the present model.
(a) Growth rate $f(z)$. 
(b) Combination value $f(z) \sigma_{8}(z)$.
The horizontal axis represents the redshift $z$.
In (a) and (b), the closed circles with error bars are observed data points taken from Refs.\ \cite{Lima2011} and \cite{SPaulDas2025}, respectively.
For the lines, see the caption of Fig.\ \ref{Fig-delta-a_present}.
}
\label{Fig-f(z)_f(z)sigma8-z_present}
\end{figure}

Finally, we discuss density perturbations.
As examined in the previous work \cite{Koma16}, density perturbations are different in dissipative and non-dissipative universes because a dissipative driving term decreases the perturbation growth factor $\delta$.
To confirm this, we observe the evolution of $\delta$ for the $\Lambda$CDM model and the combined best-fit in the present model.
As shown in Fig.\ \ref{Fig-delta-a_present}, for $\tilde{a} \lessapprox 1$, $\delta$ initially increases with $\tilde{a}$ and then the increase gradually slows. 
For $\tilde{a} \gtrapprox 1$, $\delta$ for [$\Omega_{\Lambda,\gamma}= 0.685,\beta =0$] does not decrease whereas $\delta$ for [$\Omega_{\Lambda,\gamma}= 0.600,\beta =0.025$] gradually decreases.
We can confirm that the dissipative driving term for the present model decreases $\delta$ even if $\beta$ is small.
However, $\delta$ for [$\Omega_{\Lambda,\gamma}= 0.600,\beta =0.025$], corresponding to the combined best-fit, does not decrease until $\tilde{a} = 1$, as shown in Fig.\ \ref{Fig-delta-a_present}.
Therefore, we expect that the combined best-fit is consistent with the observed data points although $\beta$ is not zero.
To examine this expectation, the evolutions of $f(z)$ and $f(z) \sigma_{8}(z)$ for the present model are plotted in Fig.\ \ref{Fig-f(z)_f(z)sigma8-z_present}.
The horizontal axis represents the redshift $z$.
As shown in Fig.\ \ref{Fig-f(z)_f(z)sigma8-z_present}, $f(z)$ and $f(z) \sigma_{8}(z)$ for the combined best-fit [$\Omega_{\Lambda,\gamma}= 0.600,\beta =0.025$] are consistent with those for the $\Lambda$CDM model [$\Omega_{\Lambda,\gamma}= 0.685,\beta =0$] and the observed data points.
Consequently, a weakly dissipative universe with $\Lambda$, close to the $\Lambda$CDM model, is favored.
This weakly dissipative universe with $\Lambda$ satisfies the transitional and thermodynamic constraints.
Therefore, both the $\Lambda$CDM model and a $\Lambda$CDM model with weak dissipation are favored from these viewpoints.
This type of weak dissipation is expected to be a key in bridging the gap between observations and standard cosmology.

It should be noted that a similar tension between density perturbations and the background evolution of the universe has been discussed from various viewpoints.
For example, it has been reported that a negative sound speed \cite{Lima2011} and the existence of clustered matter \cite{Ramos_2014} are necessary to properly describe the density perturbations in a pure BV model without $\Lambda$ \cite{Koma16}.
In addition, holographic dark energy models based on non-additive entropy \cite{Agostino2019} and a model-independent thermodynamic parametrization of dark energy \cite{Agostino2022} have been examined in detail.
Those works should also provide a key in bridging the gap between observations and standard cosmology.

In this study, we phenomenologically derived the present model from the first law of thermodynamics by applying an effective entropy proportional to the Bekenstein--Hawking entropy, given by $S_{H} =S_{\rm{BH}}(1-\beta)$.
For this, we proposed two entropies, namely a scaled Bekenstein--Hawking entropy and an extended power-law-corrected entropy.
While a microphysical derivation of these two entropies remains an open question, the present results demonstrate that they provide a physically consistent and productive framework for the model. 
This type of effective entropy may be related to, e.g., holographic entanglement entropy \cite{RyuTakayanagi2006,Relative_entropy,Arias2020}.
Further studies are needed and are left for future research.

\section{Conclusions}
\label{Conclusions}

We phenomenologically derived a modified thermodynamic relation in a flat FLRW universe using the first law of thermodynamics and the general formulation for cosmological equations.
Based on this relation, we formulated a BV model with both a constant term $f_{\Lambda}(t)$ given by $\Lambda/3$ and a dissipative driving term $h_{\textrm{B}}(t)$ given by $\beta (2 H^{2} + \dot{H})$.
Here $S_{H} =S_{\rm{BH}}(1-\beta)$ is assumed for a constant $(\partial S_{\Delta} / \partial S_{\rm{BH}})$, by applying an effective entropy such as a scaled Bekenstein--Hawking entropy and an extended power-law-corrected entropy.
The dissipative term derived is proportional to the Ricci scalar curvature, suggesting that the dynamic creation pressure is similarly related.
The present model is equivalent to the form of a dissipative $\Lambda$CDM model, although the theoretical backgrounds are different.

To clarify the fundamental properties of the present model, we examined the background evolution of the late universe.
We found that the present model for $\beta < 0.5$ satisfies a transition from a decelerating universe to an accelerating universe.
Also, we derived the irreversible entropy due to adiabatic particle creation by assuming that the irreversible entropy is related to the present model, that is, assuming holographic-like matter creation cosmology.
The form of $S_{mH}$ in the Hubble volume is similar to that of $\dot{S}_{\rm{BH}}$ on the horizon in both a non-dissipative universe and a pure dissipative universe without $\Lambda$.
This similarity may imply that $S_{mH}$ is related to $\dot{S}_{\rm{BH}}$ through the modified thermodynamic relation between quantities on the horizon and in the bulk.
The similarity is considered to be a new physical insight because it has the potential to serve as a bridge connecting the irreversible entropy and the Bekenstein--Hawking entropy through the modified thermodynamic relation.

In addition, we examined the evolution of the horizon entropy.
The present model always satisfies the second law of thermodynamics on the horizon, $\dot{S}_{\rm{BH}} \geq 0$, and satisfies maximization of entropy, $\ddot{S}_{\rm{BH}} < 0$, in the final stage.
Consequently, the universe in the present model for $\beta <0.5$ is an initially decelerating and then accelerating universe and thereafter approaches a kind of equilibrium state.
The thermodynamic constraints are consistent with constraints on a transition from a decelerating universe to an accelerating universe.

Furthermore, we examined first-order density perturbations in the linear approximation, by assuming a matter-dominated universe.
For this, a neo-Newtonian approach is applied to the present model.
The present model is consistent with the observed data points related to structure formation even when $\beta$ is not zero but is small.

Finally, we examined constraints on the present model, using three types of observational data, namely, a distance modulus $\mu(z)$, the Hubble parameter $H(z)$, and a combination value $f(z) \sigma_{8}(z)$, and the transitional and thermodynamic constraints.
We found that the present model for small $\beta$ is consistent with observations and satisfies the transitional and thermodynamic constraints.
This result implies that a weakly dissipative universe with $\Lambda$ is favored, such as the universe described by a weakly dissipative $\Lambda$CDM model.

This study provides a new interpretation of dissipative universes and reveals fundamental properties of the present model.
The assumptions used here have not been established but are a viable scenario.
The present model should contribute to a better understanding of the dissipative universe, such as matter creation cosmology, bulk viscous cosmology, and dissipative $\Lambda$CDM models.
A weakly dissipative universe with $\Lambda$ examined here is expected to be a key in bridging the gap between observations and standard cosmology.

\appendix

\section{Constant $(\partial S_{\Delta} / \partial S_{\rm{BH}})$ from a power-law-corrected entropy $S_{pl}$}
\label{The present model and a power-law-corrected entropy}

To obtain a constant $(\partial S_{\Delta} / \partial S_{\rm{BH}})$, we apply a power-law-corrected entropy $S_{pl}$ \cite{Das2008}.
The power-law-corrected entropy is based on the entanglement of quantum fields between the inside and outside of the horizon and is computed by tracing over its degrees of freedom inside its sphere \cite{Das2008}.

In this study, we extend the original form of the power-law-corrected entropy \cite{Radicella2010}.
The original form can be written as \cite{Koma11,Koma12,Koma22}
\begin{equation}
 S_{pl}  = S_{\rm{BH}}  \left [ 1-  \beta \left ( \frac{H_{0}}{H} \right )^{2- \alpha}  \right ] , 
\label{eq:Spl}     
\end{equation}
where $\alpha$ and $\beta$ are dimensionless constant parameters, with $0 < \alpha <4$ and $\beta \ge 0$ considered.
Also, $H_{0}$ represents $H$ at the present time.
Note that $\psi_{\alpha}$ used in the previous works is replaced by $\beta$, for simplicity.

To extend the original form, we assume that $\alpha$ and $\beta$ are independent free parameters \cite{Koma14,Koma22}.
When $\alpha =2$, $S_{pl}$ reduces to an extended form:
\begin{equation}
 S_{pl}  = S_{\rm{BH}}  (1- \beta)  .
\label{eq:Spl_alpha=2_app}     
\end{equation}
The above equation is considered to be an effective entropy proportional to the Bekenstein--Hawking entropy. 
Consequently, $(\partial S_{\Delta} / \partial S_{\rm{BH}})$ is constant:
\begin{align}
                    \left ( \frac{\partial S_{\Delta} }{\partial S_{\rm{BH}}}  \right )     &=   - \beta     .
\label{eq:dSpl_dSBH_cst}      
\end{align}
In this way, a constant $(\partial S_{\Delta} / \partial S_{\rm{BH}})$ should be obtained from the extended form.

The original form of the power-law-corrected entropy given by Eq.\ (\ref{eq:Spl}) is based on the entanglement of quantum fields between the inside and outside of the horizon and, therefore, $\beta$ should be related to this entanglement.
Of course, the microphysical justification of Eq.\ (\ref{eq:Spl_alpha=2_app}) has not yet been established because the original form is extended by assuming an independent parameter $\beta$.
However, this type of effective entropy based on the entanglement of quantum fields is considered to be a viable scenario.

\section{Derivation of the solution for the present model} 
\label{Solution1}

The general solution for the present model is derived applying a method examined in Refs.\ \cite{Koma45,Koma6,Koma10,Koma14}. 
The ranges of parameters used here are given by Eqs.\ (\ref{eq:Omega_Lgamma-D_range}) and \ (\ref{eq:gamma-beta_range}).
From Eq.\ (\ref{eq:Back_Present_2}), the differential equation for the present model is written as 
\begin{equation}
    \dot{H} = - \frac{3(1+w)}{2(1-\beta)} H^{2}  \left ( 1 -  \gamma - \frac{\Lambda}{3 H^{2}} \right )     ,  
\label{eq:Back2_A}
\end{equation}
where $ \gamma =4 \beta /3/(1+w)$ is given by Eq.\ (\ref{eq:gamma}).
Applying Eq.\ (\ref{eq:Back2_A}), $(dH/da) a$ can be given by 
\begin{align}
\left ( \frac{dH}{da} \right )  a     &=        \frac{dH}{dt}    \frac{dt}{da} a  
                                                   = - \frac{3(1+w)}{2(1-\beta)} H^{2}  \left ( 1-  \gamma - \frac{\Lambda}{3 H^{2}} \right )    \frac{a}{\dot{a}}                      \notag \\
                                                 &= - \frac{3(1+w)}{2(1-\beta)} H  \left ( 1-  \gamma - \frac{\Lambda}{3 H^{2}} \right )                                                      ,      
\label{eq:Back23_A}
\end{align}
using $H=\dot{a}/a$.
We consider a matter-dominated universe, i.e., $w =0$, although $w$ is retained for generality.

Substituting $ H = \tilde{H} H_{0} $ and $a = \tilde{a} a_{0} $ into Eq.\ (\ref{eq:Back23_A}) and calculating the resultant equation yields
\begin{align}
 \left ( \frac{d \tilde{H}  }{d \tilde{a}  } \right )  \tilde{a}     &=  - \frac{3 (1+w)}{2(1-\beta)} \tilde{H}  \left ( 1 - \gamma - \frac{\Lambda}{3 H_{0}^{2}}  \tilde{H}^{-2}    \right )             \notag \\
                                                                                     &=  - \frac{3 (1+w)}{2(1-\beta)} \tilde{H}  \left ( 1 - \gamma - \Omega_{\Lambda}  \tilde{H}^{-2}    \right )             ,
\label{eq:Back24_A_2}
\end{align}
where the density parameter for $\Lambda$ is given by
\begin{equation}
   \Omega_{\Lambda}  = \frac{\Lambda}{3 H_{0}^{2}} .
\label{eq:Omega_L_ap}
\end{equation}
In addition, the parameter $N$ is defined by 
\begin{equation}
   N  \equiv \ln \tilde{a}, \quad \textrm{and therefore,}  \quad   dN   = \frac{ d \tilde{a} }{ \tilde{a} }  .
\label{eq:N_A}
\end{equation}
Using Eq.\ (\ref{eq:N_A}), Eq.\ (\ref{eq:Back24_A_2}) can be written as 
\begin{equation}
    \left ( \frac{d \tilde{H}  }{d N } \right )      =   - \frac{3 (1+w)}{2(1-\beta)} \tilde{H}  \left ( 1 - \gamma - \Omega_{\Lambda}  \tilde{H}^{-2}    \right )             .
\label{eq:Back24_A_N}
\end{equation}
When $w$, $\beta$, $\gamma$, and $\Omega_{\Lambda}$ are constant, Eq.\ (\ref{eq:Back24_A_N}) can be integrated as 
\begin{equation}
    \int \frac{d \tilde{H}  }{  \tilde{H}  \left ( 1 - \gamma - \Omega_{\Lambda}  \tilde{H}^{-2}    \right )      }       =   -   \frac{3 (1+w)}{2(1-\beta)} \int dN         .        
\label{f_N_I}
\end{equation}
The solution is given by
\begin{align}
\ln  \left (  \Omega_{\Lambda}  -  (1-\gamma) \tilde{H}^{2}  \right )^{\frac{1}{2(1- \gamma)}}   
  &=   -  \frac{3 (1+w)}{2(1-\beta)}  N  + C^{\prime}                                                                                                                                         \notag \\
  &=   -  \frac{3 (1+w)}{2(1-\beta)} \ln \tilde{a} + C^{\prime}      , 
\label{f_N_Solve}
\end{align}
where $C^{\prime}$ is an integral constant.
From Eqs.\ (\ref{def_H_H0}) and (\ref{def_a_a0}), the present values of $\tilde{H}$ and $\tilde{a}$ are $1$.
Substituting $\tilde{H} =1$ and $\tilde{a} =1$ into Eq.\ (\ref{f_N_Solve}) yields 
\begin{equation}
 C^{\prime} =    \ln  \left (  \Omega_{\Lambda}  -  (1-\gamma)  \right )^{\frac{1}{2(1- \gamma)}}   .
\label{IntC}
\end{equation}
Substituting Eq.\ (\ref{IntC}) into Eq.\ (\ref{f_N_Solve}) and arranging the resultant equation yields
\begin{equation}
       \left ( \frac{  \Omega_{\Lambda}  -  (1-\gamma) \tilde{H}^{2}  }{ \Omega_{\Lambda}  -  (1-\gamma) }  \right )^{\frac{1}{2(1- \gamma)}}   = \tilde{a}^{-  \frac{3 (1+w)}{2(1-\beta)}}  , 
\label{eq:Sol_aa0}
\end{equation}
and solving this equation with respect to $\tilde{H}^{2}$ yields
\begin{align}
       \tilde{H}^{2}      &=  \left (  1-  \frac{\Omega_{\Lambda}}{1-\gamma} \right )  \tilde{a}^{ - 3 (1+w) \frac{1-\gamma}{1-\beta}  }  + \frac{\Omega_{\Lambda}}{1-\gamma}   \notag \\
                              &=          (  1-  \Omega_{\Lambda,\gamma}  )  \tilde{a}^{ - 3 (1+w) \frac{1-\gamma}{1-\beta}  }  + \Omega_{\Lambda,\gamma} \notag \\ 
                              &=          (  1-  \Omega_{\Lambda,\gamma}  )  \tilde{a}^{ - D}  + \Omega_{\Lambda,\gamma}  ,
\label{eq:Sol_HH0_ap}
\end{align}
where $\Omega_{\Lambda,\gamma}$ and $D$ are given by
\begin{equation}
   \Omega_{\Lambda,\gamma}  = \frac{\Omega_{\Lambda}}{1-\gamma}  , \quad         D = 3 (1+w) \frac{1-\gamma}{1-\beta}        .
\label{eq:Omega_Lgamma_D_ap}
\end{equation}
Equation\ (\ref{eq:Sol_HH0_ap}) is the solution for the present model.
In the present model, $\Omega_{\Lambda,\gamma}$ is an effective density parameter, related to both $\Lambda$ and $\gamma$.
We can convert $\beta$ into $\gamma = 4 \beta/3/(1+w)$, using Eq.\ (\ref{eq:gamma}).

\section{Derivation of entropy density due to adiabatic particle creation for the present model} 
\label{Derivation of entropy density}

We derive the normalized entropy density $s/s_{0}$ due to adiabatic particle creation for the present model.
Using Eqs.\ (\ref{eq:Gamma_Present}) and (\ref{eq:dotH_present}), $(\Gamma -3H) dH/\dot{H}$ is given by 
\begin{align}
\frac{\Gamma -3H}{\dot{H}} dH =&  \left ( \frac{  \beta H  [(4-D) H^{2}   +D \Omega_{\Lambda,\gamma} H_{0}^{2} ]  }{ H^{2} - \frac{\Lambda}{3}  } - 3H  \right )  \notag \\
                                            & \times  \left (   - \frac{D}{2}  ( H^{2}  - \Omega_{\Lambda,\gamma} H_{0}^{2} ) \right )^{-1} dH                                                  \notag \\
                                            =&   \left ( \frac{  \beta \tilde{H}  [(4-D)\tilde{H}^{2}   +D \Omega_{\Lambda,\gamma} ]  }{ \tilde{H}^{2} - \Omega_{\Lambda}  } - 3\tilde{H}  \right )  \notag \\
                                            & \times  \left (   - \frac{D}{2}  ( \tilde{H}^{2}  - \Omega_{\Lambda,\gamma} ) \right )^{-1}   d\tilde{H}  .
\label{eq:normalized_1}
\end{align}
From the above equation, Eq.\ (\ref{eq:dots_s_s-H}) can be calculated.
Substituting Eq.\ (\ref{eq:normalized_1}) into the integral of Eq.\ (\ref{eq:dots_s_s-H}) yields
\begin{align} 
 & \int_{H_{0}}^{{H}}  \frac{ \Gamma  -3 {H}  }  {  \dot{H}    }    dH       \notag \\
   &= \int_{1}^{{\tilde{H}}}  \left ( \frac{  \beta \tilde{H}    [(4-D)\tilde{H}^{2}   +D \Omega_{\Lambda,\gamma} ]  }{ \tilde{H}^{2} - \Omega_{\Lambda}  } - 3\tilde{H}  \right )  \notag \\
                                                & \times  \left (   - \frac{D}{2}  ( \tilde{H}^{2}  - \Omega_{\Lambda,\gamma} ) \right )^{-1}   d\tilde{H}  \notag \\
&=\left [  \frac{ \ln ( \tilde{H}^{2} - \Omega_{\Lambda})^{\beta P}  + \ln(\tilde{H}^{2}  - \Omega_{\Lambda,\gamma} )^{Q}    }{    D(\Omega_{\Lambda,\gamma} - \Omega_{\Lambda} )  }\right ]^{\tilde{H}}_{1}  \notag \\
&= \frac{ \ln \left [ \left ( \frac{ \tilde{H}^{2} - \Omega_{\Lambda}  }{1- \Omega_{\Lambda} }  \right )^{\beta P}  \left(  \frac{ \tilde{H}^{2}  - \Omega_{\Lambda,\gamma} }{  1- \Omega_{\Lambda,\gamma} } \right )^{Q}  \right ]  }{    D(\Omega_{\Lambda,\gamma} - \Omega_{\Lambda} )  }  \notag \\
&= \ln \left [  \left ( \frac{ \tilde{H}^{2} - \Omega_{\Lambda}  }{1- \Omega_{\Lambda} }  \right )^{\beta P}  \left(  \frac{ \tilde{H}^{2}  - \Omega_{\Lambda,\gamma} }{  1- \Omega_{\Lambda,\gamma} } \right )^{Q} \right ]^{\frac{1}{D(\Omega_{\Lambda,\gamma} - \Omega_{\Lambda} )} } , 
\label{eq:int01}
\end{align}
where $P$ and $Q$ are given by
\begin{align} 
P =  D (\Omega_{\Lambda,\gamma} - \Omega_{\Lambda}) + 4   \Omega_{\Lambda} ,
\label{eq:P}
\end{align}
\begin{align} 
Q  &=  (3-4\beta)\Omega_{\Lambda,\gamma} - 3 \Omega_{\Lambda}  .
\label{eq:Q}
\end{align}

In the following, several algebraic expressions used in Eq.\ (\ref{eq:int01}) are calculated.
From $\Omega_{\Lambda}=  (1-\gamma) \Omega_{\Lambda,\gamma}$ given by Eq.\ (\ref{eq:Omega_Lgamma}), $\Omega_{\Lambda,\gamma} - \Omega_{\Lambda}$ is written as
\begin{align} 
\Omega_{\Lambda,\gamma} - \Omega_{\Lambda} &=  \Omega_{\Lambda,\gamma} -  (1-\gamma) \Omega_{\Lambda,\gamma} = \gamma \Omega_{\Lambda,\gamma}  .
\label{eq:G-L}
\end{align} 
Substituting Eq.\ (\ref{eq:G-L}) into Eq.\ (\ref{eq:Q}) yields
\begin{align} 
Q  &=  (3-4\beta)\Omega_{\Lambda,\gamma} - 3 \Omega_{\Lambda}  \notag \\
    &= (3-4\beta)\Omega_{\Lambda,\gamma} - 3  (1-\gamma) \Omega_{\Lambda,\gamma}  \notag \\
    &=  (3\gamma-4\beta)\Omega_{\Lambda,\gamma} .
\label{eq:Q1}
\end{align}
From Eqs.\ (\ref{eq:G-L}) and (\ref{eq:Q1}), $Q/(\Omega_{\Lambda,\gamma} - \Omega_{\Lambda})$ is written as
\begin{align} 
\frac{Q}{\Omega_{\Lambda,\gamma} - \Omega_{\Lambda}} &= \frac{(3\gamma-4\beta)\Omega_{\Lambda,\gamma} }{\gamma \Omega_{\Lambda,\gamma}  } =3 -\frac{4\beta}{\gamma} \notag \\
                                                                                    &=3- 3(1+w) = -3w  ,
\label{eq:Q_OG-QL}
\end{align} 
where $ 4 \beta/\gamma = 3(1+w)$ given by Eq.\ (\ref{eq:gamma}) is also used.
From Eq.\ (\ref{eq:Q_OG-QL}), $Q/D/(\Omega_{\Lambda,\gamma} - \Omega_{\Lambda})$ is given by
\begin{align} 
\frac{Q}{D(\Omega_{\Lambda,\gamma} - \Omega_{\Lambda})} &= \frac{ -3w}{D}  .
\label{eq:-3w_D}
\end{align} 
Similarly, using Eqs.\ (\ref{eq:P}) and (\ref{eq:G-L}), $P/D/(\Omega_{\Lambda,\gamma} - \Omega_{\Lambda})$ can be written as
\begin{align} 
\frac{P}{D(\Omega_{\Lambda,\gamma} - \Omega_{\Lambda})} &= \frac{ D (\Omega_{\Lambda,\gamma} - \Omega_{\Lambda}) + 4   \Omega_{\Lambda} }{D(\Omega_{\Lambda,\gamma} - \Omega_{\Lambda})}  \notag \\
                                                                                         &=  1+ \frac{  4 (1-\gamma)  \Omega_{\Lambda,\gamma} }{D \gamma \Omega_{\Lambda,\gamma}}  \notag \\
                                                                                         &=  1+ \frac{  4 (1-\gamma) }{3 (1+w) \frac{1-\gamma}{1-\beta} } \frac{1}{\gamma} \notag \\
                                                                                         &=  1+ \frac{  4 (1-\beta) }{3 (1+w) } \frac{3(1+w)}{4\beta}  \notag \\
                                                                                         &=  1+ \frac{  1-\beta }{ \beta } = \frac{1}{\beta}  ,
\end{align} 
where $\Omega_{\Lambda}=  (1-\gamma) \Omega_{\Lambda,\gamma}$, $D = 3 (1+w) \frac{1-\gamma}{1-\beta}$ given by Eq.\ (\ref{eq:D}), and $\gamma = 4 \beta /3/(1+w)$ are also used.
From this equation, $\beta P/D/(\Omega_{\Lambda,\gamma} - \Omega_{\Lambda})$ is given by 
\begin{align} 
\frac{\beta P}{D(\Omega_{\Lambda,\gamma} - \Omega_{\Lambda})} &= \beta \frac{1}{\beta} =1   .
\label{eq:beta_beta_1}
\end{align} 

Substituting Eq.\ (\ref{eq:int01}) into Eq.\ (\ref{eq:dots_s_s-H}), using Eqs.\ (\ref{eq:-3w_D}) and (\ref{eq:beta_beta_1}), and calculating the resultant equation yields
\begin{align}
    \frac{s}{s_{0}}    &=  \left [  \left ( \frac{ \tilde{H}^{2} - \Omega_{\Lambda}  }{1- \Omega_{\Lambda} }  \right )^{\beta P}  \left(  \frac{ \tilde{H}^{2}  - \Omega_{\Lambda,\gamma} }{  1- \Omega_{\Lambda,\gamma} } \right )^{Q} \right ]^{\frac{1}{D(\Omega_{\Lambda,\gamma} - \Omega_{\Lambda} )} }  \notag \\
                           &=     \left ( \frac{ \tilde{H}^{2} - \Omega_{\Lambda}  }{1- \Omega_{\Lambda} }  \right )  \left(  \frac{ \tilde{H}^{2}  - \Omega_{\Lambda,\gamma} }{  1- \Omega_{\Lambda,\gamma} } \right )^{\frac{-3w}{D}}                       ,
\label{eq:dots_s_s-H_Present_0}
\end{align}
where $D = 3 (1+w) \frac{1-\gamma}{1-\beta}$ is given by Eq.\ (\ref{eq:D}), whereas $\Omega_{\Lambda} = \Lambda/(3H_{0}^{2})$ and $\Omega_{\Lambda,\gamma} = \Omega_{\Lambda}/(1-\gamma)$ are given by Eq.\ (\ref{eq:Omega_Lgamma}).
When $w=0$ is considered, the normalized entropy density is written as
\begin{align}
    \frac{s}{s_{0}}   =  \frac{ \tilde{H}^{2} - \Omega_{\Lambda}  }{1- \Omega_{\Lambda} }   \quad (w=0) .
\label{eq:dots_s_s-H_Present_w=0_app}
\end{align}
This equation is used for Eq.\ (\ref{eq:dots_s_s-H_Present_w=0}) in Sec.\ \ref{Entropy Sm}.

\section{Distance modulus $\mu(z)$ and likelihood function $L_{\textrm{SN}}$}
\label{Evolution of the distance modulus}

\begin{figure} [t] 
\begin{minipage}{0.495\textwidth}
\begin{center}
\scalebox{0.31}{\includegraphics{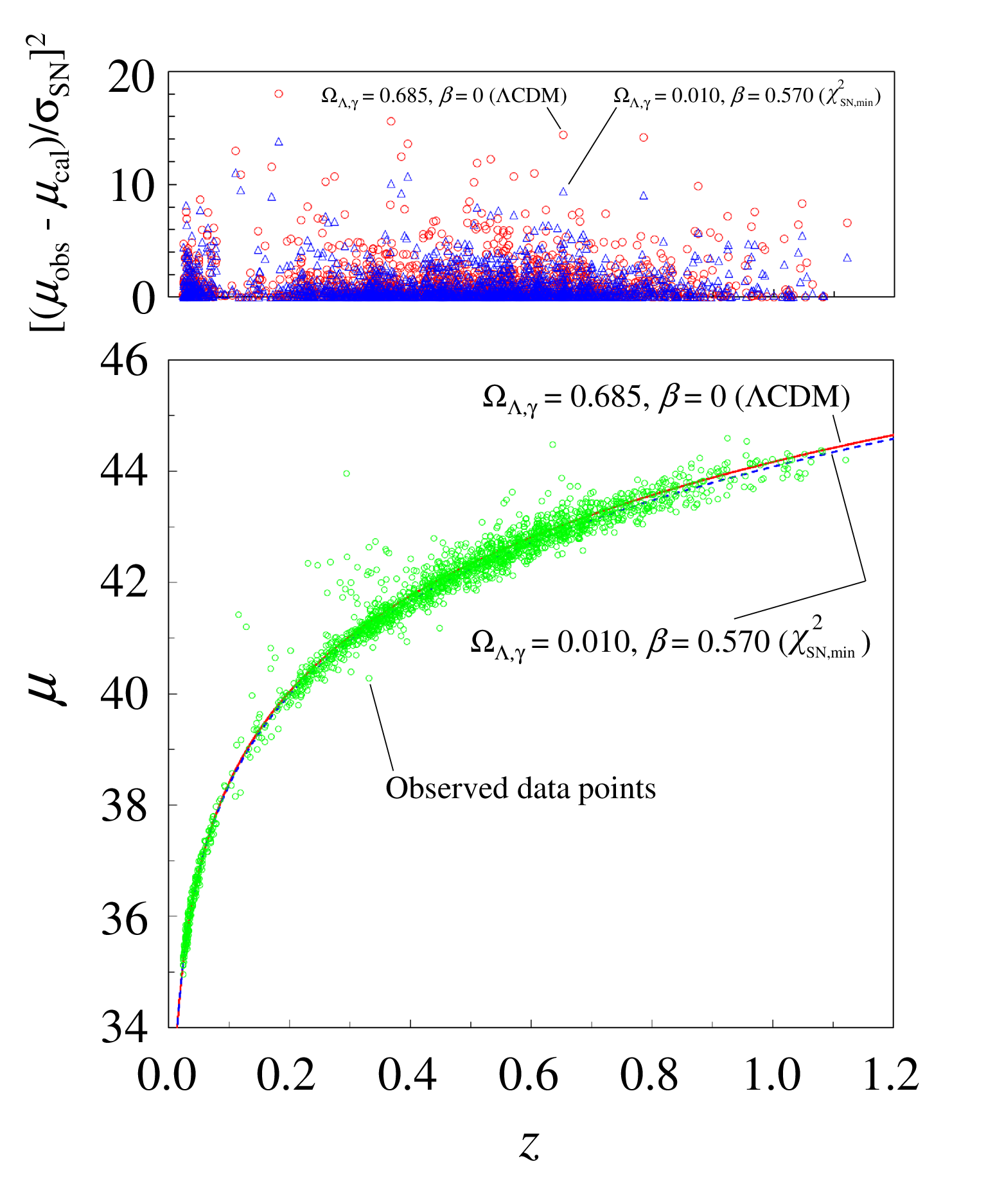}}
\end{center}
\end{minipage}
\caption{Evolution of the distance modulus $\mu(z)$.
Top: $[ (\mu_{\textrm{obs}} - \mu_{\textrm{cal}})/ \sigma_{\textrm{SN}}]^{2}$ included in $\chi_{\textrm{SN}}^{2}$ for the supernovae.
Bottom: the distance modulus $\mu$.
In the bottom panel, the solid line represents [$\Omega_{\Lambda,\gamma}= 0.685,\beta =0$], corresponding to the $\Lambda$CDM model.
The dashed line represents [$\Omega_{\Lambda,\gamma}= 0.010,\beta =0.570$], corresponding to the location of $\chi_{\textrm{SN,min}}^{2}$ in Table\ \ref{tab-chi2min}.
The open symbols are observed data points taken from Ref.\ \cite{DES-SN5YR}.
For clarity, the error bars for the observed data points are omitted.
}
\label{Fig-mu-z}
\end{figure}

\begin{figure} [b] 
\begin{minipage}{0.495\textwidth}
\begin{center}
\scalebox{0.33}{\includegraphics{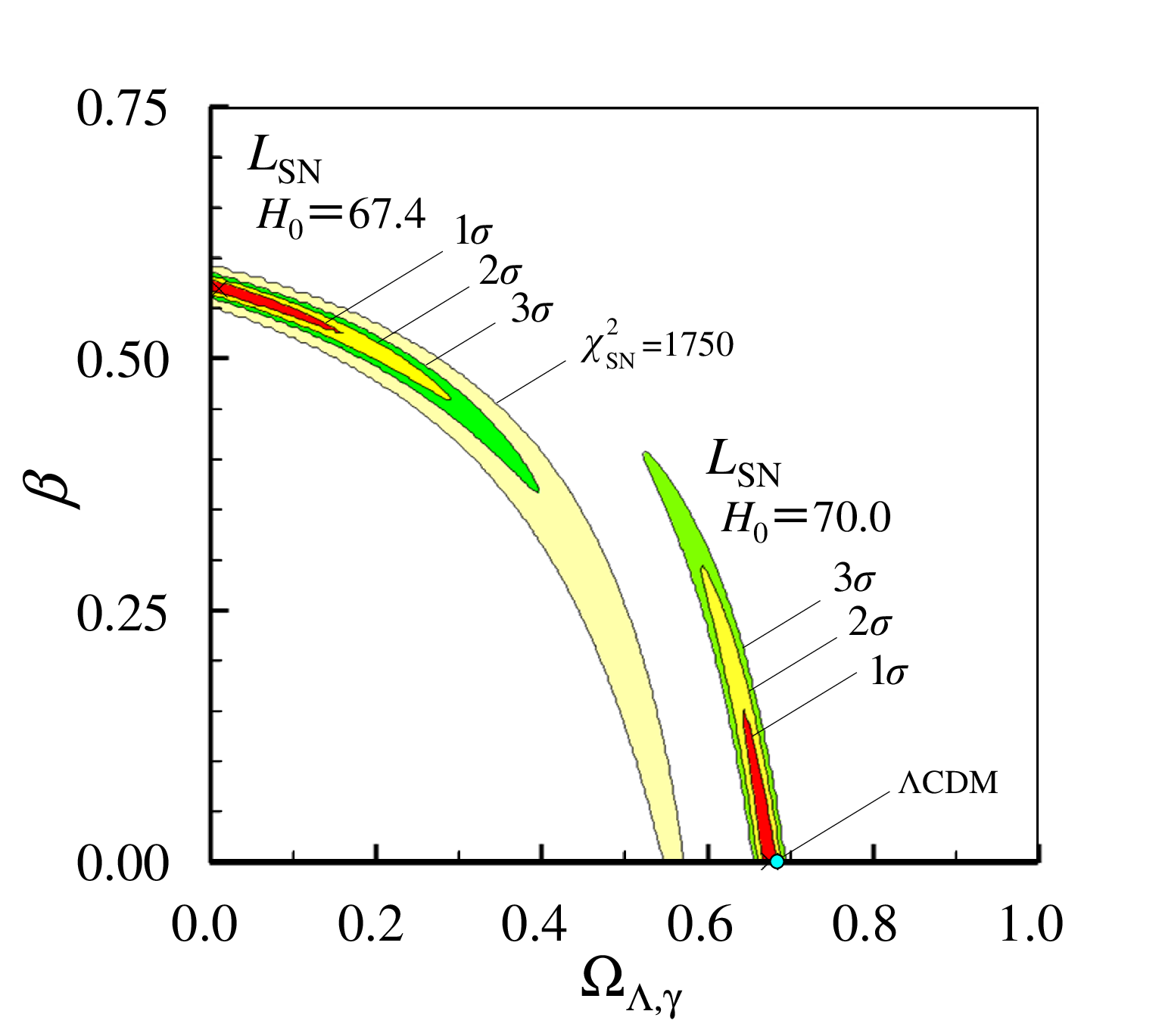}}
\end{center}
\end{minipage}
\caption{Contours of $L_{\textrm{SN}}$ in the $(\Omega_{\Lambda,\gamma}, \beta)$ plane for $H_{0}=67.4$ and $70.0$.
The contours of the $1 \sigma$, $2 \sigma$, and $3 \sigma$ confidence levels are plotted.
The contours for $H_{0}=67.4$ are plotted from Fig.\ \ref{Fig-L_plane}, where the lines outside the contours of $L_{\textrm{SN}}$ correspond to a smaller-$\chi_{\textrm{SN}}^{2}$ region $(\chi_{\textrm{SN}}^{2} \leq 1750)$.
For $H_{0}=70.0$, the smaller-$\chi_{\textrm{SN}}^{2}$ region is not plotted because the region is inside the contours of the $2 \sigma$ confidence level corresponding to $\chi_{\textrm{SN}}^{2} =1752.7$.
The closed circle on the $\Omega_{\Lambda,\gamma}$ axis indicates $(\Omega_{\Lambda,\gamma}, \beta) = (0.685, 0)$ for the $\Lambda$CDM model. 
The x mark indicates the location of the maximum value of each $L_{\textrm{SN}}$, equivalent to the location for each $\chi_{\textrm{SN,min}}^{2}$.
Note that $\chi_{\textrm{SN,min}}^{2}$ and its location $(\Omega_{\Lambda,\gamma}, \beta)$ are summarized in Table\ \ref{tab-chi2min_H0=67.4_70}. 
}
\label{Fig-L_SN_plane_H0=67.4_70.0}
\end{figure}

In this appendix, we discuss results related to the observational data for the supernovae.
As examined in Sec.\ \ref{Observational, transitional, and thermodynamic constraints}, the location of $\chi_{\textrm{SN,min}}^{2}$, $(\Omega_{\Lambda,\gamma}, \beta)=(0.010, 0.570)$, deviates from that of the $\Lambda$CDM model, $(\Omega_{\Lambda,\gamma}, \beta) = (0.685, 0)$.
This deviation (tension) may be interpreted as that higher-order terms such as $\beta (2 H^{2} + \dot{H})$ are more consistent with the dataset for the supernovae, in comparison with the constant term $\Lambda$.
To discuss the deviation and the interpretation, we observe the evolution of the distance modulus $\mu (z)$.
For this, the evolution of $\mu$ is plotted in the bottom panel of Fig.\ \ref{Fig-mu-z}.
In addition, the evolution of $[ (\mu_{\textrm{obs}} - \mu_{\textrm{cal}})/ \sigma_{\textrm{SN}}]^{2}$, which is included in $\chi_{\textrm{SN}}^{2}$ given by Eq.\ (\ref{eq:chi_SN}), is plotted in the top panel.

As shown in Fig.\ \ref{Fig-mu-z}, $\mu$ for [$\Omega_{\Lambda,\gamma}= 0.010,\beta =0.570$] is consistent with the observed data points.
At low redshift ($z<0.1$), $\mu$ for [$\Omega_{\Lambda,\gamma}= 0.685,\beta =0$] is also consistent with the observed data points.
However, $\mu$ for [$\Omega_{\Lambda,\gamma}= 0.685,\beta =0$] slightly deviates from the observed data points at high redshift ($z>0.1$).
In fact, at high redshift, $\mu$ for [$\Omega_{\Lambda,\gamma}= 0.685,\beta =0$] is slightly larger than that for [$\Omega_{\Lambda,\gamma}= 0.010,\beta =0.570$].
That is, at high redshift, the accelerated expansion of the universe for [$\Omega_{\Lambda,\gamma}= 0.010,\beta =0.570$] is slower than that for [$\Omega_{\Lambda,\gamma}= 0.685,\beta =0$].
In the present study, this difference of the accelerated expansion should lead to the deviation of the location of $\chi_{\textrm{SN,min}}^{2}$ from that of the $\Lambda$CDM model.
Consequently, higher-order terms such as $\beta (2 H^{2} + \dot{H})$ are likely more consistent with the dataset for the supernovae.
A similar result has been discussed in Ref.\ \cite{Koma16}, in which a power-law term proportional to $H^{\alpha}$ is applied to the pure $\Lambda(t)$ and BV models, respectively, where $\alpha$ is a free parameter ranging from $-2$ to $3$.
(In the previous work, higher-order terms such as $H^{2}$ and $H^{3}$ are consistent with the dataset for the supernovae, in comparison with the constant term.)

So far, $H_{0}$ has been set to $67.4$ km/s/Mpc from the Planck 2018 results \cite{Planck2018}.
As shown in Fig.\ \ref{Fig-H-a}, the observational data for the Hubble parameter \cite{57dataOdin} indicates that $H$ decreases with $\tilde{a}$.
The minimum value of $H$ \cite{57dataOdin} is $H=68.6$ at $z=0.120$.
Therefore, $H_{0}=67.4$ used in the present paper is consistent with the observational data for the Hubble parameter.
However, $H_{0}$ should also affect the tension and the location of $\chi_{\textrm{SN,min}}^{2}$, which is equivalent to the location of the maximum value of $L_{\textrm{SN}}$.
To examine this effect, we observe typical results. 
For this, the contours of $L_{\textrm{SN}}$ for $H_{0}=67.4$ and $70.0$ are plotted in Fig.\ \ref{Fig-L_SN_plane_H0=67.4_70.0}.
Here the contours for $H_{0}=67.4$ are plotted from Fig.\ \ref{Fig-L_plane}.
In addition, $\chi_{\textrm{SN,min}}^{2}$ and its location $(\Omega_{\Lambda,\gamma}, \beta)$ are summarized in Table\ \ref{tab-chi2min_H0=67.4_70}. 
When $H_{0}$ increases from $67.4$ to $70.0$, $\chi_{\textrm{SN,min}}^{2}$ increases from $1716.1$ to $1746.6$ [Table\ \ref{tab-chi2min_H0=67.4_70}].

As shown in Fig.\ \ref{Fig-L_SN_plane_H0=67.4_70.0}, the lines outside the contours of $L_{\textrm{SN}}$ for $H_{0}=67.4$ correspond to a smaller-$\chi_{\textrm{SN}}^{2}$ region $(\chi_{\textrm{SN}}^{2} \leq 1750)$ plotted from Fig.\ \ref{Fig-L_plane}.
For $H_{0}=70.0$, the smaller-$\chi_{\textrm{SN}}^{2}$ region is not plotted because the region is inside the contours of the $2 \sigma$ confidence level corresponding to $\chi_{\textrm{SN}}^{2} =1752.7$.
That is, the smaller-$\chi_{\textrm{SN}}^{2}$ region for $H_{0}=67.4$ approximately corresponds to the contours of the $2 \sigma$ confidence level for $H_{0}=70.0$.
Therefore, the tension for $H_{0}=67.4$ should be weak.
In addition, as $H_{0}$ increases from $67.4$ to $70.0$, the contours of $L_{\textrm{SN}}$ shift toward a high-$\Omega_{\Lambda,\gamma}$ and small-$\beta$ region.
Consequently, the location of $\chi_{\textrm{SN,min}}^{2}$ for $H_{0}=70.0$ is close to that of the $\Lambda$CDM model.
It is found that as $H_{0}$ increases, the tension is alleviated and the deviation decreases.
However, $\chi_{\textrm{SN,min}}^{2}$ increases with $H_{0}$ [Table\ \ref{tab-chi2min_H0=67.4_70}]. 
Therefore, the present model implies that $H_{0}=67.4$ is favored in comparison with $H_{0}=70.0$, when we consider the observational data for the supernovae at late times.
These results may provide new insights into the discussion of the $H_{0}$ tension, which is a significant tension between the early universe and the late universe \cite{Bernal2016}.
Further studies are needed and those tasks are left for future research.

\begin{table}[tb]
\caption{$\chi_{\textrm{SN,min}}^{2}$ and its location $(\Omega_{\Lambda,\gamma}, \beta)$ for $H_{0}=67.4$ and $70.0$.
In this analysis, $\Omega_{\Lambda,\gamma}$ and $\beta$ are sampled in steps of $0.005$.
For each location, $1\sigma$ ($2\sigma$) confidence levels are shown in increments of $0.005$.
Note that the contours of the $1 \sigma $, $2 \sigma$, and $3 \sigma$ confidence levels are plotted in Fig.\ \ref{Fig-L_SN_plane_H0=67.4_70.0}.
 }
\label{tab-chi2min_H0=67.4_70}
\newcommand{\m}{\hphantom{$-$}}
\newcommand{\cc}[1]{\multicolumn{1}{c}{#1}}
\renewcommand{\tabcolsep}{0.7pc} 
\renewcommand{\arraystretch}{1.9} 
\begin{tabular}{ccc}
\hline
\hline
  $H_{0}$      &  $\chi_{\textrm{SN,min}}^{2}$       &  $(\Omega_{\Lambda,\gamma}, \beta)$   \\
\hline
   $67.4$     &  $1716.1$                                   &  $(0.010^{+0.155 (0.285)}_{-0.010 (0.010)},   0.570^{+0.010 (0.015)}_{-0.050 (0.115)})$      \\ 
   $70.0$     &  $1746.6$                                   &  $(0.675^{+0.010 (0.015)}_{-0.035 (0.085)},   0.000^{+0.155 (0.295)}_{-0.000 (0.000)})$      \\ 
\hline
\hline
\end{tabular}\\
\end{table}
%

\end{document}